\documentclass[english,twocolumn, aps, prl, superscriptaddress,showpacs,floatfix]{revtex4}
\usepackage[T1]{fontenc}
\usepackage[latin9]{inputenc}
\usepackage{booktabs}
\usepackage{amsmath}
\usepackage{graphicx}
\usepackage{amssymb}

\makeatletter

\providecommand{\tabularnewline}{\\}

\@ifundefined{textcolor}{}
{%
 \definecolor{BLACK}{gray}{0}
 \definecolor{WHITE}{gray}{1}
 \definecolor{RED}{rgb}{1,0,0}
 \definecolor{GREEN}{rgb}{0,1,0}
 \definecolor{BLUE}{rgb}{0,0,1}
 \definecolor{CYAN}{cmyk}{1,0,0,0}
 \definecolor{MAGENTA}{cmyk}{0,1,0,0}
 \definecolor{YELLOW}{cmyk}{0,0,1,0}
 }

\makeatother

\usepackage{babel}

\begin{document}

\title{Synchronization of chaotic networks with time-delayed couplings:
An analytic study}

\author{A. Englert}

\affiliation{Institute for Theoretical Physics, University of Würzburg, 97074
Würzburg, Germany}

\author{S. Heiligenthal}

\affiliation{Institute for Theoretical Physics, University of Würzburg, 97074
Würzburg, Germany}

\author{W. Kinzel}

\affiliation{Institute for Theoretical Physics, University of Würzburg, 97074
Würzburg, Germany}

\author{I. Kanter}

\affiliation{Department of Physics, Bar-Ilan University, Ramat-Gan, 52900 Israel}
\begin{abstract}
Networks of nonlinear units with time-delayed couplings can synchronize
to a common chaotic trajectory. Although the delay time may be very
large, the units can synchronize completely without time shift. For
networks of coupled Bernoulli maps, analytic results are derived for
the stability of the chaotic synchronization manifold.

For a single delay time, chaos synchronization is related to the spectral
gap of the coupling matrix. For networks with multiple delay times,
analytic results are obtained from the theory of polynomials. Finally,
the analytic results are compared with networks of iterated tent maps
and Lang-Kobayashi equations which imitate the behaviour of networks
of semiconductor lasers.
\end{abstract}
\maketitle

\section{Introduction\label{sec:Introduction}}

Chaos synchronization is a phenomenon which is of fundamental scientific
interest in nonlinear dynamics and which is being investigated in
the context of secure communication and neural activity \cite{Pikovsky:Buch,BoccalettiKurths,BalanovJansonBook,MosekildeMaistrenkoBook,SchoellSchuster}.
In particular, networks of nonlinear units which relax to a common
chaotic trajectory are the focus of recent research \cite{Wu,Arenas}.

For many applications, the coupling between nonlinear dynamical units
is realized by transmitting a function of their internal variables
to their neighbours. In many cases, the transmission time is larger
than the internal time scales of the units. One example are chaotic
lasers which are coupled by their mutual laser beams \cite{SchoellSchuster,ColetRoy,UchidaRogister,VanWiggeren:1998,Argyris:2005}.
Thus, networks of nonlinear units which are coupled by their time-delayed
variables -- including time-delayed self-feedback -- are a subject
of recent research activities \cite{Wu,AtayBijikouglu,BaekOtt,BarahonaPecora,JostJoy,LiSunKurthsPRE,Pecora:1990,RomanBoccaletti,AtayJost,Choe,EarlStrogatz,hovel,KanterZigzagEnglertGeissler,KestlerKinzelKanter,KinzelEnglert,KozyreffVladimirov,MasollerMart,SchmitzerKinzelKanter,PecoraCarrollMSF,FlunkertYanchuk,KestlerKopelowitz}.

The theoretical investigations of chaotic networks are mainly based
on numerical simulations. However, there exists a powerful method
to determine the stability of the synchronization manifold (SM) of
an \emph{arbitrary} network: The master stability function (MSF) \cite{PecoraCarrollMSF}.
This method connects topology with function. Using the eigenvalues
of the connection matrix, a linear equation for a \emph{single} unit
is derived which determines the stability of chaos synchronization
for the \emph{complete} network. For the case of time-delayed couplings,
the MSF has been studied for few systems, only \cite{DhamalaJirsa,KinzelEnglert,Choe}.

The MSF is defined as the maximal Lyapunov exponent of linear equations
with time-dependent coefficients. In general it is not possible to
derive analytic results for these equations because the coefficients
are given by the chaotic trajectory of the network. Thus, one has
to rely on numerical simulations of the linear system of differential
equations (for chaotic flows) or difference equations (for coupled
map lattices).

The purpose of this paper is to derive analytic results for chaotic
networks with time-delayed couplings. Thus, we concentrate on a coupled
map lattice, a network of chaotic iterated Bernoulli maps, which allows
an analytic calculation of the MSF. The corresponding linear difference
equations have constant coefficients. Therefore, stability of the
solution is related to the roots of polynomials. Results from graph
theory, control theory and algebra will help to derive analytic statements
about chaos synchronization.

The linear stability equations of Bernoulli networks may be considered
as an approximation for linear equations with time-dependent couplings.
Thus, we compare our analytic results with numerical results for other
iterated maps and with simulations of rate equations for coupled semiconductor
lasers. In the following section the MSF for chaotic networks with
time-delayed couplings is introduced. We concentrate on networks where
chaos is generated by the couplings and/or self-feedback, similar
to semiconductor lasers. Afterwards, analytic results are derived
for Bernoulli networks with single and multiple delay times. The last
section compares these results with networks of iterated tent maps
and of laser rate equations.

\section{Master stability function (MSF)\label{sec:Master-stability-function}}

\subsection{MSF with time delay}

One powerful method to analyze the stability of synchronization in
networks of coupled systems with identical units is the master stability
function (MSF) proposed by Pecora and Caroll \cite{PecoraCarrollMSF}.
A network of coupled identical dynamical units can be analyzed by
linearizing the dynamical equation around the synchronization manifold.

In order to obtain analytic results, we restrict our investigation
to coupled map lattices, and we use an identical function $f(x)$,
$x\in[0,1]$ for the internal dynamic, the self-feedback and the couplings.
We extend the master stability function of \cite{PecoraCarrollMSF}
for a system without time delay to a system with arbitrarily many
different time delays \cite{MasollerMart,ShahverdievShore,JustReibold,ZigzagButkowskiEnglert}
where each term with delay time $\tau_{l}$ is weighted with a positive
coupling parameter $\sigma_{l}$ and a positive self-feedback parameter
$\eta_{l}$. The network consists of $N$ units with variables $x_{t}^{i}\in[0,1]$,
where $i=1,...,N$ is the index of the unit and $t$ is a discrete
time step. The system is defined by\begin{align}
x_{t}^{i} & =\eta_{0}\, f\!\left(x_{t-1}^{i}\right)+\sum_{l=1}^{M}\eta_{l}\, f\!\left(x_{t-\tau_{l}}^{i}\right)+\nonumber \\
 & \quad+\sum_{l=1}^{M}\sum_{j=1}^{N}\sigma_{l}\, G_{l,ij}\, f\!\left(x_{t-\tau_{l}}^{j}\right)\label{eq:MapDynamicSingleUnit}\end{align}
Without loss of generality we order the coupling terms with ascending
delay times, so that $\tau_{M}$ is the maximum time delay. Each coupling
delay time $\tau_{l}$ has its own coupling matrix, the normalized
weighted adjacency matrix $G_{l}$ with $\sum_{j}G_{l,ij}=1$ and
$G_{ij}\ge0$ \cite{KestlerKinzelKanter,KestlerKopelowitz}. Therefore,
the coupling is invasive and non-diffusive, it changes the trajectory
of the coupled system in comparison to a non-coupled system. We assume
that the coupling matrices $G_{l}$ commute, otherwise the MSF method
cannot be applied. The self-feedback of the system is not included
in $G_{l}$, hence $G_{l,ii}=0$.

Complete zero-lag synchronization $x_{t}^{1}=...=x_{t}^{N}=s_{t}$
is a solution of these equations. The synchronized trajectory is given
by\begin{equation}
s_{t}=\eta_{0}\, f\!\left(s_{t-1}\right)+\sum_{l=1}^{M}\left(\eta_{l}+\sigma_{l}\right)f\!\left(s_{t-\tau_{l}}\right)\label{eq:MapSM}\end{equation}

The stability of the SM is determined by linearizing Eq.~(\ref{eq:MapDynamicSingleUnit})
in the vicinity of the SM, Eq.~(\ref{eq:MapSM}). With $\delta x_{t}^{i}=x_{t}^{i}-s_{t}$
we obtain\begin{align}
\delta x_{t}^{i} & =\eta_{0}\, f'\!\left(s_{t-1}\right)\delta x_{t-1}^{i}+\sum_{l=1}^{M}\eta_{l}\, f'\!\left(s_{t-\tau_{l}}\right)\delta x_{t-\tau_{l}}^{i}+\nonumber \\
 & \quad+\sum_{l=1}^{M}\sum_{j=1}^{N}\sigma_{l}\, G_{l,ij}\, f'\!\left(s_{t-\tau_{l}}\right)\delta x_{t-\tau_{l}}^{j}\label{eq:MSFLinearizedDiscrete}\end{align}
For a network with $N$ nodes we obtain $N$ coupled linear equations
with time-dependent coefficients. Since the coupling matrices $G_{l}$
commute, we can expand the $N$-dimensional perturbation $\delta\vec{x}_{t}$
to the common eigenvectors $\vec{w}_{k}$ with eigenvalues $\gamma_{l,k}$,
$k=1,...,N$, $l=1,...,M$ of the matrices $G_{l}$. For each mode
$k$ of the perturbation, with $\delta\vec{x}_{t}=\xi_{k,t}\,\vec{w}_{k}$,
we obtain\begin{equation}
\xi_{k,t}=\sum_{l=0}^{M}\left(\eta_{l}+\sigma_{l}\,\gamma_{k,l}\right)f'\!\left(s_{t-\tau_{l}}\right)\xi_{k,t-\tau_{l}}\label{eq:MSFDiscrete}\end{equation}
where we have defined $\sigma_{0}=0$ and $\tau_{0}=1$. $\xi_{k,t}$
is the amplitude of the perturbation corresponding to the eigenvalue
$\gamma_{k,l}$ of the coupling matrix $G_{l}$.

To gain analytic results we focus on a chaotic map with constant slope,
namely the Bernoulli map which is given by\begin{equation}
f(x)=(\alpha\, x)\:\mathrm{mod}\:1\label{eq:BernoulliMap}\end{equation}
and is chaotic for $\alpha>1$. Since $f'(s_{t})=\alpha$ is constant,
Eq.~(\ref{eq:MSFDiscrete}) becomes \begin{equation}
\xi_{k,t}=\sum_{l=0}^{M}\left(\eta_{l}+\sigma_{l}\,\gamma_{k,l}\right)\alpha\,\xi_{k,t-\tau_{l}}\label{eq:MSFBernoulli}\end{equation}
With the ansatz $\xi_{k,t}=z_{k}^{t}\,\xi_{k,0}$, the whole stability
problem becomes a problem of solving the polynomial of degree $\tau_{M}$\begin{equation}
z_{k}^{\tau_{M}}=\sum_{l=0}^{M}\left(\eta_{l}+\sigma_{l}\,\gamma_{k,l}\right)\alpha\, z_{k}^{\tau_{M}-\tau_{l}}=\sum_{l=0}^{M}\beta_{k,l}\, z_{k}^{\tau_{M}-\tau_{l}}\label{eq:PolynomialFirstMentioned}\end{equation}
with $\beta_{k,l}=\left(\eta_{l}+\sigma_{l}\,\gamma_{k,l}\right)\alpha$.

For each eigenvalue $\gamma_{k}$, Eq.~(\ref{eq:PolynomialFirstMentioned})
yields $\tau_{M}$ roots which we label with $z_{k,r}$ with $r=1,...,\tau_{M}$.
Our goal is to find coupling parameters $\beta_{k,l}$ such that:
\begin{enumerate}
\item For $\gamma_{0,l}=1$, $l=1,...,M$, there exists at least one $z_{0,r_{m}}$
and $\left|z_{0,r_{m}}\right|>1$. This guarantees a chaotic dynamic
of the SM. 
\item For each $\gamma_{k}\left(k>1\right)$ all roots $z_{k,r}$ lie inside
the unit circle $\left|z_{k,r}\right|<1$. This guarantees a stable
SM. In this case, the MSF is defined as\begin{equation}
\lambda=\max_{k>0,r}\ln\!\left|z_{k,r}\right|\label{eq:MSFMap}\end{equation}

\end{enumerate}
Eq.~(\ref{eq:PolynomialFirstMentioned}) allows an analytic investigation
of chaos synchronization. Later we will discuss to which extend Eq.~(\ref{eq:PolynomialFirstMentioned})
is a good approximation for other iterated maps with time-dependent
slopes $f'$ and even for corresponding differential equations with
time-dependent Jacobi matrices.

\subsection{Concept of local Lyapunov exponents\label{sub:Concept-of-local}}

The stability of the SM of the complete network is determined by the
linear equation (\ref{eq:MSFDiscrete}). This equation yields the
Lyapunov exponents of the network parallel ($\gamma_{0}=1$) and perpendicular
($\gamma_{k}$, $k>0$) to the SM. As we will see later, it turns
out that it is useful to consider the contributions of individual
terms in Eq.~(\ref{eq:MSFDiscrete}) to the stability of the SM separately.
Hence, we define {}``local Lyapunov exponents'' to discuss these
contributions. For example, the local Lyapunov exponent is defined
as the maximal one of the equation\begin{equation}
\xi_{k,t}=\eta_{0}\, f'\!\left(s_{t-1}\right)\xi_{k,t-1}\label{eq:LocalCont}\end{equation}
 where $s_{t}$ is the trajectory of the complete network, including
time-delayed terms.

For the case of large delay times $\tau_{l}$ one finds the following
result: When the instantaneous Lyapunov exponent is positive, the
network cannot synchronize \cite{KinzelEnglert}. Note that local
Lyapunov exponents are not the Lyapunov exponents of the isolated
units, since the network changes the trajectory. Only for the Bernoulli
network the linear equations do not depend on the trajectory, in this
case the local Lyapunov exponents are identical to the ones of the
corresponding isolated units.

\subsection{Graph spectrum}

The stability of the synchronization manifold (SM) is determined by
the eigenvalues of the coupling matrices $G_{l}$, according to the
MSF Eq.~(\ref{eq:MSFMap}). We consider matrices with unit row sum,
$\sum_{j}G_{l,ij}=1$. Since we consider self-feedback separately,
we have $G_{l,ii}=0$.

We restrict our discussion to non-negative matrices \cite{BermanPlemmons},
$G_{l,ij}\ge0$, and to completely connected graphs. One eigenvalue
is unity, $\gamma_{0,l}=1$, and according to the Perron-Frobenius
theorem $\gamma_{0,l}$ is not degenerate and has the largest modulus
of all eigenvalues of $G_{l}$. Therefore, we order the eigenvalues
such that $1=\gamma_{0,l}\ge\left|\gamma_{1,l}\right|\ge\left|\gamma_{2,l}\right|\ge...\ge\left|\gamma_{N-1,l}\right|$.

The eigenvector for the largest eigenvalue $\gamma_{0}=1$ is $\vec{w}_{0}=(1,...,1)$.
It corresponds to a perturbation parallel to the SM. Since we discuss
only chaotic networks, the Lyapunov exponent $\lambda_{\mathrm{max}}$
of the mode $\gamma_{0}$ is positive; the dynamics in the SM, Eq.~(\ref{eq:MapSM})
is chaotic. As we will show later, for large single delay time $\tau$
the spectral gap $\Delta=1-\left|\gamma_{1}\right|$ determines the
stability of the SM. Complete zero-lag synchronization is possible
in the limit of weak chaos, $\lambda_{\mathrm{max}}\to0$, if and
only if the spectral gap is nonzero, $\left|\gamma_{1}\right|<1$.

The theory of nonnegative matrices \cite{BermanPlemmons} relates
the eigenvalue gap to the loop structure of the corresponding graph:
$\Delta$ is nonzero if and only if the greatest common divisor of
the length of loops of the graph is unity. Hence, for a single large
time delay, it is easy to see whether a network can synchronize in
the limit of weak chaos. In fact, this result has been extended to
networks with multiple delay times $\tau_{l}$ by a self-consistent
physical argument based on mixing information of the chaotic trajectories
of the nodes \cite{KanterZigzagEnglertGeissler}.

\begin{figure}
\hfill{}\includegraphics[width=4cm]{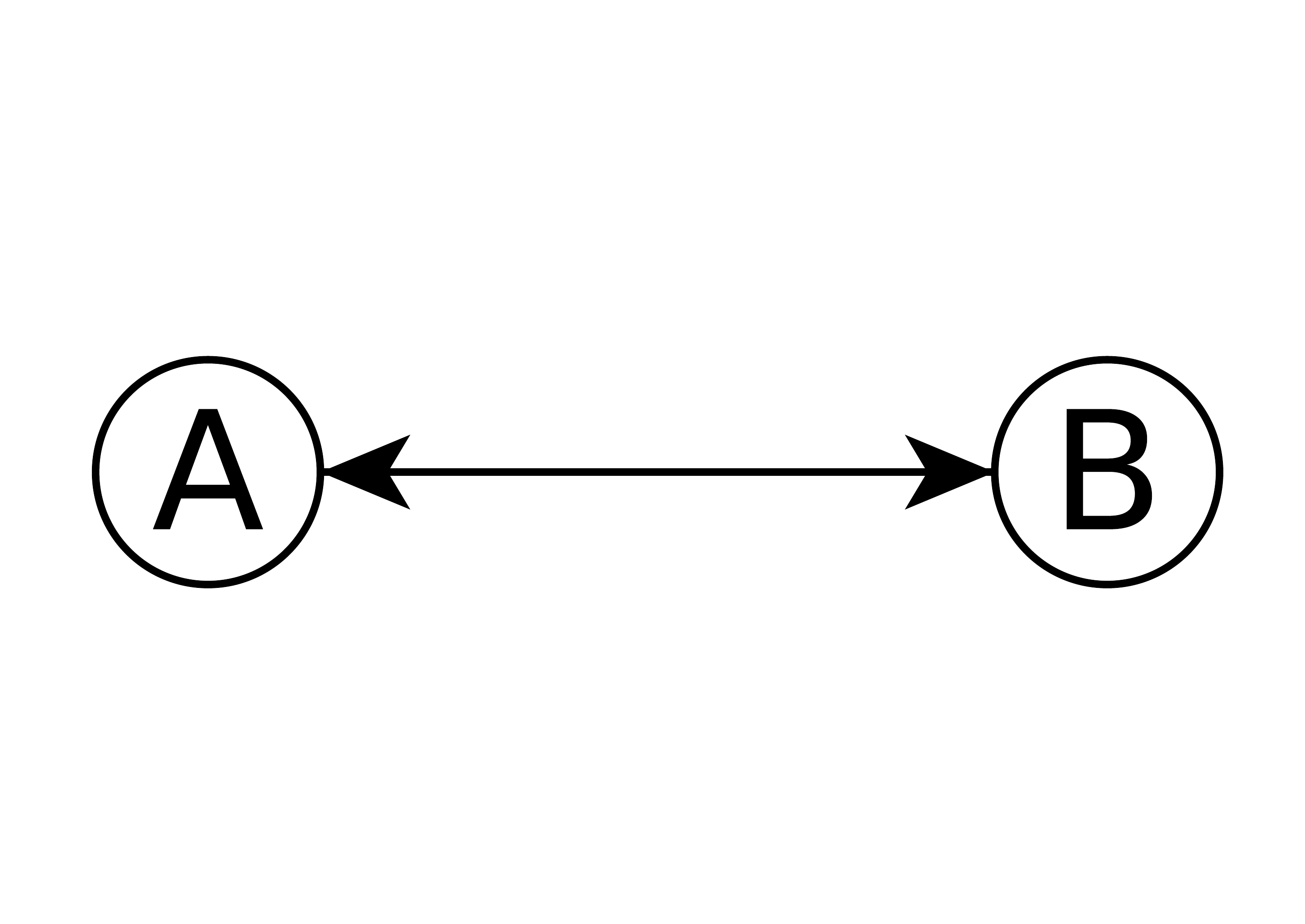}\hfill{}\includegraphics[width=4cm]{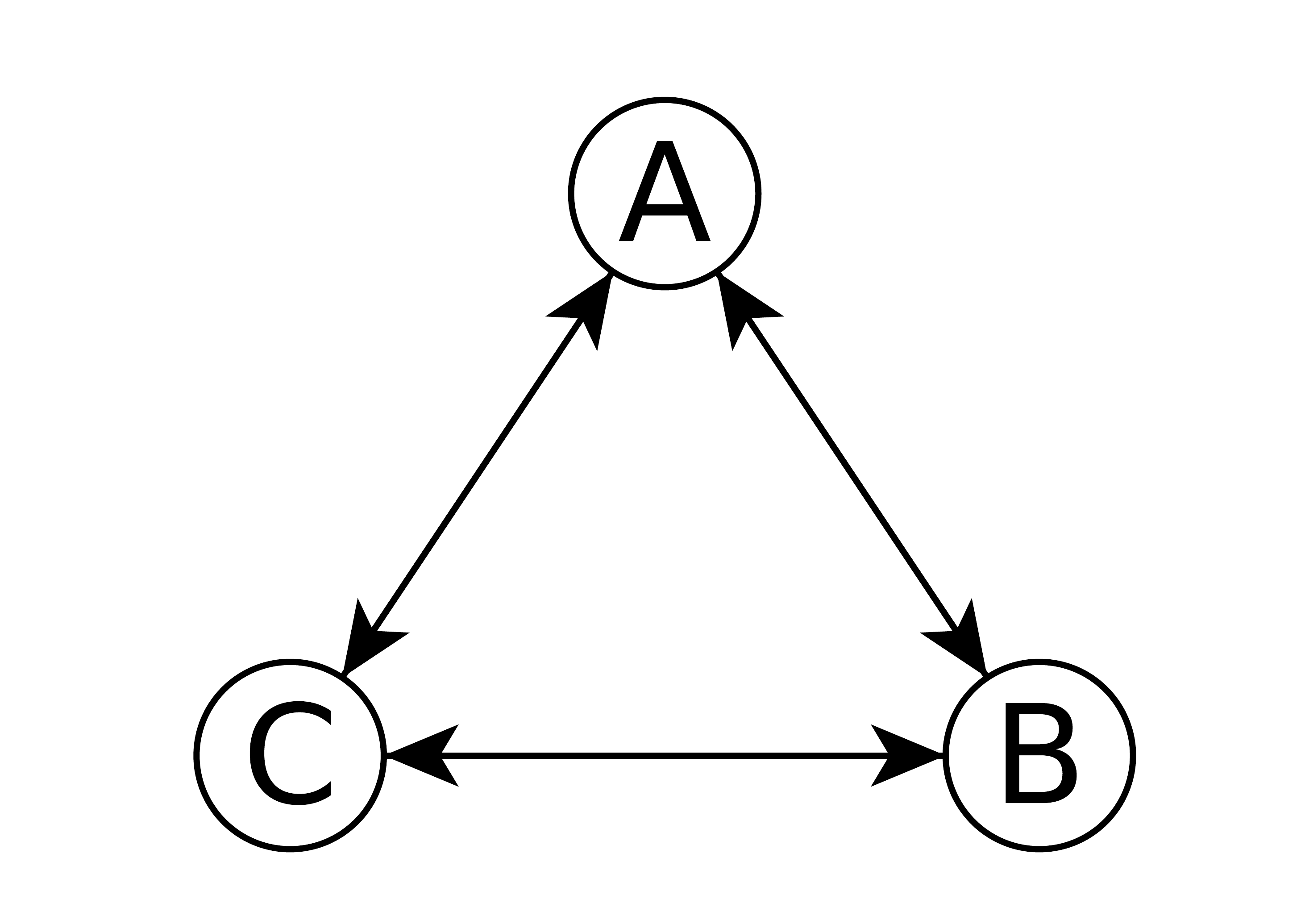}\hfill{}

\hfill{}\includegraphics[width=4cm]{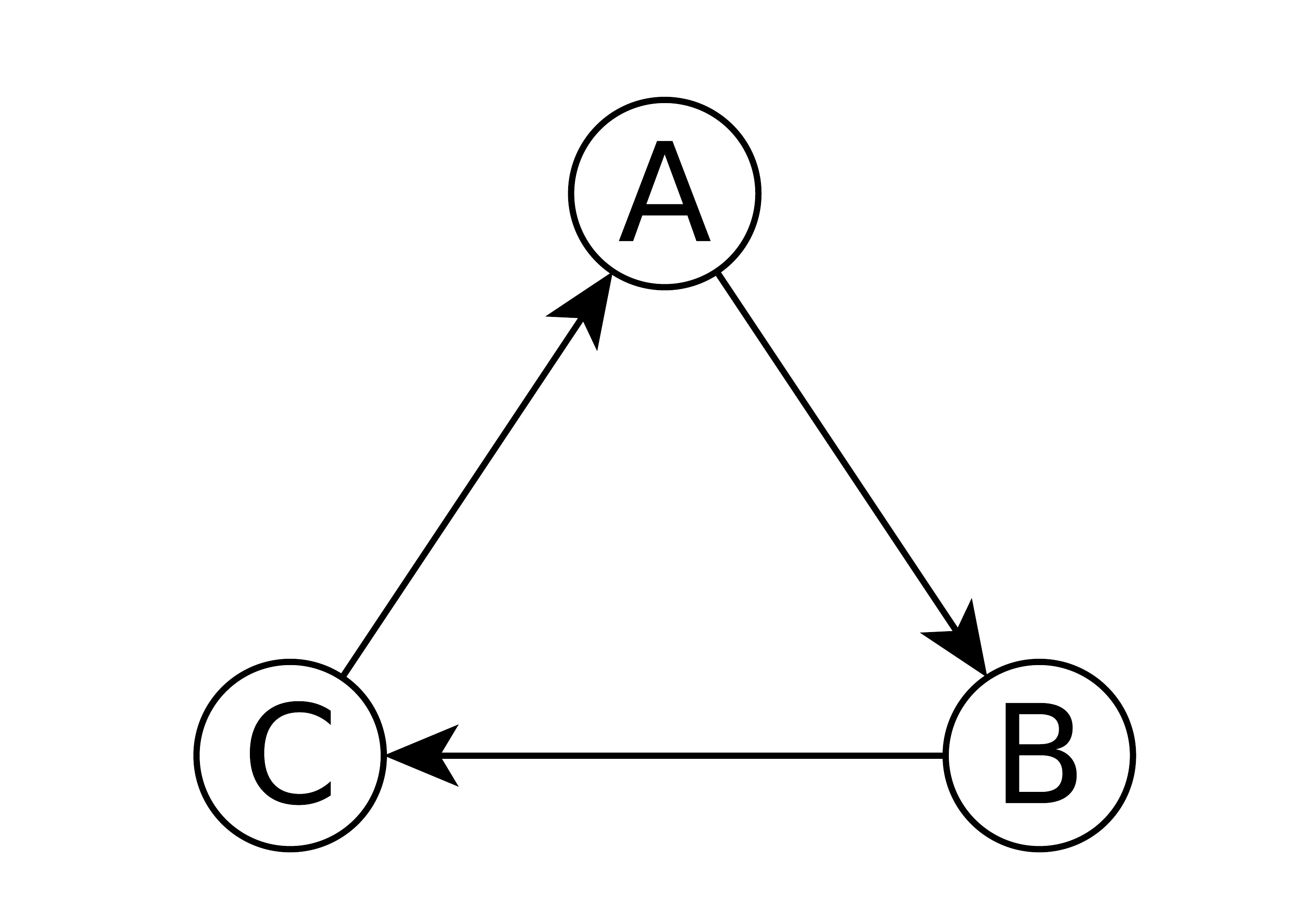}\hfill{}\includegraphics[width=4cm]{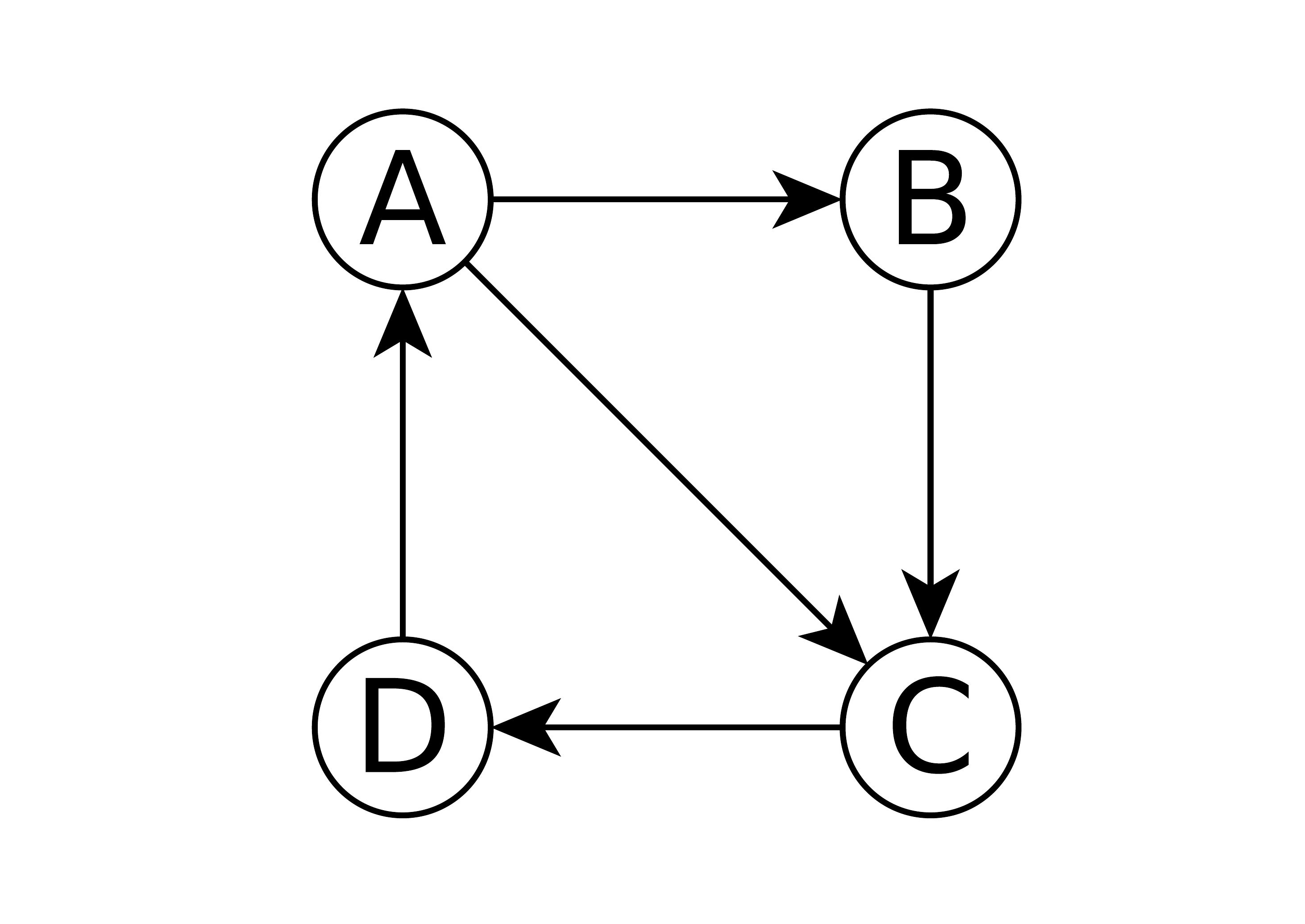}\hfill{}

\hfill{}\includegraphics[width=8cm]{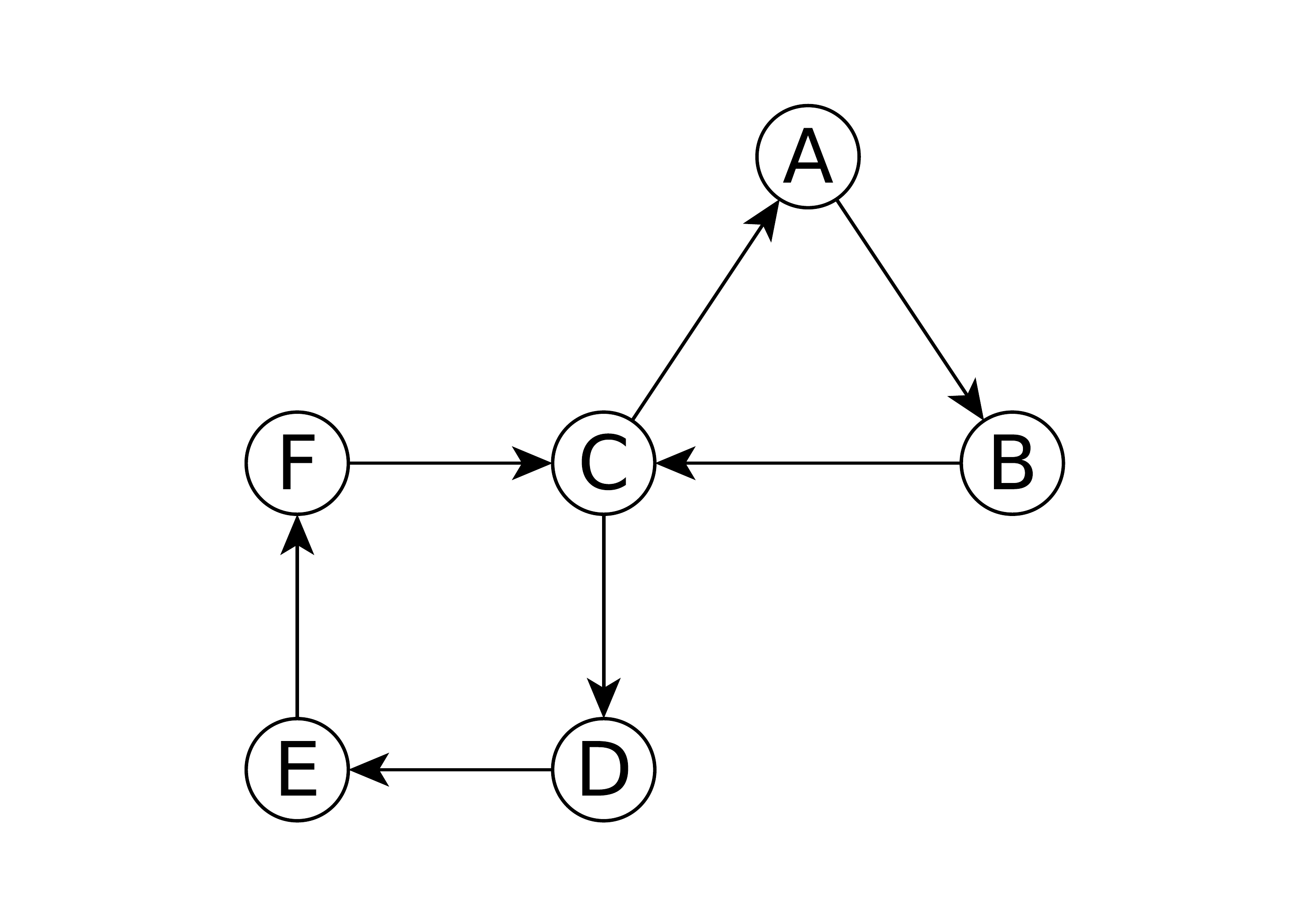}\hfill{}

\caption{Schemata of bidirectionally coupled pair, spectral gap $\Delta=0$,
bidirectionally coupled triangle, $\Delta=\frac{1}{2}$, unidirectional
triangle, $\Delta=0$, square, $\Delta=0.1215$ and combination of
triangle and square, $\Delta=0.1215$. These are the values for equal
coupling weights.\label{fig:Schemata-of-bidirectionally}}

\end{figure}

For some single graphs the eigenvalues are known analytically. Fig.~\ref{fig:Schemata-of-bidirectionally}
shows the eigenvalue gap for some graphs. Note that a directed triangle
and a directed square have zero gaps, but if one connects them, the
greatest common divisor of the loops 3 and 4 is unity and the gap
is nonzero \cite{BermanPlemmons}.

\newpage{}

\section{Analytic results\label{sec:Analytical-results}}

In this section, we present analytic results for Bernoulli networks.
We focus on complete zero-lag synchronization. According to the previous
section, the stability of the synchronization manifold (SM) is determined
by the roots of the following polynomials\begin{equation}
z^{\tau_{M}}=\sum_{l=0}^{M}\left(\eta_{l}+\sigma_{l}\,\gamma_{k,l}\right)\alpha\, z^{\tau_{M}-\tau_{l}}=\sum_{l=0}^{M}\beta_{l}\, z^{\tau_{M}-\tau_{l}}\label{eq:AllgemeinesPolynom}\end{equation}
Note that each $\beta_{l}$ includes $\gamma_{k,l}$, we omitted the
index $k$ in our notation.

The network contains $M+1$ different delay times $\tau_{l}$, and
$\tau_{M}$ is the largest one. For each mode $k$ of perturbation
we have a set of eigenvalues $\gamma_{k,l}$ of the coupling matrices
$G_{l}$ corresponding to the delay times $\tau_{l}$. Hence, we have
to calculate the roots of Eq.~(\ref{eq:AllgemeinesPolynom}) with
$\beta_{l}=\left(\eta_{l}+\sigma_{l}\,\gamma_{k,l}\right)\alpha$
for each mode of perturbation where $\sigma_{l}$ is the coupling
strength and $\alpha$ is the slope of the Bernoulli map. Eq.~(\ref{eq:AllgemeinesPolynom})
includes the local dynamics with $\tau_{0}=1$ and $\sigma_{0}=0$
and self-feedbacks with delay time $\tau_{l}$ and strength $\eta_{l}$.
The mode parallel to the SM has eigenvalues $\gamma_{0,l}=1$ for
all terms of Eq.~(\ref{eq:AllgemeinesPolynom}).

Some results can be derived immediately. The polynomial Eq.~(\ref{eq:AllgemeinesPolynom})
has $\tau_{M}$ roots $z_{r}$ and the theorem of Vieta gives $\prod_{r}\left|z_{r}\right|=\left|\beta_{M}\right|$.
Hence, for $\left|\beta_{M}\right|>1$ at least one root is outside
of the unit circle. Consequently, all modes $k$ with $\left|\left(\eta_{M}+\sigma_{M}\,\gamma_{k,M}\right)\alpha\right|>1$
are unstable. If the coupling $\sigma_{M}$ and the self-feedback
$\eta_{M}$ of the largest delay time $\tau_{M}$ are such that there
exists one eigenvalue $\gamma_{k,M}$ with $k>0$ and $\left|\eta_{M}+\sigma_{M}\,\gamma_{k,M}\right|>1/\alpha$,
the network cannot synchronize, the SM is unstable.

Eq.~(\ref{eq:AllgemeinesPolynom}) denotes the roots of a polynomial
$P(z)$ which can be written as\begin{equation}
P(z)=z^{\tau_{M}}-\sum_{l=0}^{M}\left(\eta_{l}+\sigma_{l}\,\gamma_{k,l}\right)\alpha\, z^{\tau_{M}-\tau_{l}}\label{eq:ChaoticTraj1}\end{equation}
If the mode $k$ is unstable, $P(z)$ has at least one root with $|z|>1$.
Let us assume $\sum_{l=0}^{M}\left(\eta_{l}+\sigma_{l}\,\gamma_{k,l}\right)\alpha>1$,
which gives\begin{equation}
P(1)=1-\sum_{l=0}^{M}\left(\eta_{l}+\sigma_{l}\,\gamma_{k,l}\right)\alpha<0\label{eq:ChaoticTraj2}\end{equation}
Furthermore, on the real axis we have\begin{equation}
\lim_{z\to\infty}P(z)\to\infty\label{eq:ChaoticTraj3}\end{equation}
Since $P(z)$ is continuous we can conclude that $P(z)$ has a root
on the real axis with $z_{0}>1$. Hence, the mode $k$ is unstable
if\begin{equation}
\sum_{l=0}^{M}\left(\eta_{l}+\sigma_{l}\,\gamma_{k,l}\right)\alpha>1\label{eq:ChaoticTraj4}\end{equation}
The Gershgorin's circle theorem \cite{Gershgorin} states that all
roots of $P(z)$ lie inside a circle with radius $R$ given by the
inequality\begin{equation}
R\le\max\!\left\{ 1,\sum_{l=0}^{M}\left|\left(\eta_{l}+\sigma_{l}\,\gamma_{k,l}\right)\alpha\right|\right\} \label{eq:Gershgorin1}\end{equation}
Hence, the mode $k$ is stable if\begin{equation}
\sum_{l=0}^{M}\left|\left(\eta_{l}+\sigma_{l}\,\gamma_{k,l}\right)\alpha\right|<1\label{eq:ChaoticTraj5}\end{equation}
Eqs.~(\ref{eq:ChaoticTraj4}) and (\ref{eq:ChaoticTraj5}) give the
parameter regime for chaos. Note that this is valid for all perturbation
modes $k$.

For a chaotic trajectory the perturbation mode parallel to the synchronization
manifold, $\gamma_{0,l}=1$ has to be unstable. This is the case for\begin{equation}
\sum_{l=0}^{M}\left(\eta_{l}+\sigma_{l}\right)\alpha>1\label{eq:BorderChaos1}\end{equation}
whereas for\begin{equation}
\sum_{l=0}^{M}\left(\eta_{l}+\sigma_{l}\right)\alpha<1\label{eq:BorderChaos2}\end{equation}
the perturbation mode for $k=0$ is stable. Hence, the border to a
chaotic trajectory is given by\begin{equation}
1=\sum_{l=0}^{M}\beta_{l}=\sum_{l=0}^{M}\left(\eta_{l}+\sigma_{l}\right)\alpha\label{eq:borderChaos}\end{equation}
If the sum of couplings is larger than 1 the system is chaotic. Note
that this transition to chaos does not depend on the delay times $\tau_{l}$
in contrast to the region of synchronization which is very sensitive
to the values of the delay times $\tau_{l}$. In this work we consider
only parameters where the network is chaotic. Another conclusion from
this result is the fact that for some networks it is always possible
to find coupling parameters and Bernoulli slopes for which stable
chaos synchronization is possible. This networks have to have eigenvalues
$k>0$ which fulfill Eq.~(\ref{eq:ChaoticTraj5}) for parameters
which ensure \ref{eq:BorderChaos1}. This is the case for $\left|\gamma_{k,l}\right|<1$
where $k>0$.

\subsection{Single delay time\label{sub:Single-time-delay}}

\begin{figure}
\hfill{}\includegraphics[width=4cm]{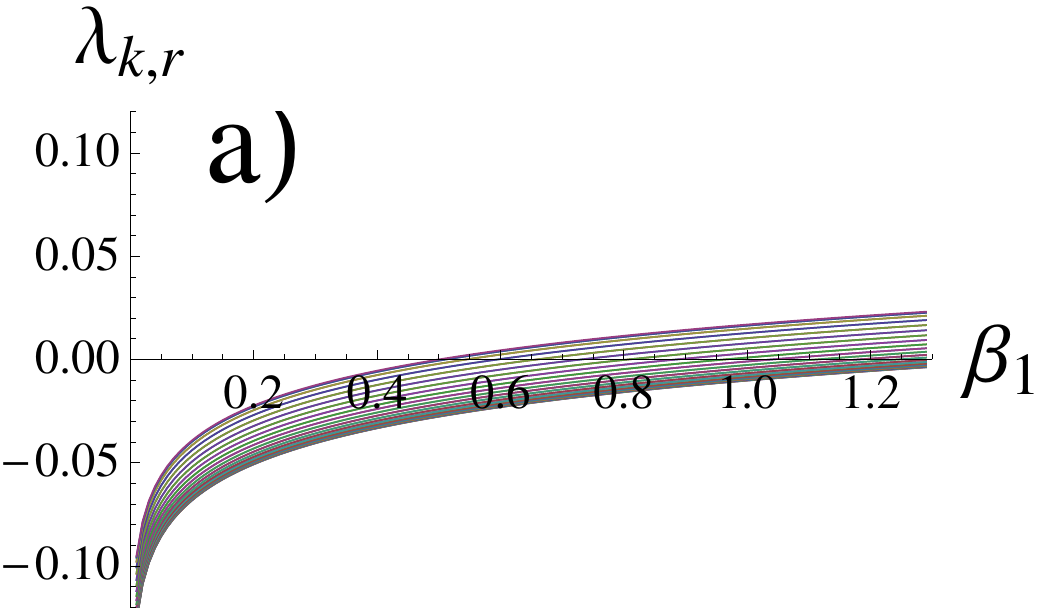}\hfill{}\includegraphics[width=4cm]{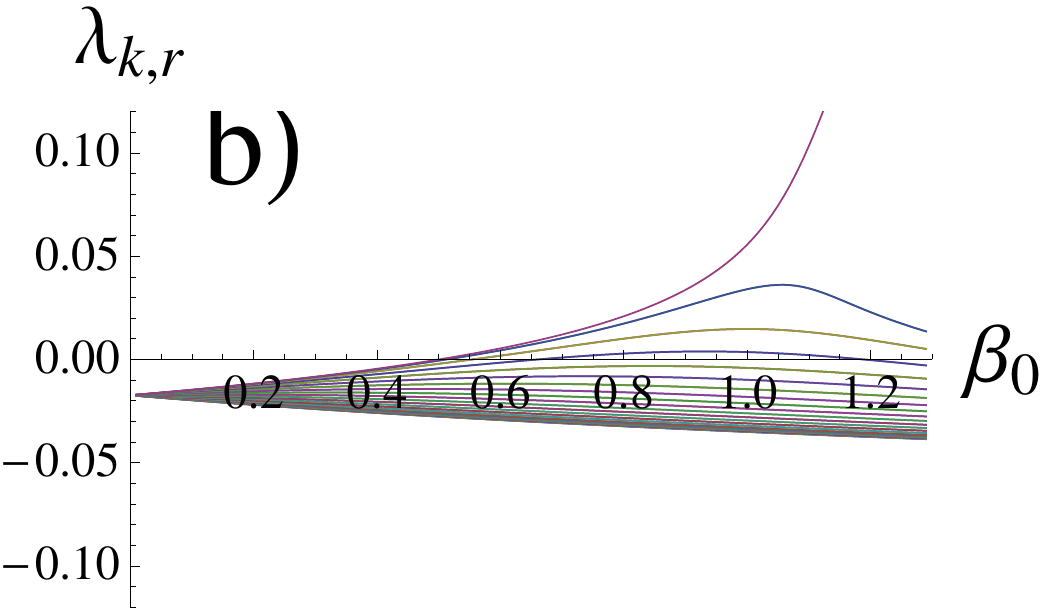}\hfill{}

\caption{Spectrum of Lyapunov exponents for the system of Eq.~(\ref{eq:OneDelayPoly})
and $\tau=40$. a) $\beta_{0}=0.5$, b) $\beta_{1}=0.5$. There is
at least one positive Lyapunov exponent for $\beta_{0}+\beta_{1}\ge1$\label{fig:Spectrum-of-Lyapunov}}

\end{figure}

\begin{figure}
\hfill{}\includegraphics[width=6cm]{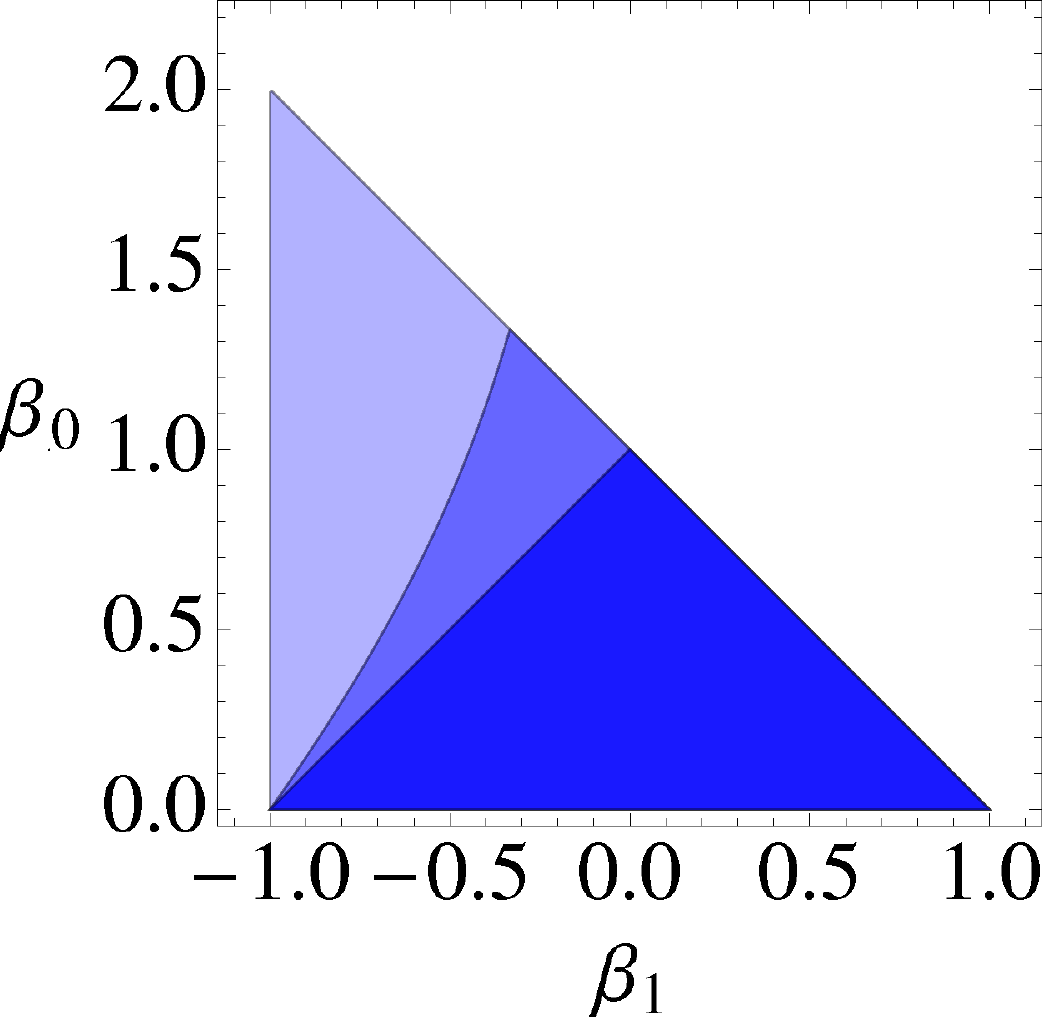}\hfill{}

\caption{Synchronization regime for $z^{\tau}=\beta_{0}\, z^{\tau-1}+\beta_{1}$
and $\tau=2$ (light blue), $\tau=4$ (blue), $\tau\to\infty$ (dark
blue). Note that $\tau\to\infty$ regime is a subset of $\tau=2$
and $\tau=4$ regime and that $\tau=4$ regime is a subset of $\tau=2$
regime.\label{fig:SynForDifferentSmallTaus}}

\end{figure}

For a network with a single delay time $\tau$, including coupling
and self-feedback we have to find the roots of\begin{equation}
z^{\tau}=\beta_{0}\, z^{\tau-1}+\beta_{1}=\eta_{0}\,\alpha\, z^{\tau-1}+\left(\eta_{1}+\sigma_{1}\,\gamma_{k}\right)\alpha\label{eq:OneDelayPoly}\end{equation}
in the chaotic region\begin{equation}
\left(\eta_{0}+\eta_{1}+\sigma_{1}\right)\alpha>1\label{eq:OneDelayChaoticRegion}\end{equation}
For each mode $k$, Eq.~(\ref{eq:OneDelayPoly}) has $\tau$ roots
$z_{k,r}$ which define a spectrum of $\tau$ Lyapunov exponents $\lambda_{k,r}=\ln\!\left|z_{k,r}\right|$.
Fig.~\ref{fig:Spectrum-of-Lyapunov} shows this spectrum as a function
of the parameters $\beta_{0}$ and $\beta_{1}$. For $\beta_{0}+\beta_{1}>0$
the maximal Lyapunov exponent is positive. Using the Schur-Cohn theorem
\cite{Schur} the region of stability can be calculated numerically.
Fig.~\ref{fig:SynForDifferentSmallTaus} shows the result for $\tau=2,4$
and $\tau\to\infty$. With increasing delay time this region shrinks
to the symmetric triangle $\left|\beta_{1}\right|<1-\beta_{0}$.

In fact, one can derive the region of stability analytically for $\tau\to\infty$.
In this limit, we have to consider two cases: a) the Lyapunov exponent
is of order 1, b) it is of order $1/\tau$ \cite{LepriGiacomelli,FarmerPhysicaD}.
We omit the index $k$ of the root $z_{k,r}$ of the perturbation
mode $k$ since $\gamma_{k}$ is included in $\beta_{1}$. From Eq.~(\ref{eq:OneDelayPoly})
we obtain\begin{equation}
\left|z_{r}\right|^{\tau}=\left|\frac{\beta_{1}\, z_{r}}{z_{r}-\beta_{0}}\right|\label{eq:4x}\end{equation}
For case a) $\left|z_{r}\right|^{\tau}$ diverges for $\tau\to\infty$
in the region of instability, hence we find $z_{r}=\beta_{0}$ for
$\tau\to\infty$. If $\beta_{0}>1$, the SM is unstable for any perturbation
mode with eigenvalue $\gamma_{k}$. But $\beta_{0}>1$ means that
the uncoupled units are chaotic. Thus, we reproduce the result already
found in \cite{KinzelEnglert}: If the local Lyapunov exponent defined
in the previous section is positive, a network cannot be synchronized
by time-delayed couplings if the delay time is much larger than the
local time scales.

For case b) we write $z_{r}=\mathrm{e}^{\lambda_{r}}\,\mathrm{e}^{\mathrm{i}\,\phi_{r}}$
with $\lambda_{r}=\Lambda_{r}/\tau$. Eq.~(\ref{eq:4x}) gives\begin{equation}
\mathrm{e}^{\Lambda_{r}}=\frac{\left|\beta_{1}\right|\mathrm{e}^{\Lambda_{r}/\tau}}{\left|\mathrm{e}^{\Lambda_{r}/\tau}\,\mathrm{\mathrm{e}}^{\mathrm{i}\,\phi_{r}}-\beta_{0}\right|}\label{eq:5x}\end{equation}
For $\tau\to\infty$ we obtain\begin{equation}
\mathrm{e}^{2\,\Lambda_{r}}=\frac{\left|\beta_{1}\right|^{2}}{\beta_{0}^{2}-2\,\beta_{0}\cos\phi_{r}+1}\label{eq:6x}\end{equation}
For $\tau\to\infty$ the phases $\phi_{r}$ are uniformly distributed
on the circle $[0,2\pi]$\cite{ErdoesTuran,Granville}, hence the
maximal Lyapunov exponent is given for $\phi_{r}=0$ which yields
the region of stability\begin{equation}
\left|\beta_{1}\right|<1-\beta_{0}\label{eq:7x}\end{equation}
Consequently, the network can synchronize if for all $k>0$\begin{equation}
\left|\left(\eta_{1}+\sigma_{1}\,\gamma_{k}\right)\alpha\right|<1-\eta_{0}\,\alpha\label{eq:8x}\end{equation}
For large delay times and without self-feedback $\eta_{1}=0$, Eq.~(\ref{eq:8x})
reduces to\begin{equation}
\left|\sigma_{1}\,\gamma_{1}\,\alpha\right|<1-\eta_{0}\,\alpha\label{eq:8xaa}\end{equation}
where $\gamma_{1}$ is the eigenvalue with the second largest absolute
value, $1\ge\left|\gamma_{1}\right|\ge\left|\gamma_{2}\right|...$
Neither the sign nor the complex phase of the eigenvalue $\gamma_{1}$
have an influence on the region of stability. This is different for
small values of $\tau$ where the region of stability depends on the
complex phase of $\gamma_{1}$. Note that this symmetry for large
delay times has recently been proven for general chaotic networks
with a continuous dynamics \cite{FlunkertYanchuk}. Eq.~(\ref{eq:8xaa})
has interesting consequences. The network is chaotic if\begin{equation}
\sigma_{1}\,\alpha>1-\eta_{0}\,\alpha\label{eq:8xa}\end{equation}
 It can synchronize if\begin{equation}
\sigma_{0}\,\alpha+\left|\sigma_{1}\,\gamma_{1}\,\alpha\right|<1\label{eq:8xb}\end{equation}
Thus, if $\left|\gamma_{1}\right|=1$, Eq.~(\ref{eq:8xaa}) is identical
to Eq.~(\ref{eq:8xa}) and the chaotic network cannot synchronize
for any set of parameters. For bipartite networks one finds $\gamma_{1}=-1$,
hence bi-partite networks cannot synchronize completely, only sublattice
or cluster synchronization is possible \cite{KestlerKopelowitz,KestlerKinzelKanter}.
According to Fig.~\ref{fig:Schemata-of-bidirectionally}, a pair,
a square without diagonals or any directed ring cannot synchronize
completely for large delay times. A sublattice or cluster synchronization
however, is possible \cite{KestlerKinzelKanter,KestlerKopelowitz,KanterZigzagEnglertGeissler}.
On the other hand, a triangle with bidirectional couplings can synchronize
completely, since $\gamma_{1}=-\frac{1}{2}$. 

We exemplify this result by a system consisting of two bidirectionally
coupled Bernoulli units with one time delay $\tau$ for the coupling
and the same time delay $\tau$ for self-feedback. By adding a self-feedback
with delay time $\tau$, one can achieve complete synchronization
in a bidirectionally coupled pair. The dynamical equation of unit
$i$ is defined as\begin{equation}
x_{t}^{i}=(1-\varepsilon)\, f\!\left(x_{t}^{i}\right)+\varepsilon\,\kappa\, f\!\left(x_{t-\tau}^{i}\right)+\varepsilon\,(1-\kappa)\, f\!\left(x_{t-\tau}^{j}\right)\label{eq:epsilonKappa}\end{equation}
with $i,j\in\{1,2\}$. According to Eq.~(\ref{eq:8xa}), the system
is chaotic for any parameters $f'(x)=\alpha>1$ and $\varepsilon,\kappa\in[0,1]$.
The pair has the eigenvalue $\gamma_{1}=-1$ which gives $\beta_{0}=(1-\varepsilon)\,\alpha$,
$\beta_{1}=\varepsilon\,(2\,\kappa-1)\,\alpha$. Eq.~(\ref{eq:7x})
yields two synchronization borders for $\tau\to\infty$\begin{align}
\kappa^{+}(\varepsilon) & =1-\frac{\alpha-1}{2\,\alpha\,\varepsilon}\nonumber \\
\kappa^{-}(\varepsilon) & =\frac{\alpha-1}{2\,\alpha\,\varepsilon}\label{eq:KappasGrenze}\end{align}
\begin{figure}
\hfill{}\includegraphics[width=5cm]{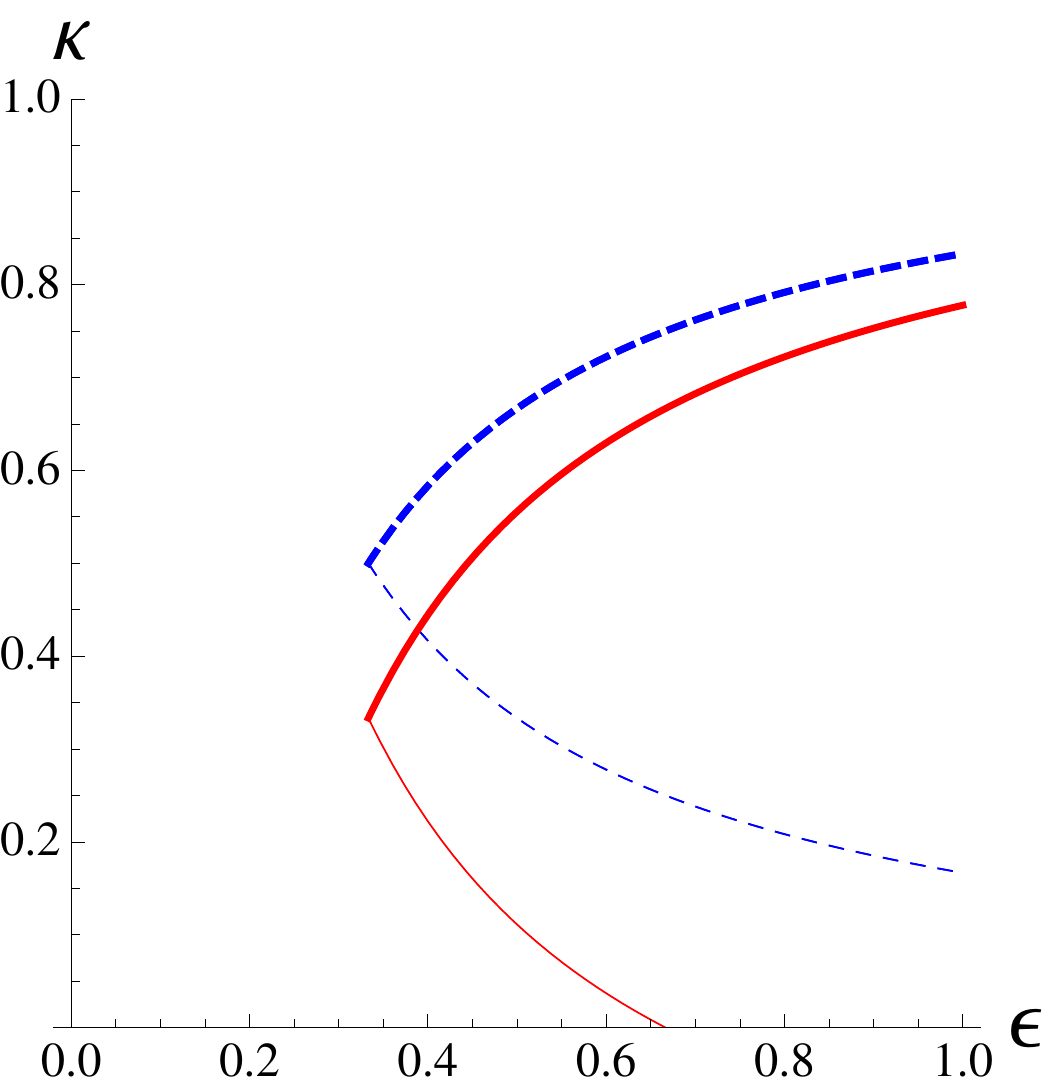}\hfill{}

\caption{Blue dashed line: border to synchronization for a bidirectionally
coupled Bernoulli pair with $\alpha=1.5$, self-feedback, $\tau\to\infty$
and $\varepsilon,\kappa$ notation. Thick line: $\kappa^{+}$, thin
line: $\kappa^{-}$. Red solid line: border to synchronization for
a bidirectionally coupled triangle, same parameters as for the pair.\label{fig:EpsKap}}

\end{figure}

For a triangle with corresponding equations we obtain with $\gamma_{1}=-\frac{1}{2}$
the coefficients $\beta_{0}=(1-\varepsilon)\,\alpha$, $\beta_{1}=\varepsilon\,\kappa\,\alpha-\frac{1}{2}\,\varepsilon\,(1-\kappa)\,\alpha=\frac{\varepsilon}{2}\,(3\,\kappa-1)\,\alpha$
which gives\begin{align}
\kappa^{+}(\varepsilon) & =1-\frac{2\,(\alpha-1)}{3\,\varepsilon\,\alpha}\nonumber \\
\kappa^{-}(\varepsilon) & =\frac{(2-\varepsilon)\,\alpha-2}{3\,\varepsilon\,\alpha}\label{eq:KappasGrenze2}\end{align}
Both results are shown in Fig.~\ref{fig:EpsKap}. Without self-feedback
($\kappa=0$), a bidirectionally coupled pair cannot synchronize whereas
a triangle synchronizes for $\varepsilon>2\,(\alpha-1)/\alpha$.

Note that the upper boundaries $\kappa^{+}$ do not depend on the
delay time $\tau$, whereas with decreasing delay time the lower boundary
moves down, the region of synchronization increases.

From Eq.~(\ref{eq:6x}) a relation between the MSF and the maximal
Lyapunov exponent in the limit of $\tau\to\infty$ can be derived.
We consider a network without self-feedback and with local stability
$\beta_{0}<1$. The MSF, i.\,e. the largest transversal Lyapunov
exponent is given by\begin{equation}
\lambda=\frac{1}{\tau}\ln\frac{\left|\sigma_{1}\,\alpha\,\gamma_{1}\right|}{1-\beta_{0}}\label{eq:lambda1}\end{equation}
whereas the maximal Lyapunov exponent, which is the Lyapunov exponent
parallel to the SM, is\begin{equation}
\lambda_{\mathrm{max}}=\frac{1}{\tau}\ln\frac{\left|\sigma_{1}\,\alpha\right|}{1-\beta_{0}}\label{eq:LambdaMaxPar}\end{equation}
Hence, one obtains\begin{equation}
\lambda=\lambda_{\mathrm{max}}+\frac{1}{\tau}\ln\!\left|\gamma_{1}\right|\label{eq:LambdaOfGamma}\end{equation}
Thus, the SM is stable if\begin{equation}
\left|\gamma_{1}\right|<\mathrm{e}^{-\lambda_{\mathrm{max}}\,\tau}\label{eq:AbsolutGamma}\end{equation}
This equation has a fundamental meaning. For any network with a stochastic
coupling matrix $G$, it relates the eigenvalue gap $1-\left|\gamma_{1}\right|$
to the synchronizability of the network. If the second largest eigenvalue
is smaller than the largest one, $\left|\gamma_{1}\right|<1$, the
SM is stable for sufficiently weak chaos inside of the SM, i.\,e.
in the limit $\lambda_{\mathrm{max}}\to0$.

We believe that Eq.~(\ref{eq:AbsolutGamma}) holds for any chaotic
network of the structure defined in the previous section, even for
corresponding delay differential equations. In fact, Eq.~(\ref{eq:AbsolutGamma})
has been derived for networks with periodic dynamics \cite{YanchukPerlikowski}
and our numerical results of laser equations confirm this condition
on chaos synchronization (see the following section).

Eq.~(\ref{eq:AbsolutGamma}) has been derived in the limit $\tau\to\infty$.
For finite $\tau$ one either has to solve Eq.~(\ref{eq:OneDelayPoly})
numerically or one can calculate regions of stability with the Schur-Cohn
theorem \cite{Schur}. But even in this case of finite $\tau$ an
analytic result is possible. The border of stability is given by $\left|z_{r}\right|=1$
where $\left|z_{r}\right|$ is the maximal root for the perturbation
modes $k$ transversal to the SM. With $z_{r}=\mathrm{e}^{\mathrm{i}\,\phi_{r}}$
one can decompose Eq.~(\ref{eq:OneDelayPoly}) into the real and
imaginary parts\begin{align}
\cos(\phi_{r}\,\tau) & =\beta_{0}\cos\!\left[\phi_{r}\,(\tau-1)\right]+\beta_{1}\nonumber \\
\sin(\phi_{r}\,\tau) & =\beta_{0}\sin\!\left[\phi_{r}\,(\tau-1)\right]\label{eq:RealIm}\end{align}
The solution of Eqs.~(\ref{eq:RealIm}) gives boundaries $\beta_{1}(\phi)$
and $\beta_{0}(\phi)$. The maximal value of $\beta_{0}(\phi)$, i.\,e.
the tip of the phase diagram of Fig.~\ref{fig:SynForDifferentSmallTaus},
is obtained in the limit $\phi\to0$, which gives\begin{align}
1 & =\beta_{0}+\beta_{1}\label{eq:TwoEqu}\\
\tau & =\beta_{0}\,(\tau-1)\end{align}
The first part agrees with Eq.~(\ref{eq:borderChaos}), the second
gives the result\begin{equation}
\max\beta_{0}=\frac{\tau}{\tau-1}\label{eq:beta0Max}\end{equation}
In the limit $\tau\to\infty$ one can synchronize only for $\beta_{0}<1$,
i.\,e. if the local Lyapunov exponent (see previous section) is negative.
But for finite delay times even chaotic units ($\beta_{0}>1$) can
synchronize if\begin{equation}
\beta_{0}<\frac{\tau}{\tau-1}\label{eq:Beta0}\end{equation}

\subsection{Two delay times\label{sub:Two-delay-times}}

\begin{figure}
\hfill{}\includegraphics[width=4cm]{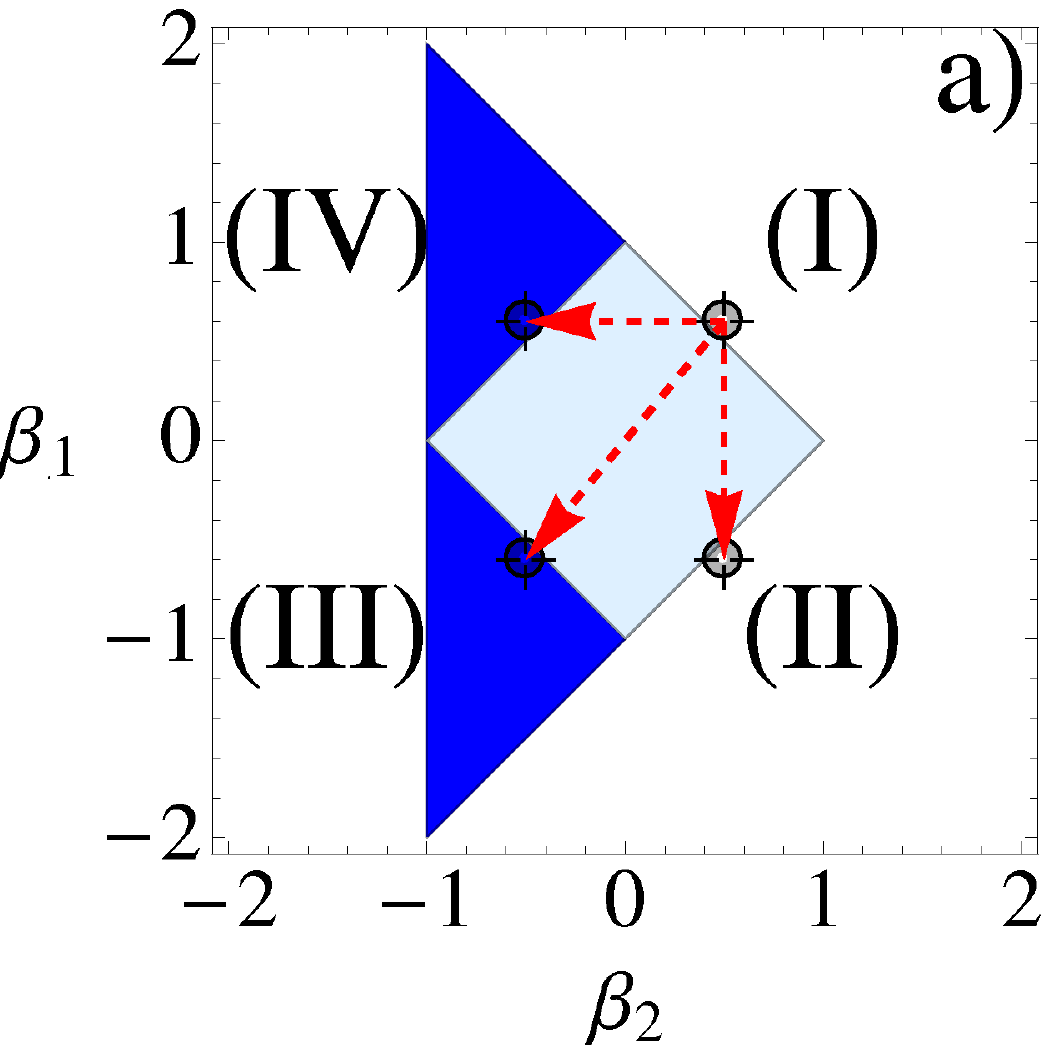}\hfill{}\includegraphics[width=4cm]{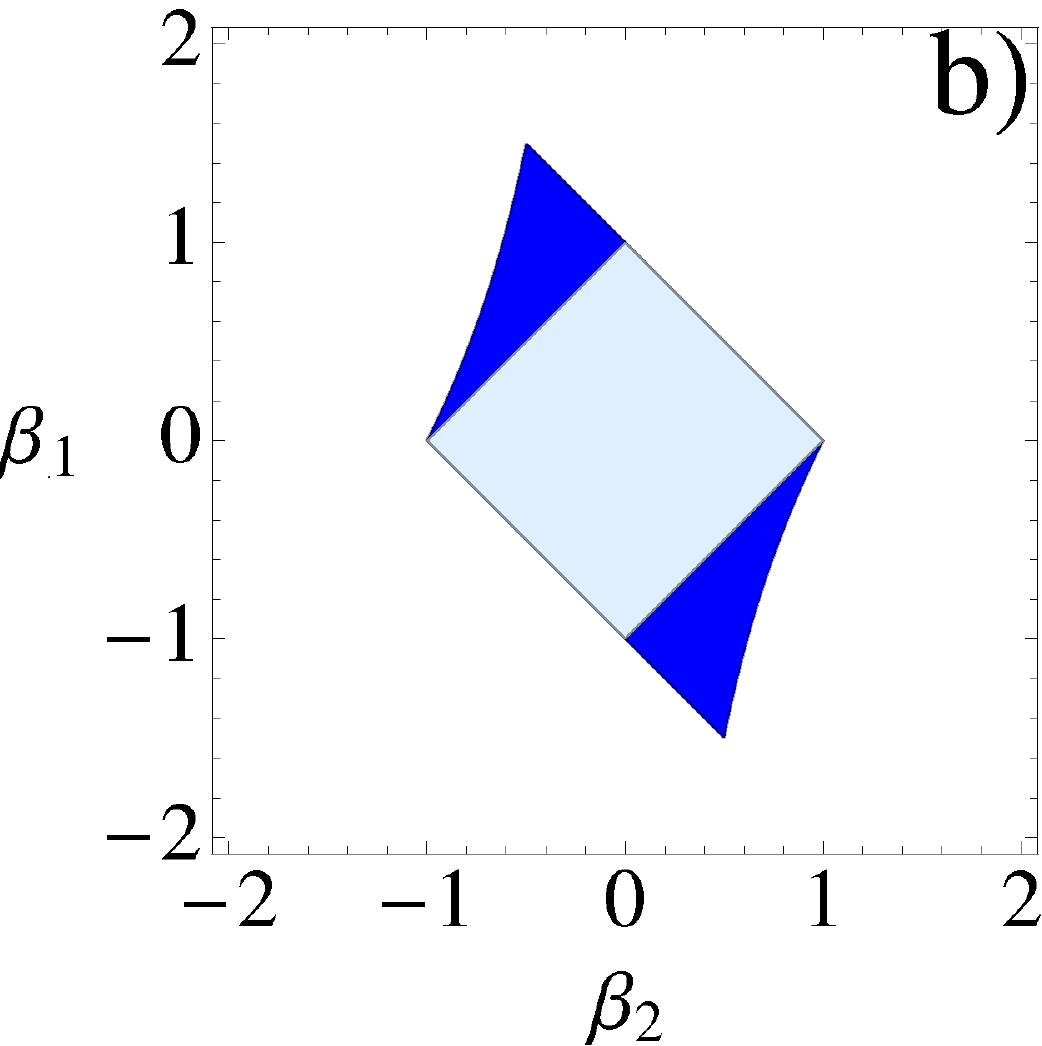}\hfill{}

\hfill{}\includegraphics[width=4cm]{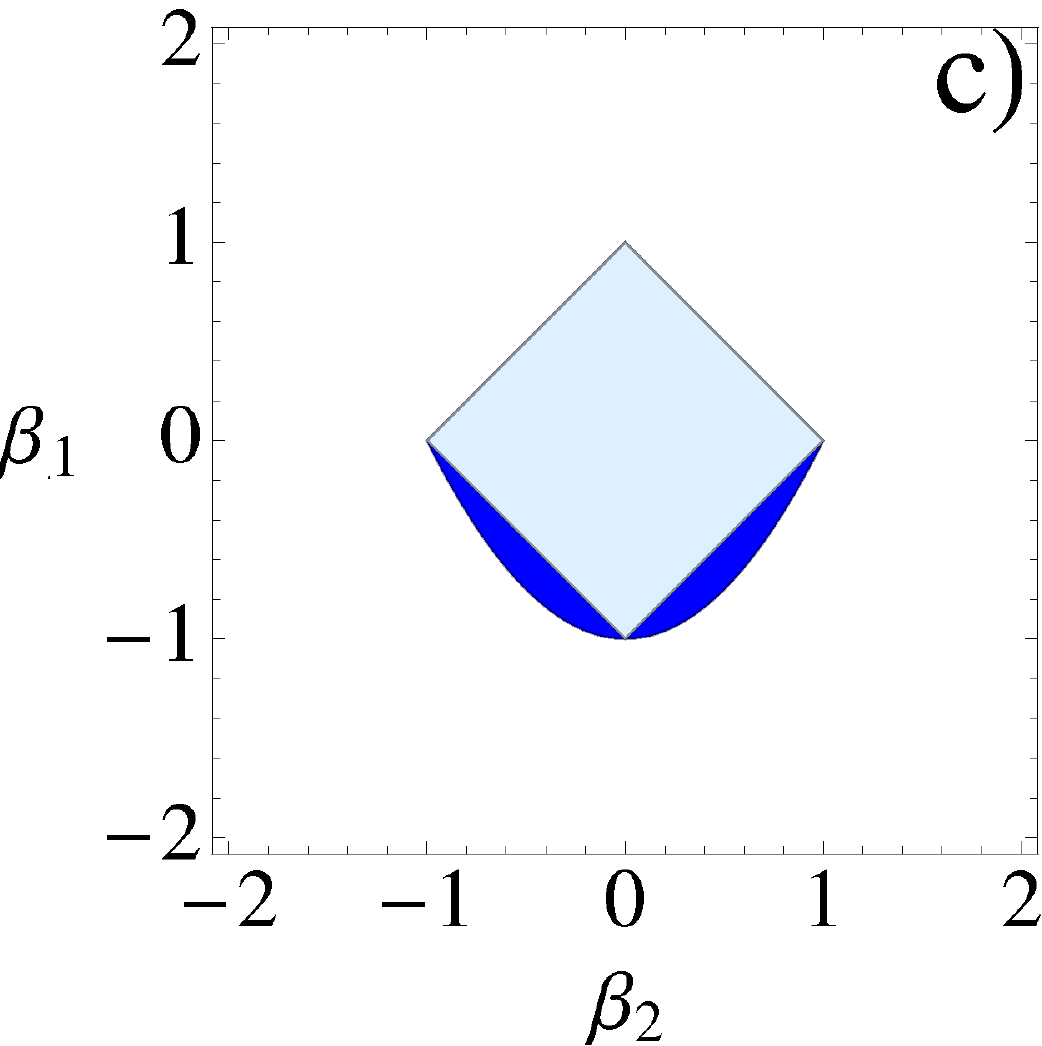}\hfill{}\includegraphics[width=4cm]{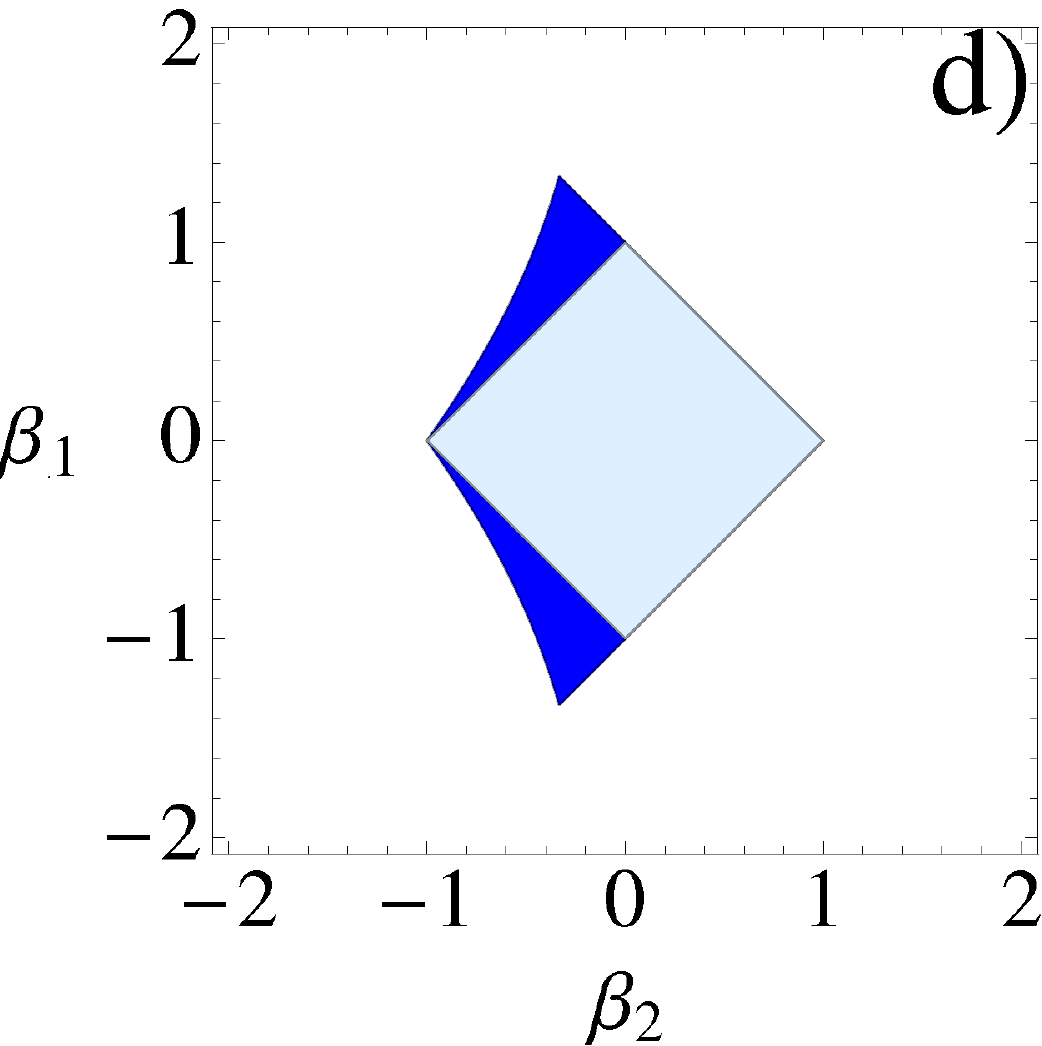}\hfill{}

\caption{Synchronization regime for $z^{\tau_{2}}=\beta_{1}\, z^{\tau_{2}-\tau_{1}}+\beta_{2}$
and $\tau_{2}/\tau_{1}=p/q$ with $p,q$ relatively prime and a) $\frac{p}{q}=2$,
b) $\frac{p}{q}=3$, c) $\frac{p}{q}=\frac{3}{2}$, d) $\frac{p}{q}=4$.
Region (I) denotes the region parallel to the synchronization manifold,
the other regions lie transversal to the synchronization manifold.\label{fig:Symmetries}}

\end{figure}

Now consider a network with two delay times $\tau_{1}$ and $\tau_{2}$
($\tau_{2}>\tau_{1}$). Each delay time belongs to a coupling matrix
$G_{1}$ and $G_{2}$ which have identical eigenvectors, i.\,e. identical
modes of perturbation. The stability of the SM is determined by the
polynomials\begin{equation}
z^{\tau_{2}}=\beta_{0}\, z^{\tau_{2}-1}+\beta_{1}\, z^{\tau_{2}-\tau_{1}}+\beta_{2}\label{eq:TwoDelayPol}\end{equation}
We consider only networks where $\beta_{1}$ and $\beta_{2}$ belong
either to a coupling or a self-feedback, therefore we rule out $\eta_{i}\neq0$
and $\sigma_{i}\neq0$ for the same system. Each mode of perturbation
has an eigenvalue $\gamma_{k,l}$ of the coupling matrices of Eq.~(\ref{eq:MapDynamicSingleUnit}),
which gives three possibilities for Eq.~(\ref{eq:TwoDelayPol}):
a system with self-feedback $\tau_{1}$ and coupling $\tau_{2}$,
a system with self-feedback $\tau_{2}$ and coupling $\tau_{1}$ and
a system with two couplings $\tau_{1}$ and $\tau_{2}$. For these
three cases we obtain the equations\begin{align}
z^{\tau_{2}} & =\eta_{0}\,\alpha\, z^{\tau_{2}-1}+\eta_{1}\,\alpha\, z^{\tau_{2}-\tau_{1}}+\sigma_{2}\,\alpha\,\gamma_{k,2}\label{eq:self-coupl}\\
z^{\tau_{2}} & =\eta_{0}\,\alpha\, z^{\tau_{2}-1}+\sigma_{1}\,\alpha\,\gamma_{k,1}\, z^{\tau_{2}-\tau_{1}}+\eta_{2}\,\alpha\label{eq:coupl-self}\\
z^{\tau_{2}} & =\eta_{0}\,\alpha\, z^{\tau_{2}-1}+\sigma_{2}\,\alpha\,\gamma_{k,1}\, z^{\tau_{2}-\tau_{1}}+\sigma_{2}\,\alpha\,\gamma_{k,2}\label{eq:coupl-coupl}\end{align}

For simplicity, we start the discussion omitting the local term, $\eta_{0}=0$,
and considering a pair of coupled units with $\gamma_{0}=1$, $\gamma_{1}=-1$.
Hence, Eq.~(\ref{eq:TwoDelayPol}) is reduced to\begin{equation}
z^{\tau_{2}}=\beta_{1}\, z^{\tau_{2}-\tau_{1}}+\beta_{2}\label{eq:TwoDelayNoLocal}\end{equation}
Let $\mu$ be the greatest common divisor of $\tau_{2}$ and $\tau_{1}$.
We can substitute $w=z^{\mu}$ to obtain\begin{equation}
w^{p}=\beta_{1}\, w^{p-q}+\beta_{2}\label{eq:TwoDelayPQ}\end{equation}
with $\tau_{2}=p\,\mu$, $\tau_{1}=q\,\mu$ and $(p,q)$ are relatively
prime. Now, all roots $z_{r}$ for $k>0$ lie inside the unit circle
if and only if all roots $w_{r}$ for $k>0$ lie in the unit circle,
hence only the ratio $\tau_{2}/\tau_{1}=p/q$ determines the stability
of Eq.~(\ref{eq:TwoDelayNoLocal}). Fig.~\ref{fig:Symmetries} shows
the regions of stability of Eq.~(\ref{eq:TwoDelayNoLocal}) for different
values of $p$ and $q$ calculated with the Schur-Cohn theorem \cite{Schur}.
As shown before, the system is chaotic if $\left(\eta_{1}+\sigma_{1}\right)\alpha+\left(\eta_{2}+\sigma_{2}\right)\alpha>1$.
For the mode $\gamma_{0}=1$ this means $\beta_{1}+\beta_{2}>1$.
For example, the point in region (I) in Fig.~\ref{fig:Symmetries}a)
belongs to a chaotic system.

The stability of the SM for a bipartite network, for example a pair
of units, is determined by the mode $\gamma_{1}=-1$, which changes
the sign of $\beta_{1}$ and/or $\beta_{2}$ of the point in region
(I), depending whether this term belongs to a coupling or to a self-feedback.
For self-feedbacks one has $\beta_{i}=\eta_{i}\,\alpha$ whereas the
coupling gives $\beta_{i}=-\sigma_{i}\,\alpha$. For example if $\tau_{1}$
belongs to a self-feedback, $\beta_{1}$ is positive, since $\eta$
and $\alpha$ are positive. For $\tau_{2}$ belonging to a bipartite
coupling, $G_{2}$ has an eigenvalue $\gamma_{1}=-1$, so $\beta_{2}$
is flipped to $-\beta_{2}$ which gives the point in region (IV) in
Fig.~\ref{fig:Symmetries}a) which is stable. When both delay times
$\tau_{1}$ and $\tau_{2}$ belong to the coupling, $G_{1}$ and $G_{2}$
are coupling matrices with $\gamma_{1}=-1$ and we obtain the point
in region (III) which is stable, as well. In both cases we find complete
zero-lag synchronization. But when $\tau_{2}$ belongs to a self-feedback
and $\tau_{1}$ to a coupling we obtain the point in region (II) which
is unstable. In this case we can never achieve synchronization for
any parameters $\sigma_{1}$ and $\sigma_{2}$ for which the system
is chaotic.

Obviously, the symmetries seen in Fig.~\ref{fig:Symmetries} prevent
synchronization for some cases depending on the values of $p$ and
$q$. These symmetries of Eq.~(\ref{eq:TwoDelayPol}) can be derived
as follows. If we change the sign of the roots, $w=-v$ the conditions
of stability do not change, but Eq.~(\ref{eq:TwoDelayPQ}) is changed
to the polynomial\begin{equation}
v^{p}=\beta_{1}\,(-1)^{q}\, v^{p-q}+\beta_{2}\,(-1)^{p}\label{eq:Symmetries}\end{equation}
 Now, we have to consider three cases:

If $q$ is even and $p$ is odd, the phase diagram has the reflection
symmetry $\beta_{2}\to-\beta_{2}$, see Fig.~\ref{fig:Symmetries}c)
for $p=3$ and $q=2$. Thus, if $\tau_{1}$ belongs to the self-feedback
and $\tau_{2}$ to the coupling, the system cannot synchronize, since
the condition for stability of the SM is identical to the condition
for chaos.

If $q$ is odd and $p$ is even, the phase diagram has the reflection
symmetry $\beta_{1}\to-\beta_{1}$, see Fig.~\ref{fig:Symmetries}a)
and \ref{fig:Symmetries}d). Thus, if $\tau_{1}$ belongs to the coupling
and $\tau_{2}$ to the self-feedback, the system cannot synchronize.

If $p$ as well as $q$ are odd, the phase diagram has the point symmetry
$\beta_{1}\to-\beta_{1}$ and $\beta_{2}\to-\beta_{2}$. Thus, if
both delay times belong to couplings, the system cannot synchronize.

Therefore, the symmetries of the roots of the polynomials Eq.~(\ref{eq:TwoDelayNoLocal})
rule out some ratios of the two delay times for which synchronization
can occur. For the ratios which are not forbidden by symmetries, synchronization
is possible in a limited paramter region, shown by the dark regions
of Fig.~\ref{fig:Symmetries} for a pair of Bernoulli units. From
numerical calculations, we observe that this region shrinks to zero
when $p$ and $q$ increase and we will later show that zero-lag synchronization
is not possible if $\tau_{1}$ and $\tau_{2}$ are large with a small
difference.

These results are in agreement with \cite{ZigzagButkowskiEnglert,ZigzagButkowskiEnglertEuro}
where a pair with multiple feedback and multiple couplings with different
delay times was analyzed. The time delays which lead to zero lag synchronization
follow \cite[Eq.~(29)]{ZigzagButkowskiEnglert}, $\sum_{i=1}^{M_{s}}l_{i}\, N_{d_{i}}+\sum_{j=1}^{M_{m}}m_{j}\, N_{c_{j}}=0$,
where $N_{d_{i}}$ are the delay times of the $M_{s}$ different self-feedbacks,
$N_{c_{j}}$ are the delay times of the $M_{m}$ different couplings,
and $l_{i},m_{j}$ are whole numbers with a restricted set of possible
values which are specific for each system. 

Note, however, that Fig.~\ref{fig:Symmetries} holds for any network
with eigenvalues $\gamma_{k}$. For example, if $\tau_{1}$ and $\tau_{2}$
belong to the mutual couplings of a triangle without self-feedback
we have $\gamma_{0}=1$ and $\gamma_{1,2}=-1/2$. Hence, the point
in region (I) of Fig.~\ref{fig:Symmetries}a) is mapped to $\beta_{1}\to-\beta_{1}/2$
and $\beta_{2}\to-\beta_{2}/2$. This means that the triangle can
sychronize for any ratio $p,q$ since the perturbation modes are located
in the interior square which is always stable. This result is due
to Gershgorin's circle theorem (\ref{eq:Gershgorin1}) and the conclusion
found earlier in this section. Symmetries rule out synchronization
for bipartite networks, only.

\begin{figure}
\hfill{}\includegraphics[width=4cm]{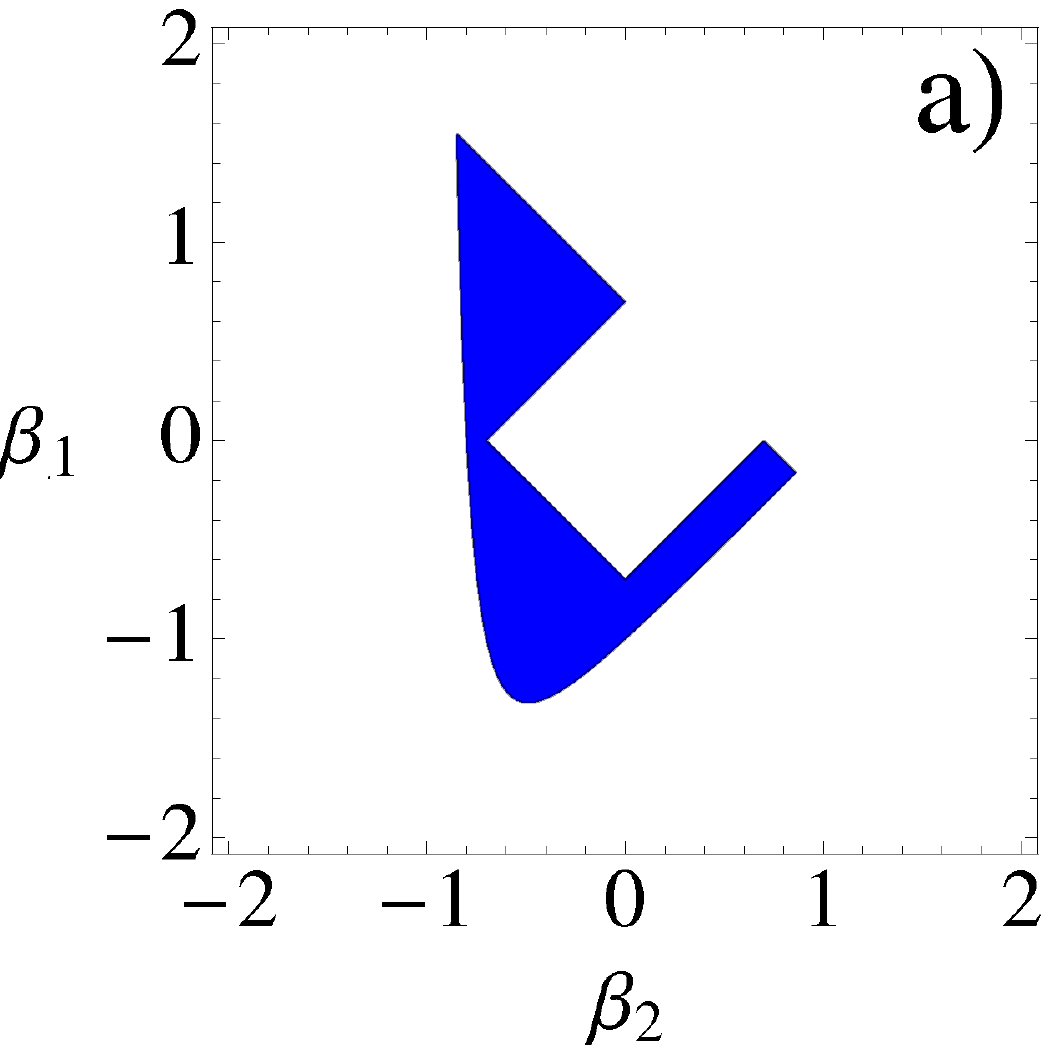}\hfill{}\includegraphics[width=4cm]{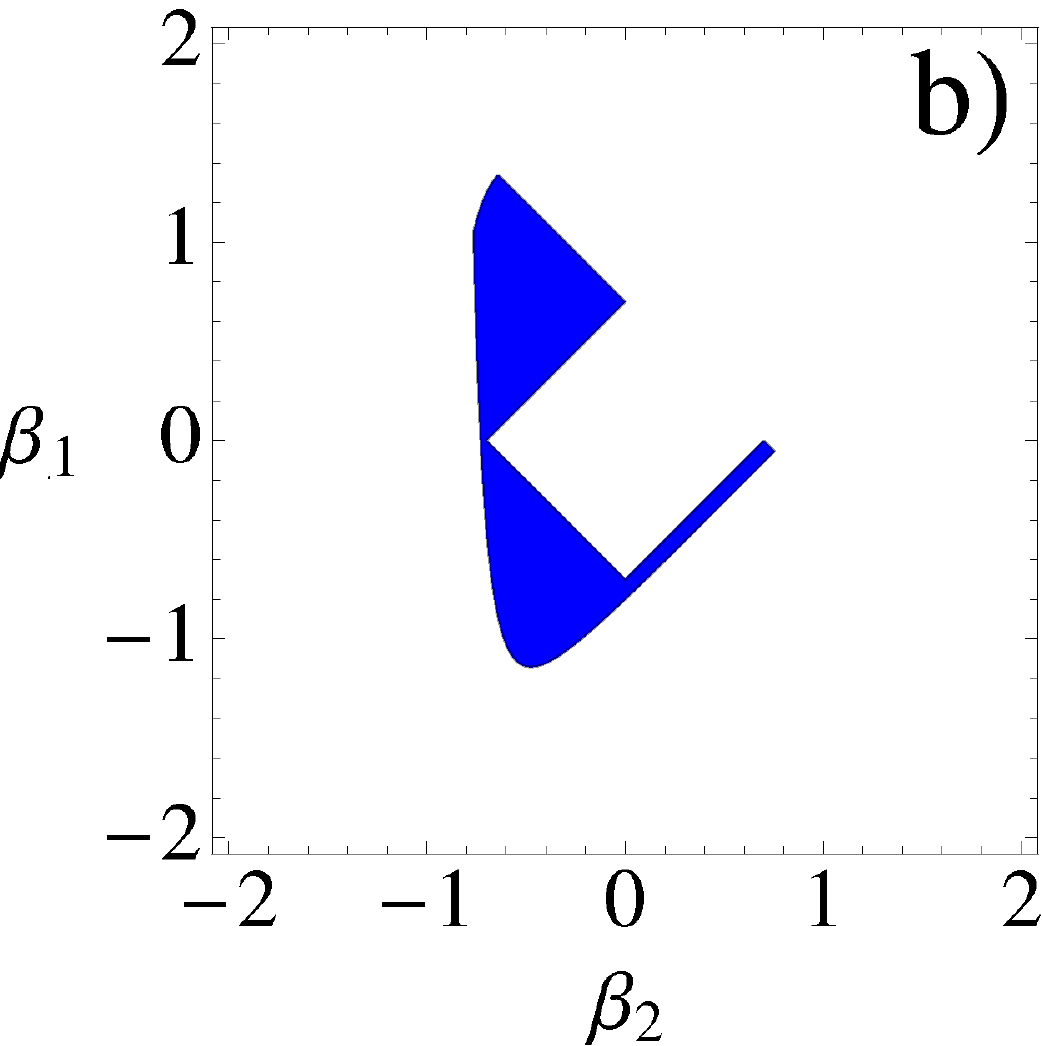}\hfill{}

\hfill{}\includegraphics[width=4cm]{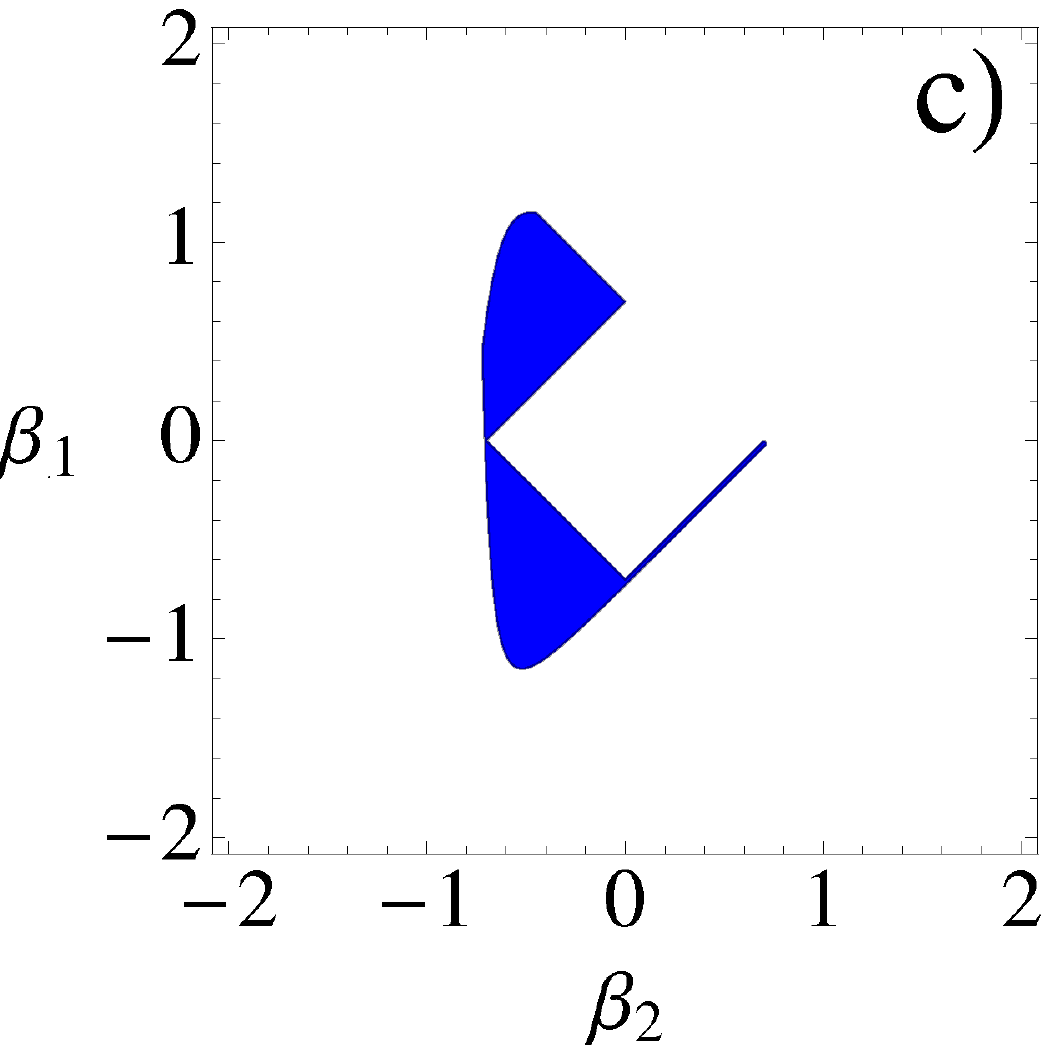}\hfill{}\includegraphics[width=4cm]{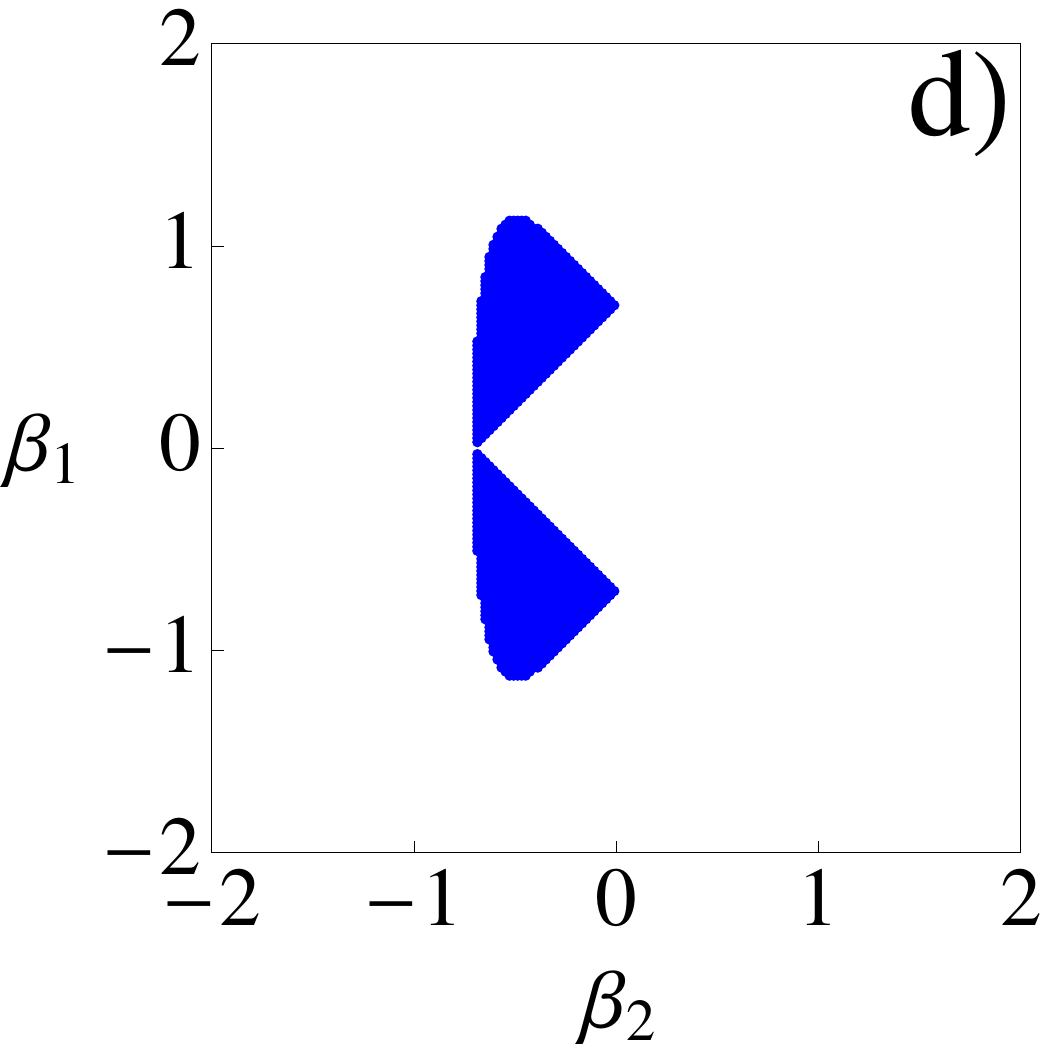}\hfill{}

\caption{Synchronization regime for $z^{\tau_{2}}=\beta_{0}\, z^{\tau_{2}-1}+\beta_{1}\, z^{\tau_{2}-\tau_{1}}+\beta_{2}$
and $\tau_{2}/\tau_{1}=2$ with $\beta_{0}=0.3$. It is a) $\tau_{2}=4$,
b) $\tau_{2}=8$, c) $\tau_{2}=16$, d) $\tau_{2}=100$. The symmetry
found for $\tau_{2}/\tau_{1}=2$ and $\beta_{0}=0$ is restored for
$\beta_{0}>0$ and large values of $\tau_{i}$.\label{fig:aEinschalten}}

\end{figure}

Up to now we have neglected the local term $\eta_{0}\,\alpha\, z^{\tau_{2}-1}$
in Eq.~(\ref{eq:TwoDelayPol}). We will show that the symmetries
still hold in the limit $\tau_{2}\to\infty$, $\tau_{1}\to\infty$,
$\tau_{2}/\tau_{1}=p/q$. The stability of the SM is determined by
the roots of\begin{equation}
z^{\tau_{2}}=\eta_{0}\,\alpha\, z^{\tau_{2}-1}+\beta_{1}\, z^{\tau_{2}-\tau_{1}}+\beta_{2}\label{eq:TwoDelaysWithLocal}\end{equation}
The border to synchronization is necessarily given by $|z|=1$ where
$|z|$ is the maximum of all roots $z_{r}$. With $z=\mathrm{e}^{\mathrm{i}\,\phi}$
we obtain\begin{equation}
1=\eta_{0}\,\alpha\,\mathrm{e}^{-\mathrm{i}\,\phi}+\beta_{1}\,\mathrm{e}^{-\mathrm{i}\,\phi\,\tau_{2}\, q/p}+\beta_{2}\,\mathrm{e}^{-\mathrm{i}\,\phi\,\tau_{2}}\label{eq:ProofLocal1}\end{equation}
In the limit of $\tau_{2}\to\infty$ we use the fact that the phases
$\phi$ of the roots are uniformly distributed in $[0,2\pi]$ \cite{ErdoesTuran,Granville}.
Hence, there exists one root which is close to $\phi=n\,\pi\frac{p}{\tau_{2}}=n\,\pi\frac{q}{\tau_{1}}$
with some integer $n$. Following this root in the limit $\tau_{2}\to\infty$
we obtain\begin{equation}
1=\eta_{0}\,\alpha+\beta_{1}\,\mathrm{e}^{-\mathrm{i}\, n\,\pi\, q}+\beta_{2}\,\mathrm{e}^{-\mathrm{i}\, n\,\pi\, p}\label{eq:ProofLocal2}\end{equation}
Now the border to chaos\begin{equation}
1=\eta_{0}\,\alpha+\left(\eta_{1}+\sigma_{1}\right)\alpha+\left(\eta_{2}+\sigma_{2}\right)\alpha\label{eq:ProofLocal3}\end{equation}
can be mapped to Eq.~(\ref{eq:ProofLocal2}), depending on whether
$\tau_{1}$ and $\tau_{2}$ belong to self-feedback or coupling delays.
For example, if $\tau_{1}$ and $\tau_{2}$ belong to a coupling,
Eq.~(\ref{eq:ProofLocal2}) becomes\begin{equation}
1=\sigma_{0}\,\alpha-\sigma_{1}\,\alpha\,\mathrm{e}^{-\mathrm{i}\, n\,\pi\, q}-\sigma_{2}\,\alpha\,\mathrm{e}^{-\mathrm{i}\, n\,\pi\, p}\label{eq:ProofLocal4}\end{equation}
If both $p$ and $q$ are odd, we choose $\mathrm{e}^{-\mathrm{i}\, n\,\pi}=-1$
and Eq.~(\ref{eq:ProofLocal3}). As synchronization regions calculated
with the Schur-Cohn theorem show, Eq.~(\ref{eq:ProofLocal2}) is
the border to synchronization. Hence, synchronization is not possible.
This shows that the ratios $\tau_{2}/\tau_{1}=p/q$ for which synchronization
is ruled out by symmetry do not depend on the local term in the limit
of large delay times $\tau_{i}\to\infty$. In fact, the numerical
simulations Fig.~\ref{fig:aEinschalten} show that the symmetries
of the complete phase diagram are the same as the ones proven before
for $\eta_{0}=0$, although the phase diagram depends on the strength
$\eta_{0}$ of the local term.

We have shown that the symmetries of the stability equation, Eq.~(\ref{eq:TwoDelaysWithLocal}),
rule out some ratios of the delay times. In fact, these results support
self-consistent arguments for general chaotic networks. These arguments
are based on the fact, that the information of each trajectory of
each unit has to mix after multiples of time intervals $\tau$ in
order to achieve zero-lag synchronization \cite{KanterZigzagEnglertGeissler}.

\begin{figure}
\hfill{}\includegraphics[width=7cm]{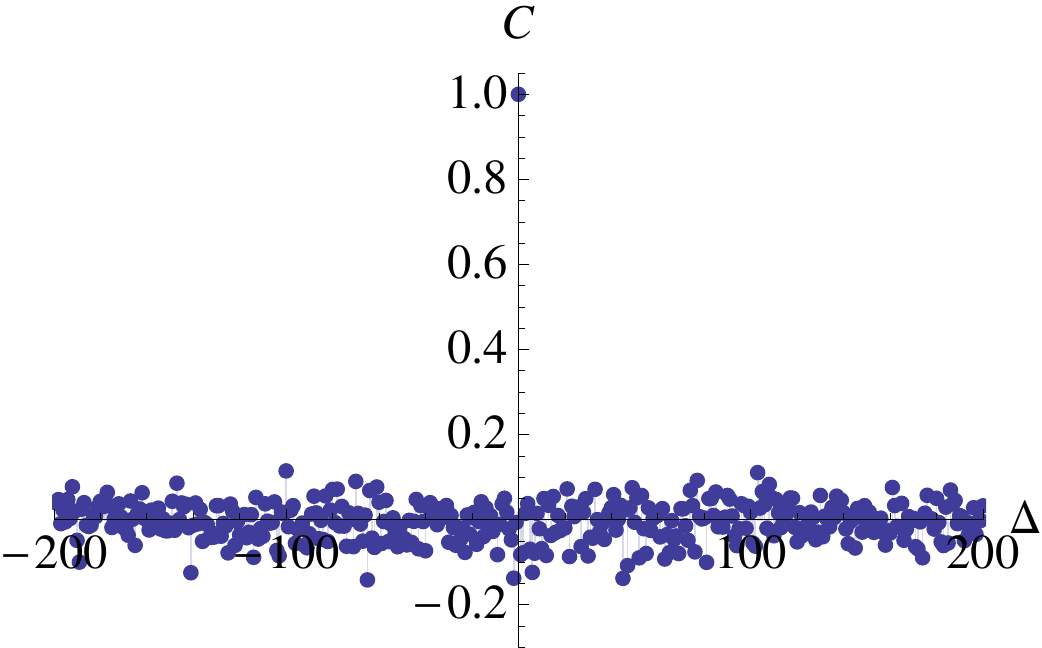}\hfill{}

\caption{Cross-correlation for a bidirectionally coupled Bernoulli pair with
self-feedback delay time $\tau_{1}$ and coupling delay time $\tau_{2}=10000$.
$\Delta=\tau_{2}-\tau_{1}$. The slope $\alpha=1.1$.\label{fig:CorrelationDelta}}

\end{figure}

Finally, we consider the question to which extend the network is sensitive
to detuning the delay times $\tau_{2}$ and $\tau_{1}$. It turns
out, that synchronization is extremly sensitive to a tiny detuning
for large delay times. For simplicity, consider the case $\eta_{0}=0$,
$\tau_{1}=\tau$, $\tau_{2}=\tau+\Delta$ with $\tau\to\infty$ where
$\Delta$ remains finite. $\tau_{1}$ belongs to a self-feedback,
and $\tau_{2}$ to the coupling of a bipartite network, for example
a pair of Bernoulli units. The boundary to chaos is determined by\begin{equation}
1\le\eta_{1}\,\alpha+\sigma_{2}\,\alpha\label{eq:SmallDelta1}\end{equation}
and the stabilty of the synchronization manifold is given by the polynomial
for $\gamma_{1}=-1$\begin{equation}
z^{\tau}=\eta_{1}\,\alpha\, z^{-\Delta}-\sigma_{2}\,\alpha\label{eq:SmallDelta2}\end{equation}
The stability is determined by the largest root $z_{k}=\mathrm{e}^{\mathrm{i}\,\phi}\,\mathrm{e}^{\Lambda/\tau}$which
gives\begin{align}
\mathrm{e}^{2\Lambda} & =\left|\eta_{1}\,\alpha\,\mathrm{e}^{-\mathrm{i}\,\phi\,\Delta}-\sigma_{2}\,\alpha\right|^{2}\nonumber \\
 & =\left(\eta_{1}\,\alpha\cos\phi\,\Delta-\sigma_{2}\,\alpha\right)^{2}+\eta_{1}^{2}\,\alpha^{2}\sin^{2}\phi\,\Delta\nonumber \\
 & =\eta_{1}^{2}\,\alpha^{2}+\sigma_{2}^{2}\,\alpha^{2}-2\,\eta_{1}\,\sigma_{2}\,\alpha^{2}\cos\phi\,\Delta\label{eq:SmallDelta3}\end{align}
In the limit $\tau\to\infty$ the distribution of the phase $\phi$
of the roots is uniform in $[0,2\pi]$, hence we always find one root
with $\cos(\phi\,\Delta)\approx-1$ which according to (\ref{eq:SmallDelta1})
gives a positive transverse Lyapunov exponent $\Lambda/\tau$. Thus,
even when the self-feedback delay differs from the coupling delay
by a single time step, $\tau_{2}=\tau_{1}\pm1$, a pair of Bernoulli
units cannot synchronize for $\tau_{2}\to\infty$. In fact, Fig.~\ref{fig:CorrelationDelta}
shows that the cross-correlation immediately decreases to zero when
the delay times are detuned. 

\subsection{Multiple delay times}

Some of the results of the previous paragraphs can immediately be
extended to a network with $M$ delay times \cite{ZigzagButkowskiEnglert,ZigzagButkowskiEnglertEuro}.
In particular, we can rule out complete zero-lag synchronization for
some networks in the limit of large delay times. 

First: If one delay time is much larger than all other ones, we find
the following result: If the network without the long delay does not
synchronize, this network cannot be synchronized by adding the long
delay. 

Second: If for a pair of coupled units all delay times are much larger
than the time scales of the isolated units, then synchronization is
ruled out for specific ratios of the delay times.

Note that we restrict our discussion to networks defined in the previous
section: To each delay time $\tau_{l}$ there exists a coupling matrix
$G_{l}$ with constant row sum, and all matrices $G_{l}$ have identical
eigenvectors with eigenvalues $\gamma_{k,l}$; $k=0,...,N-1$; $l=1,...,M$.
The stability of the SM is determined by the polynomials\begin{equation}
z^{\tau_{M}}=\sum_{l=0}^{M}\beta_{l}\, z^{\tau_{M}-\tau_{l}}\label{eq:MSF_TDelays}\end{equation}

\paragraph{Long delay time $\boldsymbol{\tau_{M}}$\protect \\
\protect \\
}

First we consider a network where the largest delay time $\tau_{M}$
is much larger than all other delays $\tau_{l}$, i.\,e. we discuss
the limit $\tau_{M}\to\infty$ with finite $\tau_{l}$, $l\neq M$.
Eq.~(\ref{eq:MSF_TDelays}) can be rewritten as\begin{equation}
z=\sum_{l=0}^{M-1}\beta_{l}\, z^{1-\tau_{l}}+\beta_{M}\, z^{1-\tau_{M}}\label{eq:MSFOhneT}\end{equation}
If this equation has a root $|z|>1$ in the limit of $\tau_{M}\to\infty$,
this root is obviously determined by the first term of Eq.~(\ref{eq:MSFOhneT}).
Hence, if the mode $k$ is unstable for the network without long delay,
the last term of Eq.~(\ref{eq:MSFOhneT}) cannot stabilize this perturbation.
If, however, the largest root of Eq.~(\ref{eq:MSFOhneT}) approaches
the unit circle for $\tau_{M}\to\infty$, the long delay has an influence.
We rewrite Eq.~(\ref{eq:MSF_TDelays}) as\begin{equation}
z^{\tau_{M}}=\frac{\beta_{M}}{1-\sum_{l=0}^{M-1}\beta_{l}\, z^{-\tau_{l}}}\label{eq:3_1}\end{equation}
In the limit of $\tau_{M}\to\infty$, we make the ansatz $z=\mathrm{e}^{\Lambda/\tau_{M}}\,\mathrm{e}^{\mathrm{i}\,\phi}$
which gives\begin{equation}
\mathrm{e^{\Lambda}}=\frac{\left|\beta_{M}\right|}{\left|1-\sum_{l=0}^{M-1}\beta_{l}\, z^{-\mathrm{i}\,\phi\,\tau_{l}}\right|}\label{eq:3_2}\end{equation}
As before, the Lyapunov exponents of the $k$-mode depend only on
the modulus of the coupling $\beta_{M}=\left(\eta_{M}+\sigma_{M}\,\gamma_{k,M}\right)\alpha$.
However, the maximal Lyapunov exponent is not given by $\phi=0$.
Only if all parameters $\beta_{l}$ are positive, we obtain the boundary
to stability as\begin{equation}
\left|\beta_{M}\right|=\left|1-\sum_{l=0}^{M-1}\beta_{l}\right|\label{eq:4_1}\end{equation}
In this case all $\beta_{l}$ belong to self-feedbacks, only $\beta_{M}$
is a coupling.

\begin{figure}
\hfill{}\includegraphics[width=6cm]{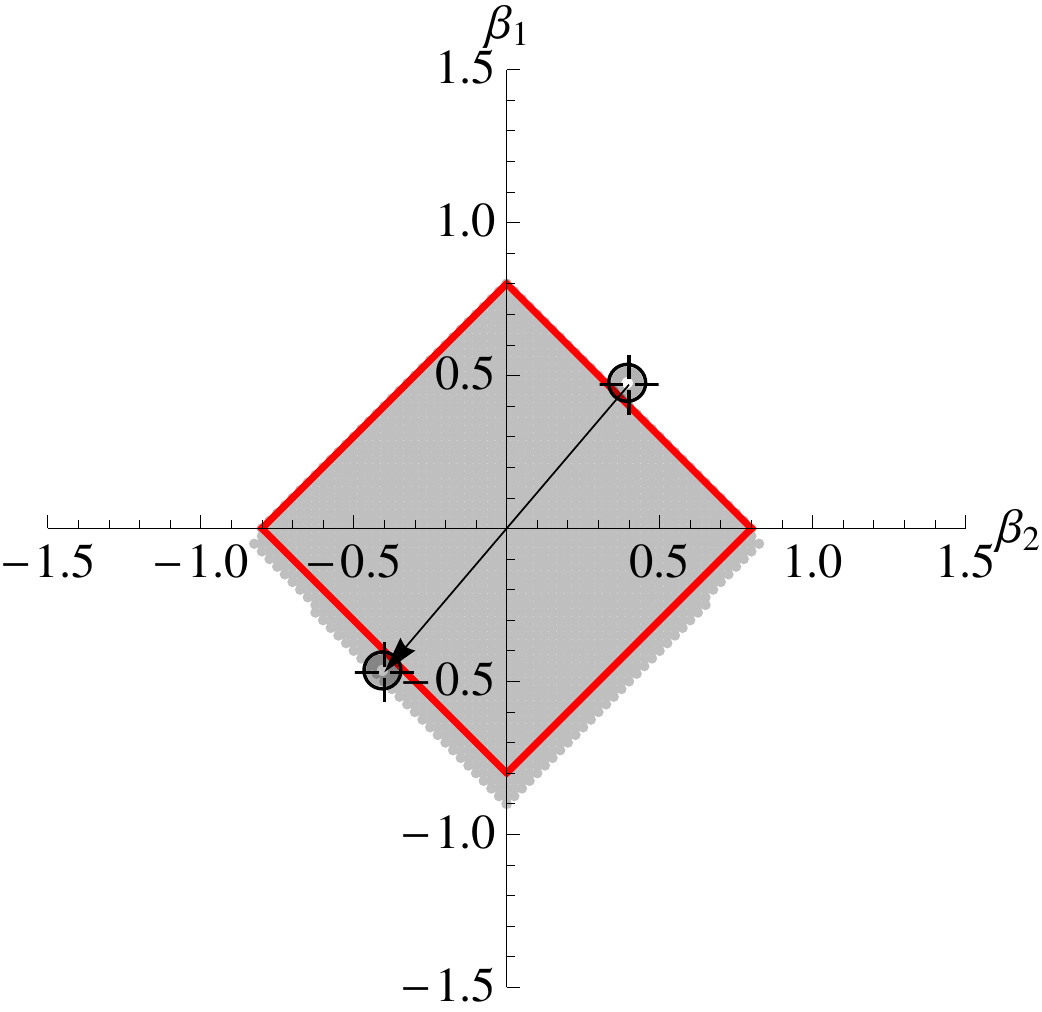}\hfill{}

\caption{Synchronization area of a system with two time delays $\tau_{1}=3$,
$\tau_{2}=300$ and $\beta_{0}=0.2$. The red lines mark the lines
$1-\beta_{0}=\left|\beta_{1}\right|+\left|\beta_{2}\right|$. The
black crosses exemplify a pair of units which is chaotic and can synchronize
if $\tau_{1}$ belongs to a coupling and $\tau_{2}$ to a coupling
or a self-feedback, but not if $\tau_{1}$ belongs to a self-feedback.
\label{fig:3_300_02}}

\end{figure}

The stability of the SM is determined by negative or complex eigenvalues
$\gamma_{k,l}$. For this case we find an interference of different
phases $\phi\tau_{l}$ in Eq.~(\ref{eq:3_2}) resulting in a non-obvious
value $\phi_{0}$ for the maximal Lyapunov exponent $\lambda_{\mathrm{max}}$.
For example, for $\tau_{1}=3$ and $\tau_{2}=300$ we obtain the phase
diagram of Fig.~\ref{fig:3_300_02}. This means, that a pair of units
which is chaotic (cross in Fig.~\ref{fig:3_300_02}) can be synchronized
if $\tau_{1}$ belongs to a coupling and $\tau_{2}$ to a coupling
or a self-feedback. If, however, $\tau_{1}$ belongs to a self-feedback,
$\tau_{2}$ cannot synchronize this pair. Note that the situation
is different for networks with $\left|\gamma_{1}\right|<1$ , e.g.
triangles. 

\pagebreak{}

\paragraph{Symmetry\protect \\
\protect \\
}

Similarly to the case of two delay times discussed before, the symmetry
of the polynomial Eq.~(\ref{eq:MSF_TDelays}) rules out synchronization
for a pair of units with specific ratios of the delay times $\tau_{l}$.
This symmetry holds for $\beta_{0}=0$ for general values of $\tau_{l}$,
but for $\beta_{0}\neq0$ only if all values of $\tau_{l}$ are large.
For simplicity, we discuss the case $\beta_{0}=0$. 

Firstly, we consider the greatest common divisor $\tau$ of all delay
times $\tau_{l}$, i.\,e. we define $p_{l}=\tau_{l}/\tau$ where
the integers $p_{l}$ are relatively prime. Secondly, we substitute
$w=z^{\tau}$ in Eq.~(\ref{eq:MSF_TDelays}) and obtain\begin{equation}
w^{p_{M}}=\sum_{l=0}^{M}\beta_{k,l}\, w^{p_{M}-p_{l}}\label{eq:wBeta}\end{equation}
If a root $z_{r}$ of Eq.~(\ref{eq:MSF_TDelays}) lies inside resp.
outside the unit circle, the root $w_{t}$ of Eq.~(\ref{eq:wBeta})
lies inside resp. outside as well. Hence, the stability of the perturbation
$k$ can be discussed using Eq.~(\ref{eq:wBeta}).

The system is chaotic, Eq.~(\ref{eq:wBeta}) has at least one root
$\left|z_{r}\right|>1$ for $\gamma_{0}=1$, i.\,e. for\begin{equation}
w^{p_{M}}=\sum_{l=0}^{M}\left(\eta_{l}+\sigma_{l}\right)\alpha\, w^{p_{M}-p_{l}}\label{eq:wParallel}\end{equation}
If we restrict our discussion to a pair of units where each $\tau_{l}$
either belongs to a self-feedback $\beta_{l}=\eta_{l}\,\alpha$ or
a coupling $\beta_{l}=-\sigma_{l}\,\alpha$, we can map the stability
of the SM to Eq.~(\ref{eq:wParallel}) for specific values of $p_{l}$.
We find the following result: Synchronization is ruled out if $p_{l}$
is odd for a coupling and if $p_{l}$ is even for a self-feedback.
These results have been previously seen in \cite{ZigzagButkowskiEnglert,ZigzagButkowskiEnglertEuro}.

\section{Comparison with other systems\label{sec:Comparison-with-other}}

The analytic results of the previous section were obtained for networks
of Bernoulli units. For this case, the linear equations describing
the stability of the synchronization manifold (SM) have constant coefficients.
This fact allows a stability analysis with polynomials of degree $\tau_{M}$,
the largest delay time.

For networks of general nonlinear units, however, the stability equations
have time-dependent coefficients generated by the chaotic trajectory
of the SM. For example, for general coupled map lattices the stability
of the SM is determined by Eq.~(\ref{eq:MSFDiscrete}), where the
time-dependent coefficients are given by the dynamics on the SM, Eq.~(\ref{eq:MapSM}).
To our knowledge, these equations cannot be solved analytically. Hence,
we compare our analytic calculations of the section {}``Master stability
function'' with numerical simulations of the skewed tent map\begin{eqnarray}
f\negthickspace:[0,1]\rightarrow[0,1],\quad f(x) & \negthickspace=\negthickspace & \begin{cases}
\frac{x}{a} & \textrm{if}\; x\leqq a\\
\frac{1-x}{1-a} & \textrm{if}\; x>a\end{cases}\label{eq:4.1}\end{eqnarray}
with $0<a<1$. We choose $a=0.86$ which results in the same Lyapunov
exponent for an isolated unit as the one of the Bernoulli map\begin{equation}
f\negthickspace:[0,1]\rightarrow[0,1],\quad f(x)=\left(\frac{3}{2}\, x\!\right)\mathrm{mod\;1}\label{eq:4.2}\end{equation}
Coupled map lattices are special dynamical systems. Thus, it will
be interesting to compare the analytic results of the section {}``Master
stability function'' with systems of nonlinear differential equations.
In particular, we consider the Lang-Kobayashi (LK) equations for coupled
semiconductor lasers \cite{LangKobayashi,KanterZigzagEnglertGeissler}.
In some cases we even compare our results with experiments on chaotic
semiconductor lasers.

For networks of differential equations, the mathematical structure
corresponding to Eq.~(\ref{eq:MapDynamicSingleUnit}) is defined
by\begin{align}
\dot{\vec{x}}_{i}(t) & =\vec{F}\!\left[\vec{x}_{i}(t)\right]+\sum_{l=1}^{M}\eta_{l}\,\vec{H}\!\left[\vec{x}_{l}\!\left(t-\tau_{l}\right)\right]+\nonumber \\
 & \quad+\sum_{l=1}^{M}\sum_{j=1}^{N}\sigma_{l}\, G_{l,ij}\,\vec{H}\!\left[\vec{x}_{j}\!\left(t-\tau_{l}\right)\right]\label{eq:4.3}\end{align}
Now $\vec{x}(t)$ is a multidimensional vector. For example, for the
LK equations $\vec{x}(t)$ contains the real and imaginary parts of
the envelope of the electric field and the population inversion of
the charge carriers. Details are given in the appendix. The dynamics
of the SM is given by\begin{equation}
\dot{\vec{s}}(t)=\vec{F}\!\left[\vec{s}(t)\right]+\sum_{l=1}^{M}\left(\eta_{l}+\sigma_{l}\right)\vec{H}\!\left[\vec{s}\!\left(t-\tau_{l}\right)\right]\label{eq:4.4}\end{equation}
The stability of the SM is described by linear equations for each
mode $k$ with eigenvalues $\gamma_{k,l}$ of the coupling matrix
$G_{l}$. The equations corresponding to Eq.~(\ref{eq:MSFDiscrete})
are\begin{align}
\dot{\vec{\xi}}_{k}(t) & =\mathrm{D}F\!\left[\vec{s}(t)\right]\vec{\xi}_{k}(t)+\nonumber \\
 & \quad+\sum_{l=1}^{M}\left(\eta_{l}+\sigma_{l}\,\gamma_{k,l}\right)\mathrm{D}H\!\left[\vec{s}\!\left(t-\tau_{l}\right)\right]\vec{\xi}_{k}\!\left(t-\tau_{l}\right)\label{eq:4.5}\end{align}
$\mathrm{D}F$ and $\mathrm{D}H$ are the Jacobian matrices of $\vec{F}$
and $\vec{H}$ respectively evaluated at the SM. The master stability
function for the LK equations is defined in the appendix. In the following
paragraphs we consider networks with a single delay time, $M=1$,
and with double delays, $M=2$.

\subsection{Networks with a single delay time\label{sec:Networks-with-a}}

In the previous section, we obtained the analytic result that the
eigenvalue gap of the coupling matrix determines the stability of
the SM in the limit of large delay times $\tau$, which will be realized
in this section by $\tau=100\,\mathrm{ns}$ for the LK equations and
$\tau=100$ for the tent map and the Bernoulli map. The relation $\left|\gamma_{1}\right|<\exp(-\lambda_{\mathrm{max}}\,\tau)$
of Eq.~(\ref{eq:AbsolutGamma}) is the condition for a stable SM;
it relates the maximal Lyapunov exponent $\lambda_{\mathrm{max}}$
of the SM to the second largest modulus of the eigenvalues of the
coupling matrix $G$.

As a consequence of Eq.~(\ref{eq:AbsolutGamma}), a pair of units
without self-feedback cannot be synchronized, since $\gamma_{1}=-1$.
This result agrees with experiments on semiconductor lasers \cite{Fischer:2006:PRL}.
Two lasers coupled by their mutual laser beams cannot be synchronized
with zero time lag, only high correlations with a time shift of $\tau$
have been observed.

With self-feedback, however, the situation is different. Eqs.~(\ref{eq:8x})
and (\ref{eq:KappasGrenze}) show that a pair of units can be synchronized
if the local Lyapunov exponent is negative. Again, this result agrees
with experiments on semiconductor lasers where the self-feedback is
realized by external mirrors \cite{Rosenbluh:046207,Klein:2006:73,Gross:2006,Klein:2006:74,Kanter:154101,KanterButkovski}.

Thus, qualitatively, the relation Eq.~(\ref{eq:AbsolutGamma}) is
in agreement with experiments on lasers. Unfortunately, up to now,
experiments on larger networks of coupled lasers are not reported.
Hence, the quantitative comparison with lasers has to rely on numerical
simulations of the LK equations.

\begin{figure}
\begin{centering}
\includegraphics[width=0.45\columnwidth]{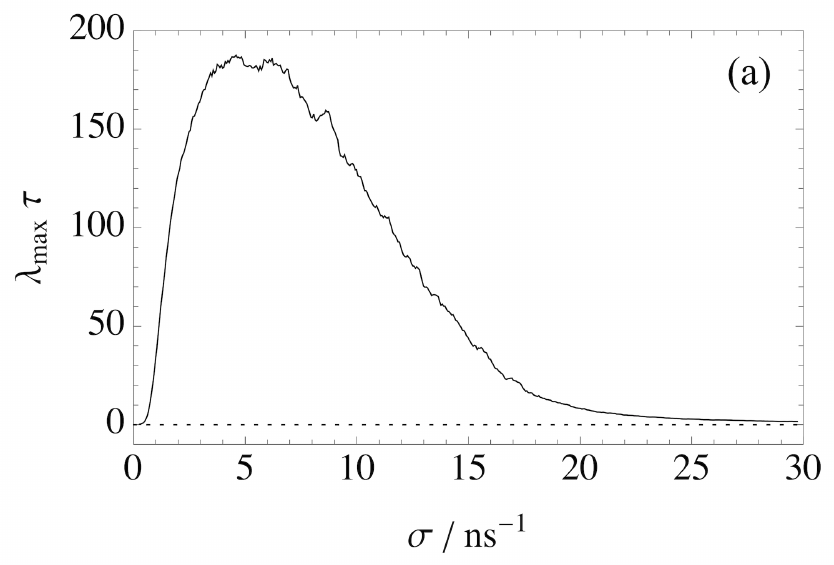}\hspace{5mm}\includegraphics[width=0.45\columnwidth]{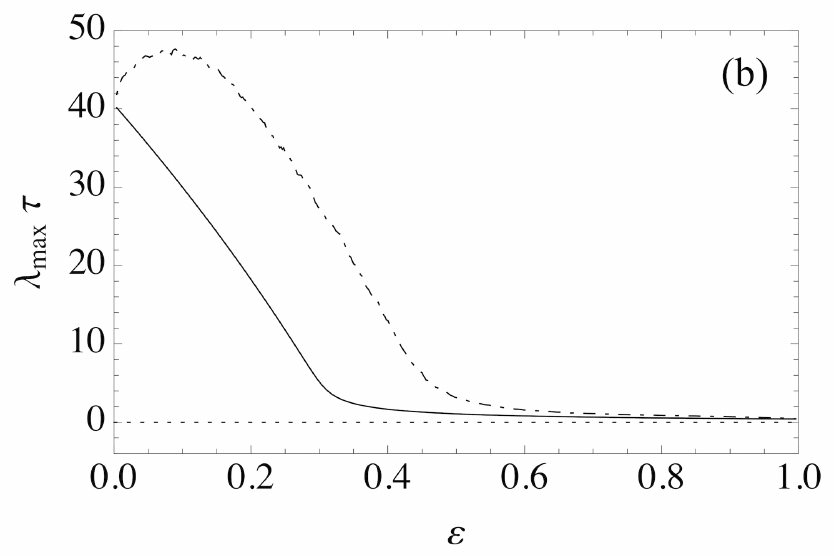}
\par\end{centering}

\caption{Maximal Lyapunov exponents of the synchronization manifold for a network
of (a) lasers modelled by Lang-Kobayashi equations and (b) tent maps
(dotdashed line) and Bernoulli maps (solid line)\label{fig:4.1}}

\end{figure}

Fig.~\ref{fig:4.1} shows the numerically calculated maximal Lyapunov
exponent $\lambda_{\mathrm{max}}$ as a function of the coupling parameter
$\sigma$ for a network of lasers and as a function of the coupling
parameter $\varepsilon$ for a network of tent maps and Bernoulli
maps. Self-feedback is suppressed ($\eta=0$). The parameters for
the LK equations are defined in the appendix.%

\begin{figure}
\hfill{}\includegraphics[width=9cm]{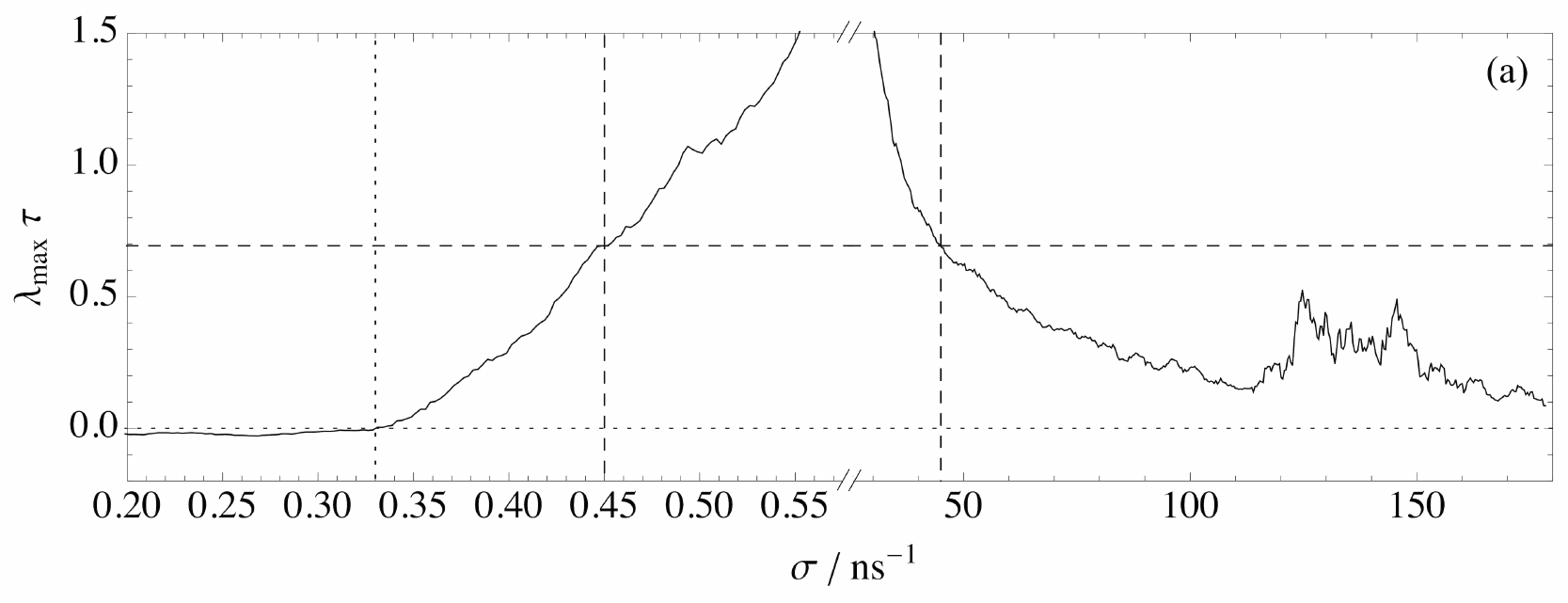}\hfill{}

\begin{centering}
\hfill{}\includegraphics[width=5cm]{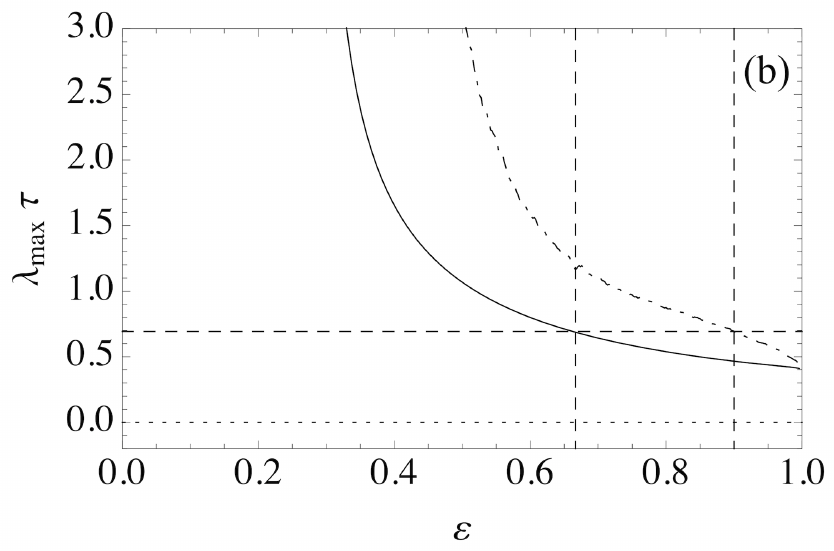}\hfill{}
\par\end{centering}

\caption{Zoom of Fig.~\ref{fig:4.1}, the critical coupling strengths (dashed
lines) according to Eq.~(\ref{eq:AbsolutGamma}), for the synchronization
of a triangle of (a) lasers modelled by Lang-Kobayashi equations and
(b) tent maps (dotdashed line) and Bernoulli maps (solid line) without
self-feedback\label{fig:4.2}}

\end{figure}

\begin{figure}
\hfill{}\includegraphics[width=9cm]{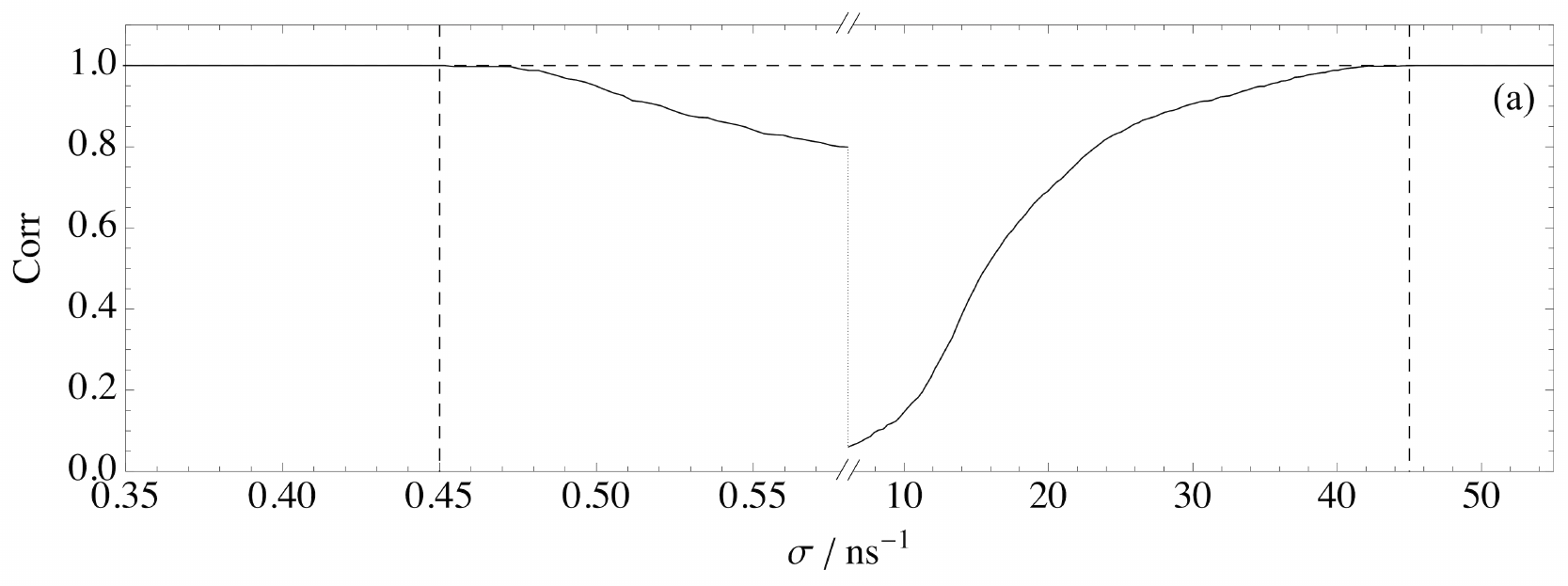}\hfill{}

\hfill{}\includegraphics[width=5cm]{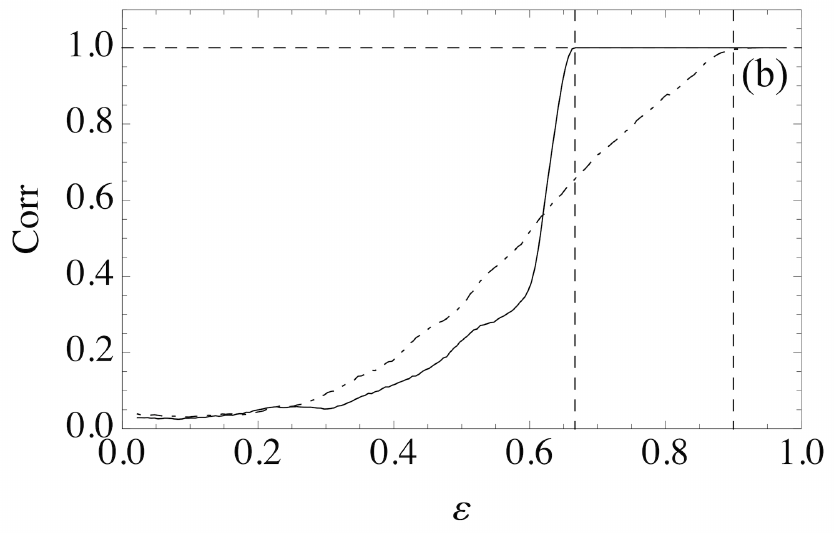}\hfill{}

\caption{Critical coupling strengths (dashed lines) for synchronization measured
in numerical simulations of a triangle of (a) lasers modelled by Lang-Kobayashi
equations and (b) tent maps (dotdashed line) and Bernoulli maps (solid
line) without self-feedback\label{fig:4.3}}

\end{figure}

For a triangle with bidirectional couplings, it is $\gamma_{1}=-\frac{1}{2}$.
Eq.~(\ref{eq:AbsolutGamma}) predicts a transition to synchronization
for $\lambda_{\mathrm{max}}\,\tau=-\ln\!\left(\frac{1}{2}\right)\approx0.69$,
the horizontal dashed line in Fig.~\ref{fig:4.2}a). The numerical
results of Fig.~\ref{fig:4.2} give a critical coupling $\varepsilon_{c}\approx0.9$
for the tent maps, vertical dashed line in Fig.~\ref{fig:4.2}b),
and two critical couplings $\sigma_{c,1}\approx0.45\,\mathrm{ns}^{-1}$
and $\sigma_{c,2}\approx45\,\mathrm{ns}^{-1}$ for the three lasers,
vertical dashed line in Fig.~\ref{fig:4.2}a).%
 Fig.~\ref{fig:4.3} shows the cross-correlations obtained from numerical
simulations of the corresponding triangle. We find the measured critical
couplings of the tent maps and the LK equations to be in good agreement
with the predictions of Eq.~(\ref{eq:AbsolutGamma}).%

\begin{figure}
\begin{centering}
\includegraphics[width=7cm]{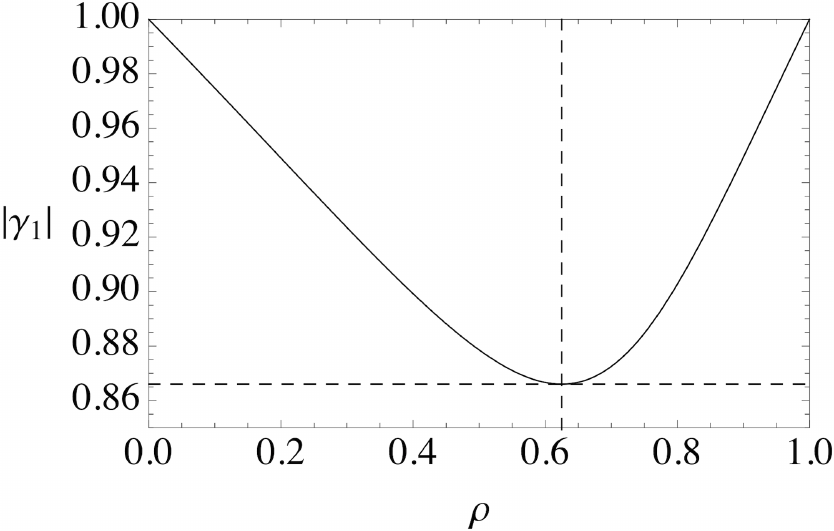}
\par\end{centering}

\caption{Eigenvalue $\left|\gamma_{1}\right|\!(\rho)$ for the square with
directed couplings and one diagonal with coupling strength $\rho$,
dashed line: position of the largest eigenvalue gap\label{fig:4.4}}

\end{figure}

A more challenging test of the condition Eq.~(\ref{eq:AbsolutGamma})
is a network with directed couplings. In this case the eigenvalue
$\gamma_{1}$ is a complex number. For example, for the square with
one diagonal of Fig.~\ref{fig:Schemata-of-bidirectionally} one can
add an additional coupling strength $\rho$ for the diagonal and obtain
the eigenvalue gap of Fig.~\ref{fig:4.4}. The largest gap is obtained
for $\rho=\frac{5}{8}$ with $\gamma_{1}=\frac{-1\pm\mathrm{i}\,\sqrt{11}}{4}$.%

\begin{figure}
\begin{centering}
\hfill{}\includegraphics[width=8cm]{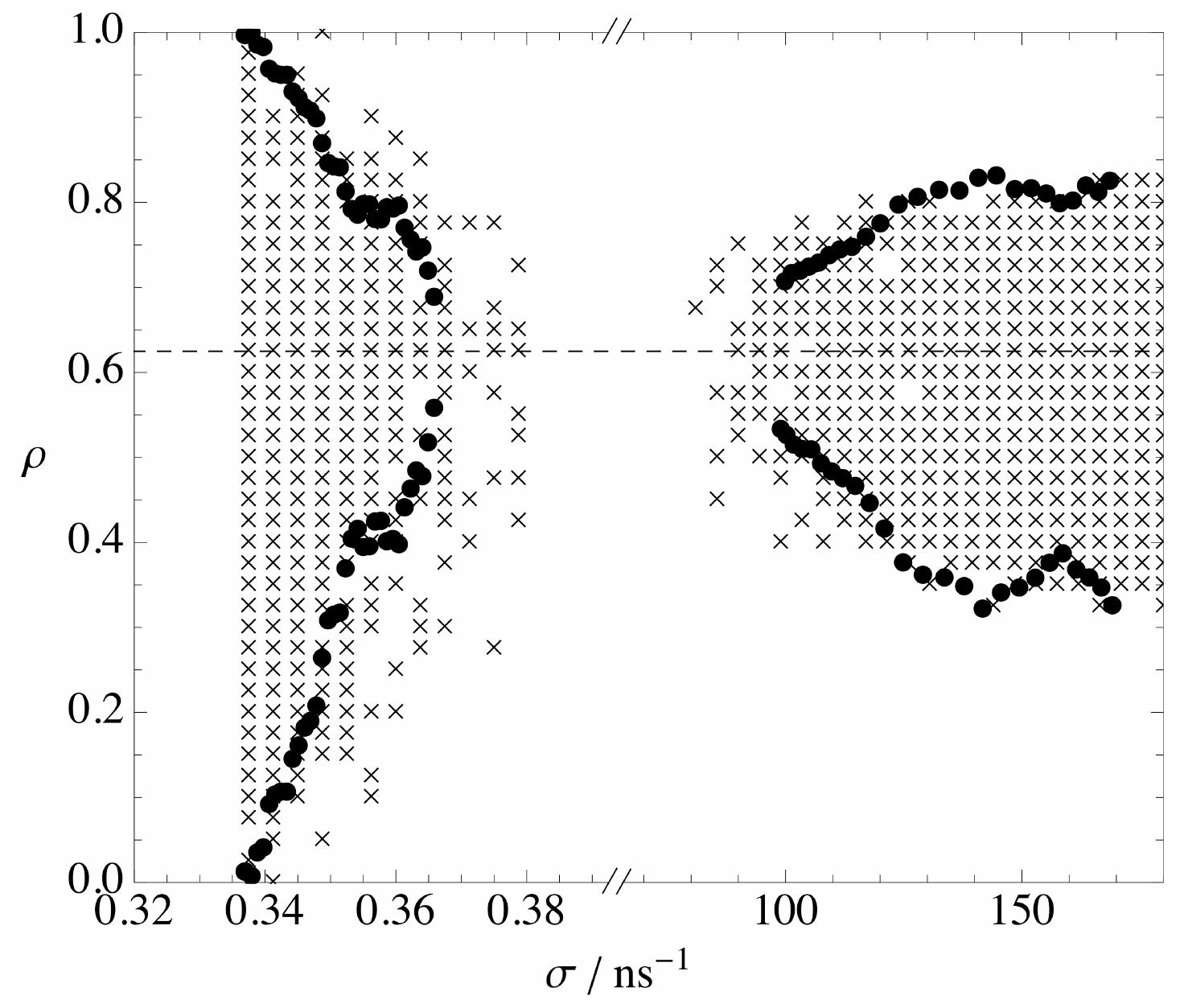}\hfill{}
\par\end{centering}

\caption{Phase diagram for a square of lasers with directed couplings and one
diagonal with coupling strength $\rho$, crosses: parameters for which
complete zero-lag synchronization is achieved in numerical simulations
(only synchronization for chaotic dynamics on the SM is shown), dots:
prediction of the stability border for synchronization, dashed line:
position of the largest eigenvalue gap\label{fig:4.5}}

\end{figure}

Combining this result with Eq.~(\ref{eq:AbsolutGamma}) and $\lambda_{\mathrm{max}}(\sigma)$
(Fig.~\ref{fig:4.1}a), we obtain the phase diagram of Fig.~\ref{fig:4.5}
for the corresponding laser network. Fig.~\ref{fig:4.5} also shows
parameters for which complete zero-lag synchronization is achieved
in numerical simulations of the complete laser network. We define
complete zero-lag synchronization between the lasers for isochronal
cross-correlations larger than 0.99 in between the power drop-outs
of the low frequency fluctuations which happen on a time scale of
the order of magnitude of $10\,\tau$. While it makes sense to speak
of complete synchronization for such values of the cross-correlations,
the maximal Lyapunov exponent is still slightly positive for correlations
around 0.99. Hence, some of the shown points of complete synchronization
in Fig.~\ref{fig:4.5} lie outside of the predicted stability border
which corresponds to a maximal Lyapunov exponent of exactly zero.
The quantitative agreement between the relation Eq.~(\ref{eq:AbsolutGamma})
and the phase diagram Fig.~\ref{fig:4.5} of complete zero-lag synchronization
is remarkable.%

\subsection{Networks with two delay times\label{sub:Networks-with-two}}

A pair of units without self-feedback cannot be synchronized if its
coupling has a single delay time. In the previous section, however,
we have shown in agreement with \cite{ZigzagButkowskiEnglert,ZigzagButkowskiEnglertEuro}
that zero-lag synchronization is possible if the coupling contains
two delay times $\tau_{1}$ and $\tau_{2}$. Only if the ratio $\frac{\tau_{2}}{\tau_{1}}$
is a ratio of odd relatively prime integers, $\frac{\tau_{2}}{\tau_{1}}=\frac{p}{q}$,
synchronization is excluded. The parameter region of synchronization
is largest for small values of $p$ and $q$.

This result is in agreement with recent experiments on semiconductor
lasers \cite{EnglertKinzelAviad}. For $\frac{\tau_{2}}{\tau_{1}}=\frac{2}{1}$
complete synchronization was observed. Cross-correlations were large
for $\frac{\tau_{2}}{\tau_{1}}\in\left\{ \frac{5}{4},\frac{4}{3},\frac{3}{2},\frac{5}{2}\right\} $
and low for $\frac{\tau_{2}}{\tau_{1}}\in\left\{ \frac{1}{1},\frac{5}{3},\frac{3}{1}\right\} $.

The analytic result was based on the symmetry of the phase diagrams,
Figs.~\ref{fig:Symmetries} and \ref{fig:aEinschalten}. Here we
show that these symmetries can be observed for the MSF of laser networks
and tent maps as well. With two delays, Eq.~(\ref{eq:4.4}) and (\ref{eq:4.5})
reduce to\begin{align}
\dot{\vec{s}}(t) & =\vec{F}\!\left[\vec{s}(t)\right]+\sigma_{1}\,\vec{H}\!\left[\vec{s}\!\left(t-\tau_{1}\right)\right]+\sigma_{2}\,\vec{H}\!\left[\vec{s}\!\left(t-\tau_{2}\right)\right]\label{eq:6.1}\\
\dot{\vec{\xi}}_{k}(t) & =\mathrm{D}F\!\left[\vec{s}(t)\right]\vec{\xi}_{k}(t)+\nonumber \\
 & \quad+\sigma_{1}\,\gamma_{k,1}\,\mathrm{D}H\!\left[\vec{s}\!\left(t-\tau_{1}\right)\right]\vec{\xi}_{k}\!\left(t-\tau_{1}\right)+\nonumber \\
 & \quad+\,\sigma_{2}\,\gamma_{k,2}\,\mathrm{D}H\!\left[\vec{s}\!\left(t-\tau_{2}\right)\right]\vec{\xi}_{k}\!\left(t-\tau_{2}\right)\label{eq:6.2}\end{align}
Because of the invasive nature of the coupling, we obtain a different
trajectory $\vec{s}(t)$ on the SM for each pair of coupling strengths
$(\sigma_{1},\sigma_{2})$ which in each case gives a different linear
stability equation. In order to compare with the previous section,
we fix $\sigma_{1}$ and $\sigma_{2}$, vary $\gamma_{k,1}$ and $\gamma_{k,2}$
and calculate the maximal Lyapunov exponent of the linear equation
Eq.~(\ref{eq:6.2}). In this interpretation of $\beta_{1}=\sigma_{1}\,\gamma_{k,1}$
and $\beta_{2}=\sigma_{2}\,\gamma_{k,2}$, the results are universal
in view of the fact that they make a statement about the stability
of all modes of every possible network with real $\gamma_{k,1}$ and
$\gamma_{k,2}$ for the chosen coupling strengths $(\sigma_{1},\sigma_{2})$.
The point $\gamma_{0,1}=\gamma_{0,2}=1$ is shared among all networks;
it determines the stability of the dynamics on the SM.

\begin{figure}
\begin{centering}
\hfill{}\includegraphics[width=4cm]{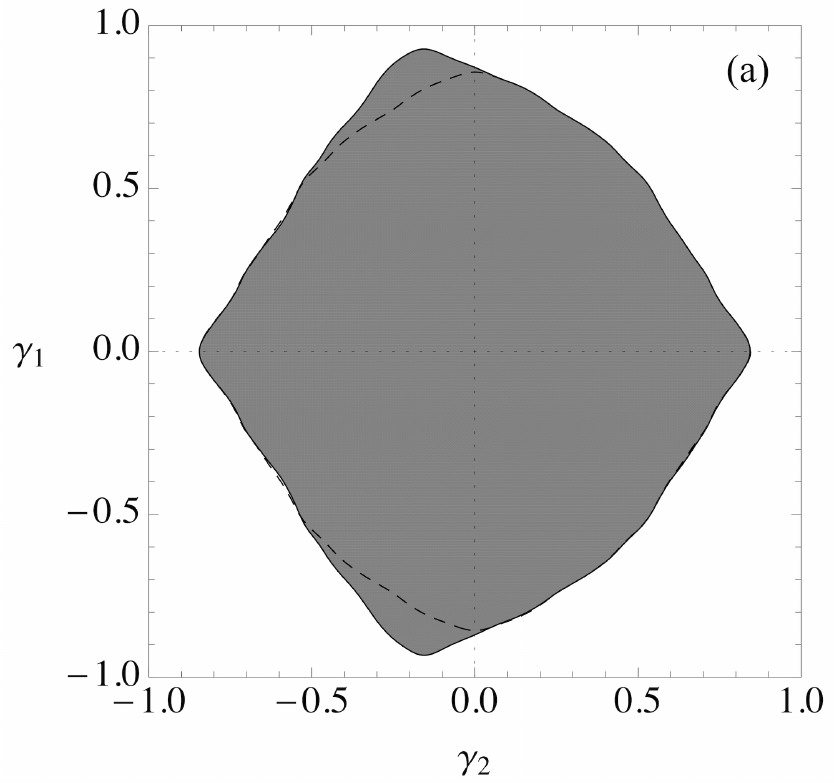}\hfill{}\includegraphics[width=4cm]{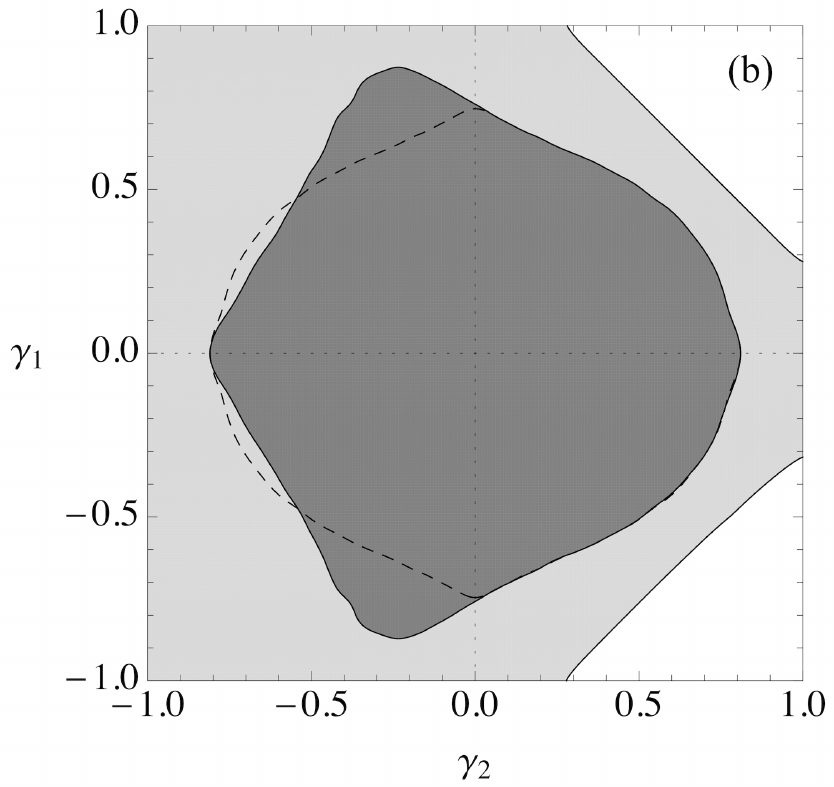}\hfill{}
\par\end{centering}

\caption{Region of stability (dark gray regime) for a pair of (a) lasers modelled
by the Lang-Kobayashi equations and (b) tent maps in comparison with
a pair of Bernoulli maps (light gray regime) with two coupling delay
times $\tau_{1}$ and $\tau_{2}$ and $\frac{\tau_{2}}{\tau_{1}}=\frac{2}{1}$,
dashed line: reflection of the top-right quadrant's stability border
into the other quadrants\label{fig:4.6}}

\end{figure}

Fig.~\ref{fig:4.6} shows the results of the numerical simulations
of Eqs.~(\ref{eq:6.1}) and (\ref{eq:6.2}) with $\tau_{1}=20\,\mathrm{ns},\tau_{2}=40\,\mathrm{ns},\sigma_{1}=\sigma_{2}=45\,\mathrm{ns}^{-1}$
for the LK equations and $\tau_{1}=200,\tau_{2}=400,\varepsilon=0.9,\kappa=0.5$
for the tent map and the Bernoulli map. As shown in the previous section,
the region of stability should have reflection symmetry at the horizontal
axis. The stability region for laser networks (Fig.~\ref{fig:4.6}a)
is in agreement with this symmetry. The corresponding stability regions
of the networks of tent maps and Bernoulli maps (Fig.~\ref{fig:4.6}b)
coupled with two delay times, show this reflection symmetry as well.%

\begin{figure}
\begin{centering}
\hfill{}\includegraphics[width=7cm]{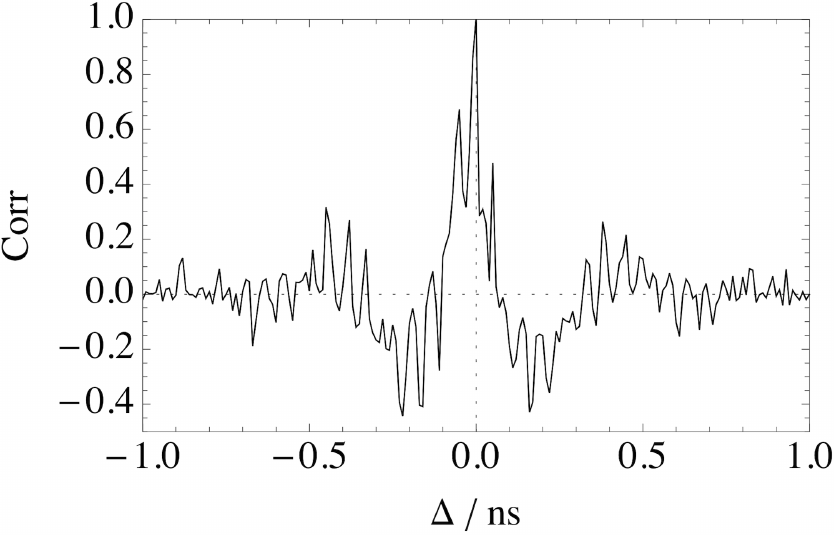}\hfill{}
\par\end{centering}

\caption{Sensitivity of cross-correlations to detuning of the delay times $\tau_{2}$
and $\tau_{1}=\tau_{2}+\Delta$ for a pair of Lang-Kobayashi equations.\label{fig:4.7}}

\end{figure}

In the previous section, we showed that synchronization is sensitive
to detuning of the ratio of the delay times, see Fig.~\ref{fig:CorrelationDelta}.
Fig.~\ref{fig:4.7} shows that synchronization of two lasers is destroyed
if $\tau_{1}$ and $\tau_{2}$ differ by about $10\,\mathrm{ps}$,
which corresponds to the coherence length of the chaotic lasers. The
coupling has a fixed delay time of $\tau_{2}=100\,\mathrm{ns}$ and
a strength of $\sigma_{2}=20\,\mathrm{ns}^{-1}$. The self-feedbacks
have a delay time of $\tau_{1}=\tau_{2}+\Delta$ and a strength of
$\sigma_{2}=30\,\mathrm{ns}^{-1}$. Thus, in agreement with the analytic
results for Bernoulli networks, lasers are sensitive to detuning of
the delay times as well \cite{AviadReidler,Klein:2006:73,Gross:2006}.
Of course, the detailed structure of the cross-correlations of Fig.~\ref{fig:4.7}
depends on the details of the laser dynamics which cannot be predicted
by iterated maps.%

\section{Summary}

Chaos synchronization of networks of identical nonlinear units with
time-delayed couplings is investigated. Although the units are coupled
with long delay times, they synchronize to a common chaotic trajectory
without time shift. For rather general networks with multiple delay
times the method of the master stability function allows to relate
the stability of the synchronization manifold (SM) to the eigenvalue
gaps of the coupling matrices.

For networks of iterated Bernoulli maps the stability of the SM is
calculated analytically in the limit of large delay times. The theory
of polynomials allows to calculate phase diagrams of chaos synchronization
and to derive their symmetries. Finally, these analytic results are
compared with numerical simulations of iterated tent maps and rate
equations for semiconductor lasers (Lang-Kobayashi (LK) equations).
Some results can even be compared with recent experiments on semiconductor
lasers.

For a single delay time, Eq.~(\ref{eq:AbsolutGamma}) is the most
important result. It relates the eigenvalue gap of the coupling matrix
of the whole network to the Lyapunov exponent of the trajectory of
a single unit with feedback. It is exact for Bernoulli networks with
a long delay time. But it compares well with our numerical results
for networks of tent maps and LK units, too. Even the phase diagram
of a directed network with complex eigenvalues, which has been calculated
for lasers, is in good agreement with this fundamental equation. For
a pair of units without self-feedback and for any bipartite network
the eigenvalue gap is zero, hence these networks cannot synchronize
completely in the limit of a large delay time, only cluster synchronization
is possible.

For networks with several delay times, we could not find a simple
relation for the stability of the SM. However, the theory of polynomials
showed some symmetries of the phase diagrams for Bernoulli networks.
Our numerical results for tent maps and LK equations showed these
symmetries, as well. For a pair of units coupled by multiple delay
times these symmetries have interesting consequences. Only for special
ratios of the delay times synchronization is possible in the limit
of long delay times, in agreement with self-consistent arguments related
to mixing of information \cite{KanterZigzagEnglertGeissler}. Again,
this analytic result is in agreement with numerical simulations of
the LK equations and even with experiments on semiconductor lasers
\cite{EnglertKinzelAviad}.

These results show that networks of iterated Bernoulli units have
universal properties. On the one hand, we have analytic tools to calculate
the stability of chaos synchronization since the linearized equations
do not contain the chaotic trajectory. On the other hand, we have
either numerical simulations of the linearized difference/differential
equations containing the chaotic trajectory as input, or we have direct
simulations of the complete network. We have found that there is good
agreement between these different systems, sometimes even on a quantitative
level.

\appendix

\section{The Lang-Kobayashi equations and their master stability equations\label{sec:AppA}}

The Lang-Kobayashi equations in their complex form are\begin{align}
\dot{\mathcal{E}}^{i}\!\left(t\right) & =\frac{1+\mathrm{i}\,\alpha}{2}\, G_{\mathrm{N}}\, n^{i}\!\left(t\right)\mathcal{E}^{i}\!\left(t\right)+\nonumber \\
 & \quad+\sigma_{1}\,\mathcal{E}^{i}\!\left(t-\tau_{1}\right)\mathrm{e}^{-\mathrm{i}\,\omega_{0}\,\tau_{1}}+\nonumber \\
 & \quad+\sigma_{2}\,\sum_{j=1}^{N}G_{ij}\,\mathcal{E}^{j}\!\left(t-\tau_{2}\right)\mathrm{e}^{-\mathrm{i}\,\omega_{0}\,\tau_{2}}\label{eq:a1}\\
\dot{n}^{i}\!\left(t\right) & =\left(p-1\right)J_{\mathrm{th}}-\gamma\, n^{i}\!\left(t\right)-\nonumber \\
 & \quad-\left[\Gamma+G_{\mathrm{N}}\, n^{i}\!\left(t\right)\right]\left|\mathcal{E}^{i}\!\left(t\right)\right|^{2}\label{eq:a2}\end{align}
where $\mathcal{E}^{i}\!\left(t\right)$ is the envelope of the complex
electric field and $n^{i}\!\left(t\right)$ is the renormalized population
inversion of the charge carriers of laser $i$. The used constants
are listed in Tab.~\ref{tab:a1}.%
\begin{table}
\caption{Used constants in the simulation of the Lang-Kobayashi equations.
Values are taken from \cite{AhlersParlitz}\label{tab:a1}.}

\centering{}\begin{tabular}{lcl}
 &  & \tabularnewline
\midrule
\midrule 
Linewidth enhancement factor & $\alpha$ & 5\tabularnewline
\addlinespace
Differential optical gain & $G_{\mathrm{N}}$ & $2.142\times10^{4}\,\mathrm{s}^{-1}$\tabularnewline
\addlinespace
Laser frequency & $\omega_{0}$ & $\frac{2\pi\, c}{635\,\mathrm{nm}}$\tabularnewline
\addlinespace
Pump current relative to $J_{\mathrm{th}}$ & $p$ & $1.02$\tabularnewline
\addlinespace
Threshold pump current of solitary laser & $J_{\mathrm{th}}$ & $\gamma\, N_{\mathrm{sol}}$\tabularnewline
\addlinespace
Carrier decay rate & $\gamma$ & $0.909\times10^{9}\,\mathrm{s}^{-1}$\tabularnewline
\addlinespace
Carrier number of solitary laser & $N_{\mathrm{sol}}$ & $1.707\times10^{8}$\tabularnewline
\addlinespace
Cavity decay rate & $\Gamma$ & $0.357\times10^{12}\,\mathrm{s}^{-1}$\tabularnewline
\midrule
\midrule 
 &  & \tabularnewline
\end{tabular}
\end{table}

For the numerical simulation of the equations, we make the ansatz\begin{equation}
\mathcal{E}^{i}\!\left(t\right)=\mathcal{R}^{i}\!\left(t\right)+\mathrm{i}\,\mathcal{I}^{i}\!\left(t\right)\label{eq:a3}\end{equation}
and thus obtain the real-valued differential equation system\pagebreak{}\begin{align}
\dot{\mathcal{R}}^{i}\!\left(t\right) & =\frac{1}{2}\, G_{\mathrm{N}}\, n^{i}\!\left(t\right)\left[\mathcal{R}^{i}\!\left(t\right)-\alpha\,\mathcal{I}^{i}\!\left(t\right)\right]+\nonumber \\
 & \quad+\sigma_{1}\,\mathcal{R}^{i}\!\left(t-\tau_{1}\right)\cos\!\left(\omega\,\tau_{1}\right)+\nonumber \\
 & \quad+\sigma_{1}\,\mathcal{I}^{i}\!\left(t-\tau_{1}\right)\sin\!\left(\omega\,\tau_{1}\right)+\nonumber \\
 & \quad+\sigma_{2}\,\sum_{j=1}^{N}G_{ij}\,\mathcal{R}^{j}\!\left(t-\tau_{2}\right)\cos\!\left(\omega\,\tau_{2}\right)+\nonumber \\
 & \quad+\sigma_{2}\,\sum_{j=1}^{N}G_{ij}\,\mathcal{I}^{j}\!\left(t-\tau_{2}\right)\sin\!\left(\omega\,\tau_{2}\right)\label{eq:a4}\\
\dot{\mathcal{I}}^{i}\!\left(t\right) & =\frac{1}{2}\, G_{\mathrm{N}}\, n^{i}\!\left(t\right)\left[\mathcal{I}^{i}\!\left(t\right)+\alpha\,\mathcal{R}^{i}\!\left(t\right)\right]-\nonumber \\
 & \quad-\sigma_{1}\,\mathcal{R}^{i}\!\left(t-\tau_{1}\right)\sin\!\left(\omega\,\tau_{1}\right)-\nonumber \\
 & \quad-\sigma_{1}\,\mathcal{I}^{i}\!\left(t-\tau_{1}\right)\cos\!\left(\omega\,\tau_{1}\right)-\nonumber \\
 & \quad-\sigma_{2}\,\sum_{j=1}^{N}G_{ij}\,\mathcal{R}^{j}\!\left(t-\tau_{2}\right)\sin\!\left(\omega\,\tau_{2}\right)-\nonumber \\
 & \quad-\sigma_{2}\,\sum_{j=1}^{N}G_{ij}\,\mathcal{I}^{j}\!\left(t-\tau_{2}\right)\cos\!\left(\omega\,\tau_{2}\right)\label{eq:a5}\\
\dot{n}^{i}\!\left(t\right) & =\left(p-1\right)J_{\mathrm{th}}-\gamma\, n^{i}\!\left(t\right)-\nonumber \\
 & \quad-\left[\Gamma+G_{\mathrm{N}}\, n^{i}\!\left(t\right)\right]\left\{ \left[\mathcal{R}^{i}\!\left(t\right)\right]^{2}+\left[\mathcal{I}^{i}\!\left(t\right)\right]^{2}\right\} \label{eq:a6}\end{align}
We integrate this differential equation system numerically using Heun's
method \cite{Heun} which is a numerical integration method of the
class of Runge-Kutta methods that is particularly suitable for delay-differential
equation systems. The used step size for integration is $\Delta t=0.1\,\mathrm{ps}$.
In order to emulate the measurement of cross-correlations with a gigahertz
oscilloscope in an experimental setup, on the one hand, we define
cross-correlations between two simulated lasers as the cross-correlations
between the absolute values of their electric fields. On the other
hand, we use a sampling time of $1\,\mathrm{ns}$ per data point during
which the absolute values of the electric field are averaged. For
the precision measurement of the sensitivity of cross-correlations
to detuning of the delay times in Fig.~\ref{fig:4.7}, we use a sampling
time of $10\,\mathrm{ps}$. Desynchronization of the simulated lasers
occurs during the power drop-outs of the low frequency fluctuations
which happen on a time scale of the order of magnitude of $10\,\tau$.
In order to avoid measuring cross-correlations during them, the cross-correlations
are calculated ten times from comparatively short time windows of
the length of $5\,\tau$, and then the five largest cross-correlations
are averaged.

The master stability equations of the Lang-Kobayashi equations are\begin{align}
\dot{\delta}_{\mathcal{R}}\!\left(t\right) & =\frac{1}{2}\, G_{\mathrm{N}}\, n\!\left(t\right)\left[\delta_{\mathcal{R}}\!\left(t\right)-\alpha\,\delta_{\mathcal{I}}\!\left(t\right)\right]+\nonumber \\
 & \quad+\frac{1}{2}\, G_{\mathrm{N}}\,\delta_{n}\!\left(t\right)\left[\mathcal{R}\!\left(t\right)-\alpha\,\mathcal{I}\!\left(t\right)\right]+\nonumber \\
 & \quad+\sigma_{1}\,\delta_{\mathcal{R}}\!\left(t-\tau_{1}\right)\cos\!\left(\omega\,\tau_{1}\right)+\nonumber \\
 & \quad+\sigma_{1}\,\delta_{\mathcal{I}}\!\left(t-\tau_{1}\right)\sin\!\left(\omega\,\tau_{1}\right)+\nonumber \\
 & \quad+\sigma_{2}\,\gamma_{k}\,\delta_{\mathcal{R}}\!\left(t-\tau_{2}\right)\cos\!\left(\omega\,\tau_{2}\right)+\nonumber \\
 & \quad+\sigma_{2}\,\gamma_{k}\,\delta_{\mathcal{I}}\!\left(t-\tau_{2}\right)\sin\!\left(\omega\,\tau_{2}\right)\label{eq:a7}\\
\dot{\delta}_{\mathcal{I}}\!\left(t\right) & =\frac{1}{2}\, G_{\mathrm{N}}\, n\!\left(t\right)\left[\delta_{\mathcal{I}}\!\left(t\right)+\alpha\,\delta_{\mathcal{R}}\!\left(t\right)\right]+\nonumber \\
 & \quad+\frac{1}{2}\, G_{\mathrm{N}}\,\delta_{n}\!\left(t\right)\left[\mathcal{I}\!\left(t\right)+\alpha\,\mathcal{R}\!\left(t\right)\right]-\nonumber \\
 & \quad-\sigma_{1}\,\delta_{\mathcal{R}}\!\left(t-\tau_{1}\right)\sin\!\left(\omega\,\tau_{1}\right)-\nonumber \\
 & \quad-\sigma_{1}\,\delta_{\mathcal{I}}\!\left(t-\tau_{1}\right)\cos\!\left(\omega\,\tau_{1}\right)-\nonumber \\
 & \quad-\sigma_{2}\,\gamma_{k}\,\delta_{\mathcal{R}}\!\left(t-\tau_{2}\right)\sin\!\left(\omega\,\tau_{2}\right)-\nonumber \\
 & \quad-\sigma_{2}\,\gamma_{k}\,\delta_{\mathcal{I}}\!\left(t-\tau_{2}\right)\cos\!\left(\omega\,\tau_{2}\right)\label{eq:a8}\\
\dot{\delta}_{n}\!\left(t\right) & =-\left(\gamma+G_{\mathrm{N}}\left\{ \left[\mathcal{R}\!\left(t\right)\right]^{2}+\left[\mathcal{I}\!\left(t\right)\right]^{2}\right\} \right)\delta_{n}\!\left(t\right)-\nonumber \\
 & \quad-2\left[\Gamma+G_{\mathrm{N}}\, n\!\left(t\right)\right]\left[\mathcal{R}\!\left(t\right)\delta_{\mathcal{R}}\!\left(t\right)+\mathcal{I}\!\left(t\right)\delta_{\mathcal{I}}\!\left(t\right)\right]\label{eq:a9}\end{align}
which are integrated numerically using Heun's method, as well. We
calculate the maximal Lyapunov exponent using Farmer's method \cite{FarmerPhysicaD}.
For correspondence to our definition of the cross-correlation above,
we define the deviation of the absolute value of the electric field
as the metric of the separation function. This is legitimate as the
Lang-Kobayashi equations form a strongly coupled differential equation
system.

\section{The Schur-Cohn theorem\label{sub:Schur-Cohn-theorem}}

In order to find the synchronization areas in parameter space, the
Schur-Cohn theorem is a possible method, see \cite{Schur}.The synchronization
area for a network of iterated one-dimensional maps with constant
slope is derived by finding the parameters for which the roots of
the characteristic polynomial lie inside the unit circle.

For a polynom with $P(x)=\sum_{i=0}^{n}a_{i}\, x^{i}$ the Schur-Cohn
theorem defines determinants{\small \begin{equation}
\delta_{\nu+1}=\left|\begin{array}{cccccccc}
a_{n} & 0 & ... & 0 & a_{0} & a_{1} & ... & a_{\nu}\\
a_{n-1} & a_{n} & ... & 0 & 0 & a_{0} & ... & a_{\nu-1}\\
... & ... & ... & ... & ... & ... & ... & ...\\
a_{n-\nu} & a_{n-\nu+1} & ... & a_{n} & 0 & 0 & ... & a_{0}\\
\overline{a_{0}} & 0 & ... & 0 & \overline{a_{n}} & \overline{a_{n-1}} & ... & \overline{a_{n-\nu}}\\
\overline{a_{1}} & \overline{a_{0}} & ... & 0 & 0 & \overline{a_{n}} & ... & \overline{a_{n-\nu+1}}\\
... & ... & ... & ... & ... & ... & ... & ...\\
\overline{a_{\nu}} & \overline{a_{\nu-1}} & ... & \overline{a_{0}} & 0 & 0 & ... & \overline{a_{n}}\end{array}\right|\label{eq:SchurCohn}\end{equation}
} with $\nu=0,1,...,n-1$.

Due to the Schur-Cohn theorem, it is $|x|<1$ when all determinants
are greater than 0.

\section{Acknowledgements}

We would like to thank the Deutsche Forschungsgemeinschaft and the
Leibniz-Rechenzentrum in Garching, Germany for their support of this
work.

\bibliography{CHAOS}

\begin{thebibliography}{56}
\expandafter\ifx\csname natexlab\endcsname\relax\def\natexlab#1{#1}\fi
\expandafter\ifx\csname bibnamefont\endcsname\relax
  \def\bibnamefont#1{#1}\fi
\expandafter\ifx\csname bibfnamefont\endcsname\relax
  \def\bibfnamefont#1{#1}\fi
\expandafter\ifx\csname citenamefont\endcsname\relax
  \def\citenamefont#1{#1}\fi
\expandafter\ifx\csname url\endcsname\relax
  \def\url#1{\texttt{#1}}\fi
\expandafter\ifx\csname urlprefix\endcsname\relax\def\urlprefix{URL }\fi
\providecommand{\bibinfo}[2]{#2}
\providecommand{\eprint}[2][]{\url{#2}}

\bibitem[{\citenamefont{Pikovsky et~al.}(2001)\citenamefont{Pikovsky,
  Rosenblum, and Kurths}}]{Pikovsky:Buch}
\bibinfo{author}{\bibfnamefont{A.}~\bibnamefont{Pikovsky}},
  \bibinfo{author}{\bibfnamefont{M.}~\bibnamefont{Rosenblum}},
  \bibnamefont{and} \bibinfo{author}{\bibfnamefont{J.}~\bibnamefont{Kurths}},
  \emph{\bibinfo{title}{Synchronization, a universal concept in nonlinear
  sciences}} (\bibinfo{publisher}{Cambridge University Press},
  \bibinfo{address}{Cambridge}, \bibinfo{year}{2001}).

\bibitem[{\citenamefont{{Boccaletti} et~al.}(2002)\citenamefont{{Boccaletti},
  {Kurths}, {Osipov}, {Valladares}, and {Zhou}}}]{BoccalettiKurths}
\bibinfo{author}{\bibfnamefont{S.}~\bibnamefont{{Boccaletti}}},
  \bibinfo{author}{\bibfnamefont{J.}~\bibnamefont{{Kurths}}},
  \bibinfo{author}{\bibfnamefont{G.}~\bibnamefont{{Osipov}}},
  \bibinfo{author}{\bibfnamefont{D.~L.} \bibnamefont{{Valladares}}},
  \bibnamefont{and} \bibinfo{author}{\bibfnamefont{C.~S.}
  \bibnamefont{{Zhou}}}, \bibinfo{journal}{Phys. Rep.}
  \textbf{\bibinfo{volume}{366}}, \bibinfo{pages}{1} (\bibinfo{year}{2002}).

\bibitem[{\citenamefont{{Balanov} et~al.}(2009)\citenamefont{{Balanov},
  {Janson}, {Postnov}, and {Sosnovtseva}}}]{BalanovJansonBook}
\bibinfo{author}{\bibfnamefont{A.}~\bibnamefont{{Balanov}}},
  \bibinfo{author}{\bibfnamefont{N.}~\bibnamefont{{Janson}}},
  \bibinfo{author}{\bibfnamefont{D.}~\bibnamefont{{Postnov}}},
  \bibnamefont{and}
  \bibinfo{author}{\bibfnamefont{O.}~\bibnamefont{{Sosnovtseva}}},
  \emph{\bibinfo{title}{{Synchronization: From Simple to Complex}}}
  (\bibinfo{year}{2009}).

\bibitem[{\citenamefont{{Mosekilde} et~al.}(2002)\citenamefont{{Mosekilde},
  {Maistrenko}, and {Postnov}}}]{MosekildeMaistrenkoBook}
\bibinfo{author}{\bibfnamefont{E.}~\bibnamefont{{Mosekilde}}},
  \bibinfo{author}{\bibfnamefont{Y.}~\bibnamefont{{Maistrenko}}},
  \bibnamefont{and}
  \bibinfo{author}{\bibfnamefont{D.}~\bibnamefont{{Postnov}}},
  \emph{\bibinfo{title}{{Chaotic Synchronization: Application to living
  systems}}} (\bibinfo{year}{2002}).

\bibitem[{\citenamefont{Sch{\"o}ll and Schuster}(2008)}]{SchoellSchuster}
\bibinfo{editor}{\bibfnamefont{E.}~\bibnamefont{Sch{\"o}ll}} \bibnamefont{and}
  \bibinfo{editor}{\bibfnamefont{H.~G.} \bibnamefont{Schuster}}, eds.,
  \emph{\bibinfo{title}{Handbook of Chaos Control}}
  (\bibinfo{publisher}{Wiley-VCH}, \bibinfo{address}{Weinheim},
  \bibinfo{year}{2008}).

\bibitem[{\citenamefont{Wu}(2007)}]{Wu}
\bibinfo{author}{\bibfnamefont{C.~W.} \bibnamefont{Wu}},
  \emph{\bibinfo{title}{Synchronization in complex networks of nonlinear
  dynamical systems}} (\bibinfo{publisher}{World Scientific, Singapore},
  \bibinfo{year}{2007}).

\bibitem[{\citenamefont{{Arenas} et~al.}(2008)\citenamefont{{Arenas},
  {D{\'\i}az-Guilera}, {Kurths}, {Moreno}, and {Zhou}}}]{Arenas}
\bibinfo{author}{\bibfnamefont{A.}~\bibnamefont{{Arenas}}},
  \bibinfo{author}{\bibfnamefont{A.}~\bibnamefont{{D{\'\i}az-Guilera}}},
  \bibinfo{author}{\bibfnamefont{J.}~\bibnamefont{{Kurths}}},
  \bibinfo{author}{\bibfnamefont{Y.}~\bibnamefont{{Moreno}}}, \bibnamefont{and}
  \bibinfo{author}{\bibfnamefont{C.}~\bibnamefont{{Zhou}}},
  \bibinfo{journal}{Physical Reports} \textbf{\bibinfo{volume}{469}},
  \bibinfo{pages}{93} (\bibinfo{year}{2008}).

\bibitem[{\citenamefont{{Colet} and {Roy}}(1994)}]{ColetRoy}
\bibinfo{author}{\bibfnamefont{P.}~\bibnamefont{{Colet}}} \bibnamefont{and}
  \bibinfo{author}{\bibfnamefont{R.}~\bibnamefont{{Roy}}},
  \bibinfo{journal}{Optics Letters} \textbf{\bibinfo{volume}{19}},
  \bibinfo{pages}{2056} (\bibinfo{year}{1994}).

\bibitem[{\citenamefont{Uchida et~al.}(2005)\citenamefont{Uchida, Rogister,
  Garcfa-Ojalvo, and Roy}}]{UchidaRogister}
\bibinfo{author}{\bibfnamefont{A.}~\bibnamefont{Uchida}},
  \bibinfo{author}{\bibfnamefont{F.}~\bibnamefont{Rogister}},
  \bibinfo{author}{\bibfnamefont{J.}~\bibnamefont{Garcfa-Ojalvo}},
  \bibnamefont{and} \bibinfo{author}{\bibfnamefont{R.}~\bibnamefont{Roy}},
  \bibinfo{journal}{Prog. Optics} \textbf{\bibinfo{volume}{48}},
  \bibinfo{pages}{203} (\bibinfo{year}{2005}).

\bibitem[{\citenamefont{VanWiggeren and Roy}(1998)}]{VanWiggeren:1998}
\bibinfo{author}{\bibfnamefont{G.~D.} \bibnamefont{VanWiggeren}}
  \bibnamefont{and} \bibinfo{author}{\bibfnamefont{R.}~\bibnamefont{Roy}},
  \bibinfo{journal}{Science} \textbf{\bibinfo{volume}{279}},
  \bibinfo{pages}{1198} (\bibinfo{year}{1998}).

\bibitem[{\citenamefont{Argyris et~al.}(2005)\citenamefont{Argyris, Syvridis,
  Larger, Annovazzi-Lodi, Colet, Fischer, Garc{\'\i}a-Ojalvo, Mirasso,
  Pesquera, and Shore}}]{Argyris:2005}
\bibinfo{author}{\bibfnamefont{A.}~\bibnamefont{Argyris}},
  \bibinfo{author}{\bibfnamefont{D.}~\bibnamefont{Syvridis}},
  \bibinfo{author}{\bibfnamefont{L.}~\bibnamefont{Larger}},
  \bibinfo{author}{\bibfnamefont{V.}~\bibnamefont{Annovazzi-Lodi}},
  \bibinfo{author}{\bibfnamefont{P.}~\bibnamefont{Colet}},
  \bibinfo{author}{\bibfnamefont{I.}~\bibnamefont{Fischer}},
  \bibinfo{author}{\bibfnamefont{J.}~\bibnamefont{Garc{\'\i}a-Ojalvo}},
  \bibinfo{author}{\bibfnamefont{C.~R.} \bibnamefont{Mirasso}},
  \bibinfo{author}{\bibfnamefont{L.}~\bibnamefont{Pesquera}}, \bibnamefont{and}
  \bibinfo{author}{\bibfnamefont{K.~A.} \bibnamefont{Shore}},
  \bibinfo{journal}{Nature} \textbf{\bibinfo{volume}{437}},
  \bibinfo{pages}{343} (\bibinfo{year}{2005}).

\bibitem[{\citenamefont{{Atay} and
  {B{\i}y{\i}ko\u{g}lu}}(2005)}]{AtayBijikouglu}
\bibinfo{author}{\bibfnamefont{F.~M.} \bibnamefont{{Atay}}} \bibnamefont{and}
  \bibinfo{author}{\bibfnamefont{T.}~\bibnamefont{{B{\i}y{\i}ko\u{g}lu}}},
  \bibinfo{journal}{Phys. Rev. E} \textbf{\bibinfo{volume}{72}},
  \bibinfo{pages}{016217} (\bibinfo{year}{2005}).

\bibitem[{\citenamefont{Baek and Ott}(2004)}]{BaekOtt}
\bibinfo{author}{\bibfnamefont{S.-J.} \bibnamefont{Baek}} \bibnamefont{and}
  \bibinfo{author}{\bibfnamefont{E.}~\bibnamefont{Ott}},
  \bibinfo{journal}{Phys. Rev. E} \textbf{\bibinfo{volume}{69}},
  \bibinfo{pages}{066210} (\bibinfo{year}{2004}).

\bibitem[{\citenamefont{Barahona and Pecora}(2002)}]{BarahonaPecora}
\bibinfo{author}{\bibfnamefont{M.}~\bibnamefont{Barahona}} \bibnamefont{and}
  \bibinfo{author}{\bibfnamefont{L.~M.} \bibnamefont{Pecora}},
  \bibinfo{journal}{Phys. Rev. Lett.} \textbf{\bibinfo{volume}{89}},
  \bibinfo{pages}{054101} (\bibinfo{year}{2002}).

\bibitem[{\citenamefont{Jost and Joy}(2001)}]{JostJoy}
\bibinfo{author}{\bibfnamefont{J.}~\bibnamefont{Jost}} \bibnamefont{and}
  \bibinfo{author}{\bibfnamefont{M.~P.} \bibnamefont{Joy}},
  \bibinfo{journal}{Phys. Rev. E} \textbf{\bibinfo{volume}{65}},
  \bibinfo{pages}{016201} (\bibinfo{year}{2001}).

\bibitem[{\citenamefont{Li et~al.}(2007)\citenamefont{Li, Sun, and
  Kurths}}]{LiSunKurthsPRE}
\bibinfo{author}{\bibfnamefont{C.}~\bibnamefont{Li}},
  \bibinfo{author}{\bibfnamefont{W.}~\bibnamefont{Sun}}, \bibnamefont{and}
  \bibinfo{author}{\bibfnamefont{J.}~\bibnamefont{Kurths}},
  \bibinfo{journal}{Phys. Rev. E} \textbf{\bibinfo{volume}{76}},
  \bibinfo{eid}{046204} (\bibinfo{year}{2007}).

\bibitem[{\citenamefont{Pecora and Carroll}(1990)}]{Pecora:1990}
\bibinfo{author}{\bibfnamefont{L.~M.} \bibnamefont{Pecora}} \bibnamefont{and}
  \bibinfo{author}{\bibfnamefont{T.~L.} \bibnamefont{Carroll}},
  \bibinfo{journal}{Phys. Rev. Lett.} \textbf{\bibinfo{volume}{64}},
  \bibinfo{pages}{821} (\bibinfo{year}{1990}).

\bibitem[{\citenamefont{{de San Roman} et~al.}(1998)\citenamefont{{de San
  Roman}, Boccaletti, Maza, and Mancini}}]{RomanBoccaletti}
\bibinfo{author}{\bibfnamefont{F.~S.} \bibnamefont{{de San Roman}}},
  \bibinfo{author}{\bibfnamefont{S.}~\bibnamefont{Boccaletti}},
  \bibinfo{author}{\bibfnamefont{D.}~\bibnamefont{Maza}}, \bibnamefont{and}
  \bibinfo{author}{\bibfnamefont{H.}~\bibnamefont{Mancini}},
  \bibinfo{journal}{Phys. Rev. Lett.} \textbf{\bibinfo{volume}{81}},
  \bibinfo{pages}{3639} (\bibinfo{year}{1998}).

\bibitem[{\citenamefont{{Atay} et~al.}(2004)\citenamefont{{Atay}, {Jost}, and
  {Wende}}}]{AtayJost}
\bibinfo{author}{\bibfnamefont{F.~M.} \bibnamefont{{Atay}}},
  \bibinfo{author}{\bibfnamefont{J.}~\bibnamefont{{Jost}}}, \bibnamefont{and}
  \bibinfo{author}{\bibfnamefont{A.}~\bibnamefont{{Wende}}},
  \bibinfo{journal}{Phys. Rev. Lett.} \textbf{\bibinfo{volume}{92}},
  \bibinfo{pages}{144101} (\bibinfo{year}{2004}).

\bibitem[{\citenamefont{{Choe} et~al.}(2010)\citenamefont{{Choe}, {Dahms},
  {H{\"o}vel}, and {Sch{\"o}ll}}}]{Choe}
\bibinfo{author}{\bibfnamefont{C.}~\bibnamefont{{Choe}}},
  \bibinfo{author}{\bibfnamefont{T.}~\bibnamefont{{Dahms}}},
  \bibinfo{author}{\bibfnamefont{P.}~\bibnamefont{{H{\"o}vel}}},
  \bibnamefont{and}
  \bibinfo{author}{\bibfnamefont{E.}~\bibnamefont{{Sch{\"o}ll}}},
  \bibinfo{journal}{\pre} \textbf{\bibinfo{volume}{81}},
  \bibinfo{pages}{025205} (\bibinfo{year}{2010}), \eprint{0908.3984}.

\bibitem[{\citenamefont{Earl and Strogatz}(2003)}]{EarlStrogatz}
\bibinfo{author}{\bibfnamefont{M.~G.} \bibnamefont{Earl}} \bibnamefont{and}
  \bibinfo{author}{\bibfnamefont{S.~H.} \bibnamefont{Strogatz}},
  \bibinfo{journal}{Phys. Rev. E} \textbf{\bibinfo{volume}{67}},
  \bibinfo{pages}{036204} (\bibinfo{year}{2003}).

\bibitem[{\citenamefont{{H{\"o}vel} et~al.}(2010)\citenamefont{{H{\"o}vel},
  {Dahlem}, and {Sch{\"o}ll}}}]{hovel}
\bibinfo{author}{\bibfnamefont{P.}~\bibnamefont{{H{\"o}vel}}},
  \bibinfo{author}{\bibfnamefont{M.~A.} \bibnamefont{{Dahlem}}},
  \bibnamefont{and}
  \bibinfo{author}{\bibfnamefont{E.}~\bibnamefont{{Sch{\"o}ll}}},
  \bibinfo{journal}{Int. J. Bifurcation Chaos Appl. Sci. Eng.}
  \textbf{\bibinfo{volume}{20}}, \bibinfo{pages}{813} (\bibinfo{year}{2010}).

\bibitem[{\citenamefont{Kanter et~al.}(2010{\natexlab{a}})\citenamefont{Kanter,
  Zigzag, Englert, Geissler, and Kinzel}}]{KanterZigzagEnglertGeissler}
\bibinfo{author}{\bibfnamefont{I.}~\bibnamefont{Kanter}},
  \bibinfo{author}{\bibfnamefont{M.}~\bibnamefont{Zigzag}},
  \bibinfo{author}{\bibfnamefont{A.}~\bibnamefont{Englert}},
  \bibinfo{author}{\bibfnamefont{F.}~\bibnamefont{Geissler}}, \bibnamefont{and}
  \bibinfo{author}{\bibfnamefont{W.}~\bibnamefont{Kinzel}},
  \bibinfo{journal}{arXiv:1012.0990}  (\bibinfo{year}{2010}{\natexlab{a}}).

\bibitem[{\citenamefont{Kestler et~al.}(2007)\citenamefont{Kestler, Kinzel, and
  Kanter}}]{KestlerKinzelKanter}
\bibinfo{author}{\bibfnamefont{J.}~\bibnamefont{Kestler}},
  \bibinfo{author}{\bibfnamefont{W.}~\bibnamefont{Kinzel}}, \bibnamefont{and}
  \bibinfo{author}{\bibfnamefont{I.}~\bibnamefont{Kanter}},
  \bibinfo{journal}{Phys. Rev. E} \textbf{\bibinfo{volume}{76}},
  \bibinfo{eid}{035202} (\bibinfo{year}{2007}).

\bibitem[{\citenamefont{{Kinzel} et~al.}(2009)\citenamefont{{Kinzel},
  {Englert}, {Reents}, {Zigzag}, and {Kanter}}}]{KinzelEnglert}
\bibinfo{author}{\bibfnamefont{W.}~\bibnamefont{{Kinzel}}},
  \bibinfo{author}{\bibfnamefont{A.}~\bibnamefont{{Englert}}},
  \bibinfo{author}{\bibfnamefont{G.}~\bibnamefont{{Reents}}},
  \bibinfo{author}{\bibfnamefont{M.}~\bibnamefont{{Zigzag}}}, \bibnamefont{and}
  \bibinfo{author}{\bibfnamefont{I.}~\bibnamefont{{Kanter}}},
  \bibinfo{journal}{Phys. Rev. E} \textbf{\bibinfo{volume}{79}},
  \bibinfo{pages}{056207} (\bibinfo{year}{2009}).

\bibitem[{\citenamefont{Kozyreff et~al.}(2000)\citenamefont{Kozyreff,
  Vladimirov, and Mandel}}]{KozyreffVladimirov}
\bibinfo{author}{\bibfnamefont{G.}~\bibnamefont{Kozyreff}},
  \bibinfo{author}{\bibfnamefont{A.~G.} \bibnamefont{Vladimirov}},
  \bibnamefont{and} \bibinfo{author}{\bibfnamefont{P.}~\bibnamefont{Mandel}},
  \bibinfo{journal}{Phys. Rev. Lett.} \textbf{\bibinfo{volume}{85}},
  \bibinfo{pages}{3809} (\bibinfo{year}{2000}).

\bibitem[{\citenamefont{{Masoller} and A.{Marti}}(2005)}]{MasollerMart}
\bibinfo{author}{\bibfnamefont{C.}~\bibnamefont{{Masoller}}} \bibnamefont{and}
  \bibinfo{author}{\bibnamefont{A.{Marti}}}, \bibinfo{journal}{Phys. Rev.
  Lett.} \textbf{\bibinfo{volume}{94}}, \bibinfo{pages}{134102}
  (\bibinfo{year}{2005}).

\bibitem[{\citenamefont{Schmitzer et~al.}(2009)\citenamefont{Schmitzer, Kinzel,
  and Kanter}}]{SchmitzerKinzelKanter}
\bibinfo{author}{\bibfnamefont{B.}~\bibnamefont{Schmitzer}},
  \bibinfo{author}{\bibfnamefont{W.}~\bibnamefont{Kinzel}}, \bibnamefont{and}
  \bibinfo{author}{\bibfnamefont{I.}~\bibnamefont{Kanter}},
  \bibinfo{journal}{Phys. Rev. E} \textbf{\bibinfo{volume}{80}},
  \bibinfo{pages}{047203} (\bibinfo{year}{2009}).

\bibitem[{\citenamefont{Pecora and Carroll}(1998)}]{PecoraCarrollMSF}
\bibinfo{author}{\bibfnamefont{L.~M.} \bibnamefont{Pecora}} \bibnamefont{and}
  \bibinfo{author}{\bibfnamefont{T.~L.} \bibnamefont{Carroll}},
  \bibinfo{journal}{Phys. Rev. Lett.} \textbf{\bibinfo{volume}{80}},
  \bibinfo{pages}{2109} (\bibinfo{year}{1998}).

\bibitem[{\citenamefont{{Flunkert} et~al.}(2010)\citenamefont{{Flunkert},
  {Yanchuk}, {Dahms}, and {Schoell}}}]{FlunkertYanchuk}
\bibinfo{author}{\bibfnamefont{V.}~\bibnamefont{{Flunkert}}},
  \bibinfo{author}{\bibfnamefont{S.}~\bibnamefont{{Yanchuk}}},
  \bibinfo{author}{\bibfnamefont{T.}~\bibnamefont{{Dahms}}}, \bibnamefont{and}
  \bibinfo{author}{\bibfnamefont{E.}~\bibnamefont{{Schoell}}},
  \bibinfo{journal}{to be published}  (\bibinfo{year}{2010}).

\bibitem[{\citenamefont{Kestler et~al.}(2008)\citenamefont{Kestler, Kopelowitz,
  Kanter, and Kinzel}}]{KestlerKopelowitz}
\bibinfo{author}{\bibfnamefont{J.}~\bibnamefont{Kestler}},
  \bibinfo{author}{\bibfnamefont{E.}~\bibnamefont{Kopelowitz}},
  \bibinfo{author}{\bibfnamefont{I.}~\bibnamefont{Kanter}}, \bibnamefont{and}
  \bibinfo{author}{\bibfnamefont{W.}~\bibnamefont{Kinzel}},
  \bibinfo{journal}{Phys. Rev. E} \textbf{\bibinfo{volume}{77}},
  \bibinfo{eid}{046209} (\bibinfo{year}{2008}).

\bibitem[{\citenamefont{Dhamala et~al.}(2004)\citenamefont{Dhamala, Jirsa, and
  Ding}}]{DhamalaJirsa}
\bibinfo{author}{\bibfnamefont{M.}~\bibnamefont{Dhamala}},
  \bibinfo{author}{\bibfnamefont{V.~K.} \bibnamefont{Jirsa}}, \bibnamefont{and}
  \bibinfo{author}{\bibfnamefont{M.}~\bibnamefont{Ding}},
  \bibinfo{journal}{Phys. Rev. Lett.} \textbf{\bibinfo{volume}{92}},
  \bibinfo{pages}{074104} (\bibinfo{year}{2004}).

\bibitem[{\citenamefont{Shahverdiev and Shore}(2008)}]{ShahverdievShore}
\bibinfo{author}{\bibfnamefont{E.~M.} \bibnamefont{Shahverdiev}}
  \bibnamefont{and} \bibinfo{author}{\bibfnamefont{K.~A.} \bibnamefont{Shore}},
  \bibinfo{journal}{Phys. Rev. E} \textbf{\bibinfo{volume}{77}},
  \bibinfo{pages}{057201} (\bibinfo{year}{2008}).

\bibitem[{\citenamefont{{Just} et~al.}(1999)\citenamefont{{Just}, {Reibold},
  {Benner}, {Kacperski}, {Fronczak}, and {Ho{\l}yst}}}]{JustReibold}
\bibinfo{author}{\bibfnamefont{W.}~\bibnamefont{{Just}}},
  \bibinfo{author}{\bibfnamefont{E.}~\bibnamefont{{Reibold}}},
  \bibinfo{author}{\bibfnamefont{H.}~\bibnamefont{{Benner}}},
  \bibinfo{author}{\bibfnamefont{K.}~\bibnamefont{{Kacperski}}},
  \bibinfo{author}{\bibfnamefont{P.}~\bibnamefont{{Fronczak}}},
  \bibnamefont{and}
  \bibinfo{author}{\bibfnamefont{J.}~\bibnamefont{{Ho{\l}yst}}},
  \bibinfo{journal}{Phys. Lett. A} \textbf{\bibinfo{volume}{254}},
  \bibinfo{pages}{158} (\bibinfo{year}{1999}).

\bibitem[{\citenamefont{Zigzag et~al.}(2010)\citenamefont{Zigzag, Butkowski,
  Englert, Kinzel, and Kanter}}]{ZigzagButkowskiEnglert}
\bibinfo{author}{\bibfnamefont{M.}~\bibnamefont{Zigzag}},
  \bibinfo{author}{\bibfnamefont{M.}~\bibnamefont{Butkowski}},
  \bibinfo{author}{\bibfnamefont{A.}~\bibnamefont{Englert}},
  \bibinfo{author}{\bibfnamefont{W.}~\bibnamefont{Kinzel}}, \bibnamefont{and}
  \bibinfo{author}{\bibfnamefont{I.}~\bibnamefont{Kanter}},
  \bibinfo{journal}{Phys. Rev. E} \textbf{\bibinfo{volume}{81}},
  \bibinfo{pages}{036215} (\bibinfo{year}{2010}).

\bibitem[{\citenamefont{Berman and Plemmons}(1979)}]{BermanPlemmons}
\bibinfo{author}{\bibfnamefont{A.}~\bibnamefont{Berman}} \bibnamefont{and}
  \bibinfo{author}{\bibfnamefont{R.}~\bibnamefont{Plemmons}},
  \emph{\bibinfo{title}{Nonnegative matrices in the mathematical sciences}}
  (\bibinfo{publisher}{Academic Press}, \bibinfo{address}{New York},
  \bibinfo{year}{1979}).

\bibitem[{\citenamefont{Gershgorin}(1931)}]{Gershgorin}
\bibinfo{author}{\bibfnamefont{S.}~\bibnamefont{Gershgorin}},
  \bibinfo{journal}{Bulletin de l'acad{/'e}mie des Sciences de l'URSS. Classe
  des sciences mathematiques et na} \textbf{\bibinfo{volume}{6}},
  \bibinfo{pages}{749} (\bibinfo{year}{1931}).

\bibitem[{\citenamefont{Schur}(1916)}]{Schur}
\bibinfo{author}{\bibfnamefont{I.}~\bibnamefont{Schur}}, \bibinfo{journal}{J.
  Math.} \textbf{\bibinfo{volume}{148}}, \bibinfo{pages}{205}
  (\bibinfo{year}{1916}).

\bibitem[{\citenamefont{{Lepri} et~al.}(1994)\citenamefont{{Lepri},
  {Giacomelli}, {Politi}, and {Arecchi}}}]{LepriGiacomelli}
\bibinfo{author}{\bibfnamefont{S.}~\bibnamefont{{Lepri}}},
  \bibinfo{author}{\bibfnamefont{G.}~\bibnamefont{{Giacomelli}}},
  \bibinfo{author}{\bibfnamefont{A.}~\bibnamefont{{Politi}}}, \bibnamefont{and}
  \bibinfo{author}{\bibfnamefont{F.~T.} \bibnamefont{{Arecchi}}},
  \bibinfo{journal}{Physica D} \textbf{\bibinfo{volume}{70}},
  \bibinfo{pages}{235} (\bibinfo{year}{1994}).

\bibitem[{\citenamefont{Farmer}(1982)}]{FarmerPhysicaD}
\bibinfo{author}{\bibfnamefont{J.~D.} \bibnamefont{Farmer}},
  \bibinfo{journal}{Physica D} \textbf{\bibinfo{volume}{4}},
  \bibinfo{pages}{366} (\bibinfo{year}{1982}).

\bibitem[{\citenamefont{{Erd{\"o}s} and {Turan}}(1950)}]{ErdoesTuran}
\bibinfo{author}{\bibfnamefont{P.}~\bibnamefont{{Erd{\"o}s}}} \bibnamefont{and}
  \bibinfo{author}{\bibfnamefont{P.}~\bibnamefont{{Turan}}},
  \bibinfo{journal}{Ann. Math.} \textbf{\bibinfo{volume}{51}},
  \bibinfo{pages}{105} (\bibinfo{year}{1950}).

\bibitem[{\citenamefont{Granville}(2007)}]{Granville}
\bibinfo{author}{\bibfnamefont{A.}~\bibnamefont{Granville}}, in
  \emph{\bibinfo{booktitle}{Equidistribution in Number Theory, An
  Introduction}}, edited by
  \bibinfo{editor}{\bibfnamefont{A.}~\bibnamefont{Granville}} \bibnamefont{and}
  \bibinfo{editor}{\bibfnamefont{Z.}~\bibnamefont{Rudnick}}
  (\bibinfo{publisher}{Springer Netherlands}, \bibinfo{year}{2007}).

\bibitem[{\citenamefont{{Yanchuk} and
  {Perlikowski}}(2009)}]{YanchukPerlikowski}
\bibinfo{author}{\bibfnamefont{S.}~\bibnamefont{{Yanchuk}}} \bibnamefont{and}
  \bibinfo{author}{\bibfnamefont{P.}~\bibnamefont{{Perlikowski}}},
  \bibinfo{journal}{Phys. Rev. E.} \textbf{\bibinfo{volume}{79}},
  \bibinfo{pages}{046221} (\bibinfo{year}{2009}).

\bibitem[{\citenamefont{Zigzag et~al.}(2009)\citenamefont{Zigzag, Butkowski,
  Englert, Kinzel, and Kanter}}]{ZigzagButkowskiEnglertEuro}
\bibinfo{author}{\bibfnamefont{M.}~\bibnamefont{Zigzag}},
  \bibinfo{author}{\bibfnamefont{M.}~\bibnamefont{Butkowski}},
  \bibinfo{author}{\bibfnamefont{A.}~\bibnamefont{Englert}},
  \bibinfo{author}{\bibfnamefont{W.}~\bibnamefont{Kinzel}}, \bibnamefont{and}
  \bibinfo{author}{\bibfnamefont{I.}~\bibnamefont{Kanter}},
  \bibinfo{journal}{Europhys. Lett.} \textbf{\bibinfo{volume}{85}},
  \bibinfo{pages}{60005} (\bibinfo{year}{2009}).

\bibitem[{\citenamefont{{Lang} and {Kobayashi}}(1980)}]{LangKobayashi}
\bibinfo{author}{\bibfnamefont{R.}~\bibnamefont{{Lang}}} \bibnamefont{and}
  \bibinfo{author}{\bibfnamefont{K.}~\bibnamefont{{Kobayashi}}},
  \bibinfo{journal}{IEEE Journal of Quantum Electronics}
  \textbf{\bibinfo{volume}{16}}, \bibinfo{pages}{347} (\bibinfo{year}{1980}).

\bibitem[{\citenamefont{Fischer et~al.}(2006)\citenamefont{Fischer, Vicente,
  Buldu, Peil, Mirasso, Torrent, and Garc{\'\i}a-Ojalvo}}]{Fischer:2006:PRL}
\bibinfo{author}{\bibfnamefont{I.}~\bibnamefont{Fischer}},
  \bibinfo{author}{\bibfnamefont{R.}~\bibnamefont{Vicente}},
  \bibinfo{author}{\bibfnamefont{J.~M.} \bibnamefont{Buldu}},
  \bibinfo{author}{\bibfnamefont{M.}~\bibnamefont{Peil}},
  \bibinfo{author}{\bibfnamefont{C.~R.} \bibnamefont{Mirasso}},
  \bibinfo{author}{\bibfnamefont{M.~C.} \bibnamefont{Torrent}},
  \bibnamefont{and}
  \bibinfo{author}{\bibfnamefont{J.}~\bibnamefont{Garc{\'\i}a-Ojalvo}},
  \bibinfo{journal}{Phys. Rev. Lett.} \textbf{\bibinfo{volume}{97}},
  \bibinfo{eid}{123902} (\bibinfo{year}{2006}).

\bibitem[{\citenamefont{Rosenbluh et~al.}(2007)\citenamefont{Rosenbluh, Aviad,
  Cohen, Khaykovich, Kinzel, Kopelowitz, Yoskovits, and
  Kanter}}]{Rosenbluh:046207}
\bibinfo{author}{\bibfnamefont{M.}~\bibnamefont{Rosenbluh}},
  \bibinfo{author}{\bibfnamefont{Y.}~\bibnamefont{Aviad}},
  \bibinfo{author}{\bibfnamefont{E.}~\bibnamefont{Cohen}},
  \bibinfo{author}{\bibfnamefont{L.}~\bibnamefont{Khaykovich}},
  \bibinfo{author}{\bibfnamefont{W.}~\bibnamefont{Kinzel}},
  \bibinfo{author}{\bibfnamefont{E.}~\bibnamefont{Kopelowitz}},
  \bibinfo{author}{\bibfnamefont{P.}~\bibnamefont{Yoskovits}},
  \bibnamefont{and} \bibinfo{author}{\bibfnamefont{I.}~\bibnamefont{Kanter}},
  \bibinfo{journal}{Phys. Rev. E} \textbf{\bibinfo{volume}{76}},
  \bibinfo{eid}{046207} (\bibinfo{year}{2007}).

\bibitem[{\citenamefont{Klein et~al.}(2006{\natexlab{a}})\citenamefont{Klein,
  Gross, Rosenbluh, Kinzel, Khaykovich, and Kanter}}]{Klein:2006:73}
\bibinfo{author}{\bibfnamefont{E.}~\bibnamefont{Klein}},
  \bibinfo{author}{\bibfnamefont{N.}~\bibnamefont{Gross}},
  \bibinfo{author}{\bibfnamefont{M.}~\bibnamefont{Rosenbluh}},
  \bibinfo{author}{\bibfnamefont{W.}~\bibnamefont{Kinzel}},
  \bibinfo{author}{\bibfnamefont{L.}~\bibnamefont{Khaykovich}},
  \bibnamefont{and} \bibinfo{author}{\bibfnamefont{I.}~\bibnamefont{Kanter}},
  \bibinfo{journal}{Phys. Rev. E} \textbf{\bibinfo{volume}{73}},
  \bibinfo{eid}{066214} (\bibinfo{year}{2006}{\natexlab{a}}).

\bibitem[{\citenamefont{Gross et~al.}(2006)\citenamefont{Gross, Kinzel, Kanter,
  Rosenbluh, and Khaykovich}}]{Gross:2006}
\bibinfo{author}{\bibfnamefont{N.}~\bibnamefont{Gross}},
  \bibinfo{author}{\bibfnamefont{W.}~\bibnamefont{Kinzel}},
  \bibinfo{author}{\bibfnamefont{I.}~\bibnamefont{Kanter}},
  \bibinfo{author}{\bibfnamefont{M.}~\bibnamefont{Rosenbluh}},
  \bibnamefont{and}
  \bibinfo{author}{\bibfnamefont{L.}~\bibnamefont{Khaykovich}},
  \bibinfo{journal}{Optics Comm.} \textbf{\bibinfo{volume}{267}},
  \bibinfo{pages}{464} (\bibinfo{year}{2006}).

\bibitem[{\citenamefont{Klein et~al.}(2006{\natexlab{b}})\citenamefont{Klein,
  Gross, Kopelowitz, Rosenbluh, Khaykovich, Kinzel, and
  Kanter}}]{Klein:2006:74}
\bibinfo{author}{\bibfnamefont{E.}~\bibnamefont{Klein}},
  \bibinfo{author}{\bibfnamefont{N.}~\bibnamefont{Gross}},
  \bibinfo{author}{\bibfnamefont{E.}~\bibnamefont{Kopelowitz}},
  \bibinfo{author}{\bibfnamefont{M.}~\bibnamefont{Rosenbluh}},
  \bibinfo{author}{\bibfnamefont{L.}~\bibnamefont{Khaykovich}},
  \bibinfo{author}{\bibfnamefont{W.}~\bibnamefont{Kinzel}}, \bibnamefont{and}
  \bibinfo{author}{\bibfnamefont{I.}~\bibnamefont{Kanter}},
  \bibinfo{journal}{Phys. Rev. E} \textbf{\bibinfo{volume}{74}},
  \bibinfo{pages}{046201} (\bibinfo{year}{2006}{\natexlab{b}}).

\bibitem[{\citenamefont{Kanter et~al.}(2007)\citenamefont{Kanter, Gross, Klein,
  Kopelowitz, Yoskovits, Khaykovich, Kinzel, and Rosenbluh}}]{Kanter:154101}
\bibinfo{author}{\bibfnamefont{I.}~\bibnamefont{Kanter}},
  \bibinfo{author}{\bibfnamefont{N.}~\bibnamefont{Gross}},
  \bibinfo{author}{\bibfnamefont{E.}~\bibnamefont{Klein}},
  \bibinfo{author}{\bibfnamefont{E.}~\bibnamefont{Kopelowitz}},
  \bibinfo{author}{\bibfnamefont{P.}~\bibnamefont{Yoskovits}},
  \bibinfo{author}{\bibfnamefont{L.}~\bibnamefont{Khaykovich}},
  \bibinfo{author}{\bibfnamefont{W.}~\bibnamefont{Kinzel}}, \bibnamefont{and}
  \bibinfo{author}{\bibfnamefont{M.}~\bibnamefont{Rosenbluh}},
  \bibinfo{journal}{Phys. Rev. Lett.} \textbf{\bibinfo{volume}{98}},
  \bibinfo{eid}{154101} (\bibinfo{year}{2007}).

\bibitem[{\citenamefont{Kanter et~al.}(2010{\natexlab{b}})\citenamefont{Kanter,
  Butkovski, Peleg, Zigzag, Aviad, Reidler, Rosenbluh, and
  Kinzel}}]{KanterButkovski}
\bibinfo{author}{\bibfnamefont{I.}~\bibnamefont{Kanter}},
  \bibinfo{author}{\bibfnamefont{M.}~\bibnamefont{Butkovski}},
  \bibinfo{author}{\bibfnamefont{Y.}~\bibnamefont{Peleg}},
  \bibinfo{author}{\bibfnamefont{M.}~\bibnamefont{Zigzag}},
  \bibinfo{author}{\bibfnamefont{Y.}~\bibnamefont{Aviad}},
  \bibinfo{author}{\bibfnamefont{I.}~\bibnamefont{Reidler}},
  \bibinfo{author}{\bibfnamefont{M.}~\bibnamefont{Rosenbluh}},
  \bibnamefont{and} \bibinfo{author}{\bibfnamefont{W.}~\bibnamefont{Kinzel}},
  \bibinfo{journal}{Opt. Express} \textbf{\bibinfo{volume}{18}},
  \bibinfo{pages}{18292} (\bibinfo{year}{2010}{\natexlab{b}}).

\bibitem[{\citenamefont{{Englert} et~al.}(2010)\citenamefont{{Englert},
  {Kinzel}, {Aviad}, {Butkovski}, {Reidler}, {Zigzag}, {Kanter}, and
  {Rosenbluh}}}]{EnglertKinzelAviad}
\bibinfo{author}{\bibfnamefont{A.}~\bibnamefont{{Englert}}},
  \bibinfo{author}{\bibfnamefont{W.}~\bibnamefont{{Kinzel}}},
  \bibinfo{author}{\bibfnamefont{Y.}~\bibnamefont{{Aviad}}},
  \bibinfo{author}{\bibfnamefont{M.}~\bibnamefont{{Butkovski}}},
  \bibinfo{author}{\bibfnamefont{I.}~\bibnamefont{{Reidler}}},
  \bibinfo{author}{\bibfnamefont{M.}~\bibnamefont{{Zigzag}}},
  \bibinfo{author}{\bibfnamefont{I.}~\bibnamefont{{Kanter}}}, \bibnamefont{and}
  \bibinfo{author}{\bibfnamefont{M.}~\bibnamefont{{Rosenbluh}}},
  \bibinfo{journal}{Phys. Rev. Lett.} \textbf{\bibinfo{volume}{104}},
  \bibinfo{pages}{114102} (\bibinfo{year}{2010}).

\bibitem[{\citenamefont{{Aviad} et~al.}(2008)\citenamefont{{Aviad}, {Reidler},
  {Kinzel}, {Kanter}, and {Rosenbluh}}}]{AviadReidler}
\bibinfo{author}{\bibfnamefont{Y.}~\bibnamefont{{Aviad}}},
  \bibinfo{author}{\bibfnamefont{I.}~\bibnamefont{{Reidler}}},
  \bibinfo{author}{\bibfnamefont{W.}~\bibnamefont{{Kinzel}}},
  \bibinfo{author}{\bibfnamefont{I.}~\bibnamefont{{Kanter}}}, \bibnamefont{and}
  \bibinfo{author}{\bibfnamefont{M.}~\bibnamefont{{Rosenbluh}}},
  \bibinfo{journal}{Phys. Rev. E} \textbf{\bibinfo{volume}{78}},
  \bibinfo{pages}{025204} (\bibinfo{year}{2008}).

\bibitem[{\citenamefont{{Ahlers} et~al.}(1998)\citenamefont{{Ahlers},
  {Parlitz}, and {Lauterborn}}}]{AhlersParlitz}
\bibinfo{author}{\bibfnamefont{V.}~\bibnamefont{{Ahlers}}},
  \bibinfo{author}{\bibfnamefont{U.}~\bibnamefont{{Parlitz}}},
  \bibnamefont{and}
  \bibinfo{author}{\bibfnamefont{W.}~\bibnamefont{{Lauterborn}}},
  \bibinfo{journal}{Phys. Rev. E} \textbf{\bibinfo{volume}{58}},
  \bibinfo{pages}{7208} (\bibinfo{year}{1998}).

\bibitem[{\citenamefont{{Heun}}(1900)}]{Heun}
\bibinfo{author}{\bibfnamefont{K.}~\bibnamefont{{Heun}}}, \bibinfo{journal}{Z.
  Math. Phys} \textbf{\bibinfo{volume}{45}}, \bibinfo{pages}{23}
  (\bibinfo{year}{1900}).

\end{thebibliography}

\end{document}